\documentclass{aa}

\usepackage{graphicx}
\usepackage[varg]{txfonts}
\usepackage{lscape}

\usepackage{natbib,twoopt}
\usepackage[breaklinks=true]{hyperref} 
\bibpunct{(}{)}{;}{a}{}{,} 
\makeatletter
\newcommandtwoopt{\citeads}[3][][]{\href{https://adsabs.harvard.edu/abs/#3}%
{\def\hyper@linkstart##1##2{}%
\let\hyper@linkend\@empty\citealp[#1][#2]{#3}}}
\newcommandtwoopt{\citetads}[3][][]{\href{https://adsabs.harvard.edu/abs/#3}%
{\def\hyper@linkstart##1##2{}%
\let\hyper@linkend\@empty\citet[#1][#2]{#3}}}
\newcommandtwoopt{\citepads}[3][][]{\href{https://adsabs.harvard.edu/abs/#3}%
{\def\hyper@linkstart##1##2{}%
\let\hyper@linkend\@empty\citep[#1][#2]{#3}}}
\newcommandtwoopt{\citeyearads}[3][][]%
{\href{http://adsabs.harvard.edu/abs/#3}
{\def\hyper@linkstart##1##2{}%
\let\hyper@linkend\@empty\citeyear[#1][#2]{#3}}}
\newcommandtwoopt{\citealtads}[3][][]{\href{https://adsabs.harvard.edu/abs/#3}%
{\def\hyper@linkstart##1##2{}%
\let\hyper@linkend\@empty#1\citealt{#3}#2}}
\makeatother

\begin{document}

   \title{The \textit{XMM-Newton} serendipitous survey}

   \subtitle{X: The second source catalogue from overlapping
     \textit{XMM-Newton} observations and its long-term variable
     content\thanks{Based on observations obtained with \textit{XMM-Newton},
       an ESA science mission with instruments and contributions directly
       funded by ESA Member States and NASA.}\fnmsep\thanks{The catalogue is
       available in FITS format via the SSC web pages at
       \url{http://xmmssc.irap.omp.eu} and \url{https://xmmssc.aip.de} and
       searchable via CDS
       \url{https://vizier.u-strasbg.fr/viz-bin/VizieR?-source=IX/}\emph{will-be-added},
       XSA \url{https://www.cosmos.esa.int/web/xmm-newton/xsa}, and HEASARC
       \url{https://heasarc.gsfc.nasa.gov/W3Browse/xmm-newton/xmmstack.html}
       data services.}}

   \titlerunning{The \textit{XMM-Newton} serendipitous survey. X: The second
     source catalogue from overlapping observations}

   \author{I.\ Traulsen\inst{1}
          \and A.\ D.\ Schwope\inst{1}
          \and G.\ Lamer\inst{1}
          \and J.\ Ballet\inst{2}
          \and F.\ J.\ Carrera\inst{3}
          \and M.\ T.\ Ceballos\inst{3}
          \and M.\ Coriat\inst{4}
          \and M.\ J.\ Freyberg\inst{5}
          \and F.\ Koliopanos\inst{4}
          \and J.\ Kurpas\inst{1}
          \and L.\ Michel\inst{6}
          \and C.\ Motch\inst{6}
          \and M.\ J.\ Page\inst{7}
          \and M.\ G.\ Watson\inst{8}
          \and N.\ A.\ Webb\inst{4}
          }

   \institute{Leibniz-Institut f\"ur Astrophysik Potsdam (AIP), An der
             Sternwarte 16, 14482 Potsdam, Germany\\
              \email{itraulsen@aip.de}
         \and    
             AIM, CEA, CNRS, Universit\'e Paris-Saclay, Universit\'e Paris
             Diderot, Sorbonne Paris Cit\'e, 91191 Gif-sur-Yvette, France
         \and    
             Instituto de F\'isica de Cantabria (CSIC-UC), Avenida de los
             Castros, 39005 Santander, Spain
         \and    
             IRAP, Universit\'e de Toulouse, CNRS, UPS, CNES, 9 Avenue du
             Colonel Roche, BP 44346, 31028 Toulouse Cedex 4, France
         \and    
             Max-Planck-Institut f\"ur extraterrestrische Physik,
             Giessenbachstra{\ss}e 1, 85748 Garching, Germany
         \and    
             Observatoire astronomique, Universit\'e de Strasbourg, CNRS, UMR
             7550, 11 rue de l’Universit\'e, 67000 Strasbourg, France
         \and    
              Mullard Space Science Laboratory, University College London,
              Holbury St Mary, Dorking, Surrey RH5 6NT, UK
         \and    
             Department of Physics \& Astronomy, University of Leicester,
             Leicester, LE1 7RH, UK
             }

   \date{Received February 11, 2020; accepted June 26, 2020}


\abstract
  {The \textit{XMM-Newton} Survey Science Centre Consortium (SSC) develops
    software in close collaboration with the Science Operations Centre to
    perform a pipeline analysis of all \textit{XMM-Newton} observations. In
    celebration of the twentieth anniversary of the \textit{XMM-Newton}
    launch, the SSC has compiled the fourth generation of serendipitous source
    catalogues, 4XMM.}
  {The catalogue described here, 4XMM-DR9s, explores sky areas that were
    observed more than once by \textit{XMM-Newton}. These observations are
    bundled in groups referred to as stacks. Stacking leads to a higher
    sensitivity, resulting in newly discovered sources and better constrained
    source parameters, and unveils long-term brightness variations.}
  {The 4XMM-DR9s catalogue was constructed from simultaneous source detection
    on overlapping observations. As a novel feature, positional rectification
    was applied beforehand. Observations with all filters and suitable camera
    settings were included. Exposures with a high background were
    discarded. The high-background thresholds were determined through a
    statistical analysis of all exposures in each instrument
    configuration. The X-ray background maps used in source detection were
    modelled via an adaptive smoothing procedure with newly determined
    parameters. Source fluxes were derived for all contributing observations,
    irrespective of whether the source would be detectable in an individual
    observation.}
  {The new catalogue lists the X-ray sources detected in 1\,329 stacks with
    6\,604 contributing observations over repeatedly covered 300 square
    degrees in the sky. Most stacks are composed of two observations, the
    largest one comprises 352 observations. We find 288\,191 sources of which
    218\,283 were observed several times. The number of observations of a
    source ranges from 1 to 40. Auxiliary products, like X-ray full-band and
    false-colour images, long-term X-ray light curves, and optical finding
    charts, are published as well.}
  {4XMM-DR9s contains new detections and is considered a prime resource to
    explore long-term variability of X-ray sources discovered by
    \textit{XMM-Newton}. Regular incremental releases, including new public
    observations, are planned.}


\maketitle


\section{Introduction}
\label{sec:introduction}

  Since its launch in December 1999, ESA's space-based X-ray telescope
  \textit{XMM-Newton} \citepads{2001A&A...365L...1J} has observed about 1\,150
  square degrees of the sky in pointed and mosaic mode, and about a third of
  that area several times. From all public X-ray data taken with its
  EPIC\footnote{European Photon Imaging Camera} CCD instruments MOS1, MOS2
  \citepads{2001A&A...365L..27T}, and pn \citepads{2001A&A...365L..18S}, the
  \textit{XMM-Newton} Survey Science Centre Consortium \citepads[SSC,
  ][]{2001A&A...365L..51W} compiles the serendipitous source catalogues. Their
  third generation named 3XMM \citepads{2016A&A...590A...1R} was augmented by
  the first source catalogue from overlapping observations, 3XMM-DR7s
  \citepads{2019A&A...624A..77T}. Other major source catalogues based on
  \textit{XMM-Newton} observations include the \textit{XMM-Newton} Slew Survey
  Source Catalogue \citepads{2008A&A...480..611S} of X-ray sources observed by
  EPIC pn during telescope slews with its latest edition
  XMMSL2\footnote{\url{https://www.cosmos.esa.int/web/xmm-newton/xmmsl2-ug}}
  and the OM Serendipitous Ultraviolet Source Survey Catalogue
  \citepads{2012MNRAS.426..903P} of sources observed by the Optical Monitor
  \citepads{2001A&A...365L..36M} with its latest edition
  SUSS4.1\footnote{\url{https://www.cosmos.esa.int/web/xmm-newton/om-catalogue}}.

  After upgrades to the catalogue pipelines and the underlying
  SAS\footnote{\textit{XMM-Newton} Science Analysis System,
    \url{https://www.cosmos.esa.int/web/xmm-newton/sas}.} tasks, all
  \textit{XMM-Newton} EPIC data were fully re-processed\footnote{The pipeline
    version ppsprod-18.0 incorporates tasks from SAS 18.0.0.} by the
  \textit{XMM-Newton} Science Operations Centre (SOC) at the European Space
  Astronomy Centre (ESAC). Based on these data, the SSC compiled the fourth
  generation of serendipitous source catalogues. The new 4XMM-DR9 catalogue
  from all individual observations is described by \citet{2020dr9} and the
  second catalogue from overlapping observations, named 4XMM-DR9s, is explored
  in this work. To produce 4XMM-DR9s, observations that overlap by at least
  1\arcmin\ in radius are grouped into stacks and processed together. Source
  detection is performed simultaneously in all images of all observations in a
  stack. The number of input images for a detection run is thus given by the
  total number of exposures\footnote{The term `observation' is used for a full
    \textit{XMM-Newton} pointing and all its data products. It comprises one
    or more `exposure(s)'. An exposure is taken by one of the EPIC instruments
    pn, MOS1, MOS2.} for this stack times the number of energy bands (which is
  five in the serendipitous source catalogues).

  This paper on the second catalogue from overlapping \textit{XMM-Newton}
  observations describes the updates with respect to
  \citetads{2019A&A...624A..77T} to the underlying software and the respective
  processing parameters in Section~\ref{sec:software}. In particular, an
  astrometric field rectification prior to stacked source detection is new to
  this catalogue and introduced in Section~\ref{sec:astcorr}. Updates to the
  field selection and the catalogue processing are described in
  Section~\ref{sec:obssel}. Section~\ref{sec:catalogue} is dedicated to the
  structure of 4XMM-DR9s, the revised and newly introduced columns, and the
  cross-match with 4XMM-DR9. Positional accuracy, source content and
  information in comparison to 4XMM-DR9, and long-term variability of
  catalogue sources are studied in
  Section~\ref{sec:results}. Section~\ref{sec:summary} summarises the results.

\section{Updates to data processing and source detection}
\label{sec:software}

  \begin{figure}
    \centering
    \includegraphics[width=\linewidth]{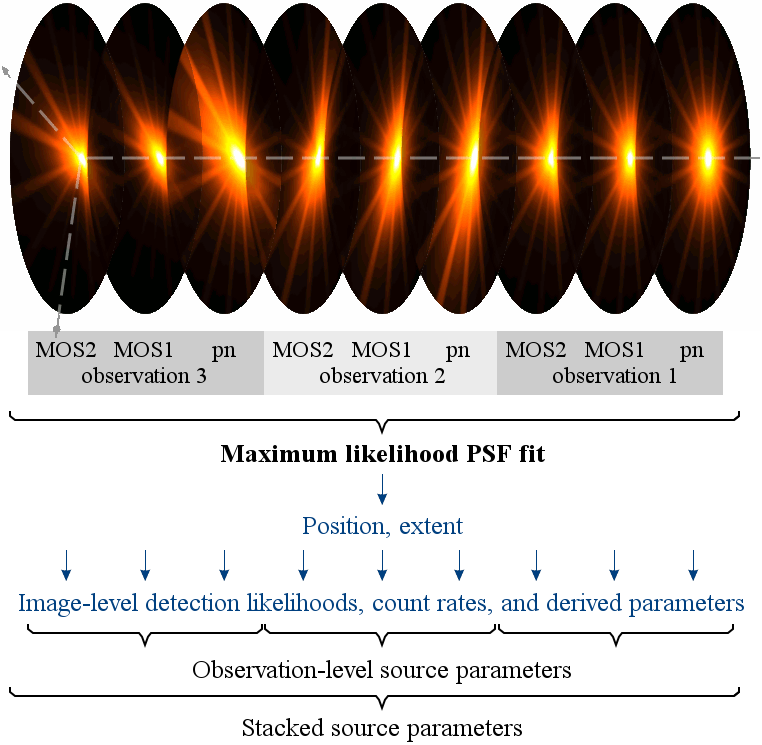}
    \caption{Illustration of the PSF shapes in the three EPIC instruments at
      different off-axis and azimuth angles in overlapping observations and of
      the parameter determination in the simultaneous maximum-likelihood PSF
      fit. The normalised PSFs -- shown here for 1\,keV in a linear intensity
      scale -- are scaled to the count rate in the respective image within the
      detection fit. Position and extent of a source are the same in all
      observations, while the count rates are fitted in each
      image. Observation-level and stacked parameters are derived from the
      parameters of the contributing images \citepads{2019A&A...624A..77T}.}
    \label{fig:psfstack}
  \end{figure}

  The data processing for 4XMM-DR9s was based on the event lists and attitude
  files produced for 4XMM-DR9 and essentially followed the strategy of
  \citetads{2019A&A...624A..77T}. Mosaic-mode observations, which consist of a
  series of exposures, were split into several observation identifiers by the
  pipeline, creating one for each sub-pointing
  \citepads[cf.][]{2016A&A...590A...1R}. They were thus treated like
  independently overlapping observations by stacked source detection. The data
  of each stack were projected tangentially onto the same reference
  coordinates with respect to the centre of the full area covered by all
  observations in the stack. The centre was calculated from the attitude
  coordinates of the contributing observations as the mean of minimum and
  maximum right ascension and declination. Input products for source detection
  were created for each observation, consisting of images, background maps,
  exposure maps, and detection masks.

  The source-detection method was described in detail by
  \citetads{2019A&A...624A..77T}. Images, exposure and background maps were
  created for each exposure and energy band, and detection masks were created
  for each exposure. They were used in the two-step source detection process,
  which was simultaneously run on all images in the stack. Sliding-box source
  detection was performed to create an input list of tentative source positions
  for the main step, a maximum-likelihood PSF fit in all images of the
  contributing observations, instruments, and energy bands. Using the three
  EPIC CCD instruments pn, MOS1, MOS2, and the five \textit{XMM-Newton}
  standard energy bands\footnote{Band 1: $0.2-0.5$\,keV, band 2:
    $0.5-1.0$\,keV, band 3: $1.0-2.0$\,keV, band 4: $2.0-4.5$\,keV, band 5:
    $4.5-12.0$\,keV, full band: $0.2-12.0$\,keV.}, the fit thus involved five,
  ten, or fifteen images for each observation. The point-spread functions
  (PSFs) at the tentative source position and mid-band energy were read
  (cf.\ sketch in Fig.~\ref{fig:psfstack}), taking the different PSF shapes
  and distortions in the individual contributing images into account. The 2d
  model of the point-spread functions was derived by
  \citetads{2011A&A...534A..34R} and involves a Gaussian core, a King
  component, and the primary and secondary spokes that arise from the mirror
  structure and scattering. In the simultaneous maximum-likelihood PSF fit,
  the count rate in each image and a common source position and source extent
  were determined. Position and extent were considered to be the same in all
  images since they are not expected to vary between observations, while the
  count rate and all derived parameters were determined in each image
  individually. A detection likelihood was calculated from the count rate under
  the PSF using Cash statistics \citepads{1979ApJ...228..939C} and the null
  hypothesis that the measured signal arises from pure background
  fluctuations. Source extent was tested by convolving the point-like PSF with
  a beta model, whose radius was a free parameter of the fit. A minimum extent
  radius of 6\arcsec\ and a log-likelihood difference of at least four between
  the fits with the extended and the point-like PSF were required to accept the
  extended fit. Otherwise, the source was considered point-like. All source
  parameters, in particular the detection likelihood, were derived from the
  combined fit. The fit results of all images were used to calculate the
  all-stack source parameters. Observation-level parameters were calculated
  from the same fit results by using the subset of the images of each
  observation separately (Fig.~\ref{fig:psfstack}). Comprehensive parameter
  lists are given in the Appendix of \citetads{2019A&A...624A..77T} and of
  this paper. Its structure is also shown in
  Fig.~\ref{fig:catview_topcat}. The catalogue includes sources which have a
  detection likelihood of at least six in the total stack or in a contributing
  observation. The likelihood threshold of six is used in all
  \textit{XMM-Newton} serendipitous source catalogues.

  \subsection{Task updates}
  \label{sec:taskupdates}

  Several source-detection SAS tasks have been revised since publication of
  the first catalogue 3XMM-DR7s, in particular
  \texttt{esplinemap}\footnote{The task is still named after its initial
    functionality, but used in its new smoothing mode
    \citepads{2019A&A...624A..77T}.} producing the background maps,
  \texttt{eboxdetect} for the first sliding-box detection step,
  \texttt{emldetect} performing the maximum-likelihood source detection, and
  \texttt{srcmatch}, which was used to create merged 4XMM-DR9s input source
  lists and the final 4XMM-DR9 source lists. The most relevant changes are
  described in this Section. More details on all updates are included in the
  task history of the SAS packages, which is for example part of the SAS
  release
  notes\footnote{\url{https://www.cosmos.esa.int/web/xmm-newton/sas-release-notes/}}.

  For all 4XMM catalogues (DR9 and DR9s), a pile-up estimate has been
  introduced \citep{2020dr9}. It is calculated for each detection and EPIC
  instrument within \texttt{emldetect}. Also, \texttt{emldetect} gives the
  extent likelihood now for all sources, including those fitted with a
  point-like PSF (cf.\ Sect.~\ref{sec:newcolumns} and \citealt{2020dr9}).

  To handle large stacks, runtime and memory requirements of several tasks
  were reduced, in particular for \texttt{esplinemap} and
  \texttt{emldetect}. Runtime is crucial for very deep stacks with many
  directly overlapping observations. Memory is crucial for very extended
  stacks containing observations that cover several square degrees on the sky
  in total. Depending on the depth and size of the stack, \texttt{emldetect}
  for example consumes up to 30\,\% less memory after the revision and becomes
  faster by 20\,\% to 90\,\%. These enhancements and new functionalities
  become public with the release of SAS\,19.

  In addition, SSC-internal versions of \texttt{eboxdetect},
  \texttt{emldetect}, and \texttt{srcmatch} were established which can process
  large stacks of up to 400 observations. They are automatically chosen by the
  catalogue pipeline whenever necessary. These updates and the pipeline are
  not part of the public SAS releases, since compiler options deviating from
  SAS standards are used to produce them and hardware beyond standard PCs is
  required to run them.


  \subsection{Background modelling: new default parameters for all
    4XMM catalogues}
  \label{sec:bkgparams}

  To determine the background at the tentative source positions during source
  detection, a background map was created for each input image and the
  modelled signal of out-of-time events added to the EPIC-pn
  maps\footnote{\url{https://xmm-tools.cosmos.esa.int/external/sas/current/doc/esplinemap}}.
  The background structure was modelled based on the source-free image
  regions. For catalogues prior to 3XMM-DR7s, it was approximated with a
  spline fit. \citetads{2019A&A...624A..77T} employed an adaptive smoothing
  technique for the first catalogue from overlapping observations, which was
  created from good-quality data. The exposure-normalised images were
  convolved with a Gaussian kernel and divided by the detection masks which
  were smoothed using the same kernel. For increasing kernel widths, the
  signal-to-noise ratio in each pixel was derived from the counts under the
  kernel. The background value for which the intended signal-to-noise was
  reached was written to the background map. Parameters of the adaptive
  smoothing method are the signal-to-noise value, the minimum kernel width,
  and the brightness threshold above which sources are excised from the input
  image. This approach has now been used for all 4XMM catalogues
  consistently. To work with single and stacked observations of very different
  background levels, the smoothing parameters were re-investigated as
  described in \citet{2020dr9}. A brightness threshold for the source cut-out
  radius of $2\times 10^{-4}\,\textrm{counts}\,\textrm{arcsec}^{-2}$, a
  minimum smoothing radius of 10 pixels (40\arcsec\ in default image binning),
  and a signal-to-noise ratio of 12 were established for 4XMM-DR9 and for
  4XMM-DR9s.

  \subsection{Event-based astrometric correction}
  \label{sec:astcorr}

  \begin{figure}
    \centering
    \includegraphics[height=.592\linewidth]{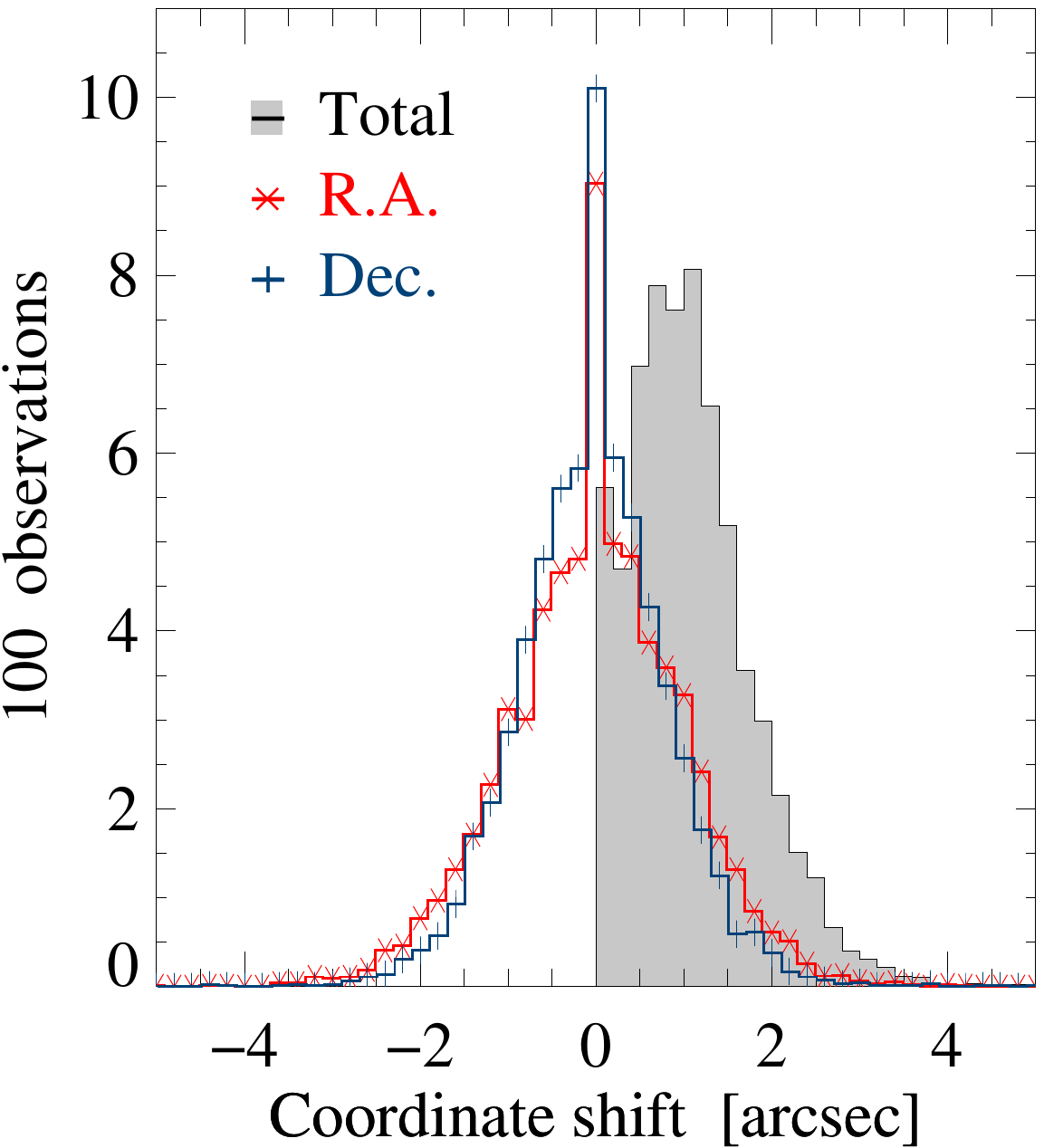}\hfill%
    \includegraphics[height=.592\linewidth]{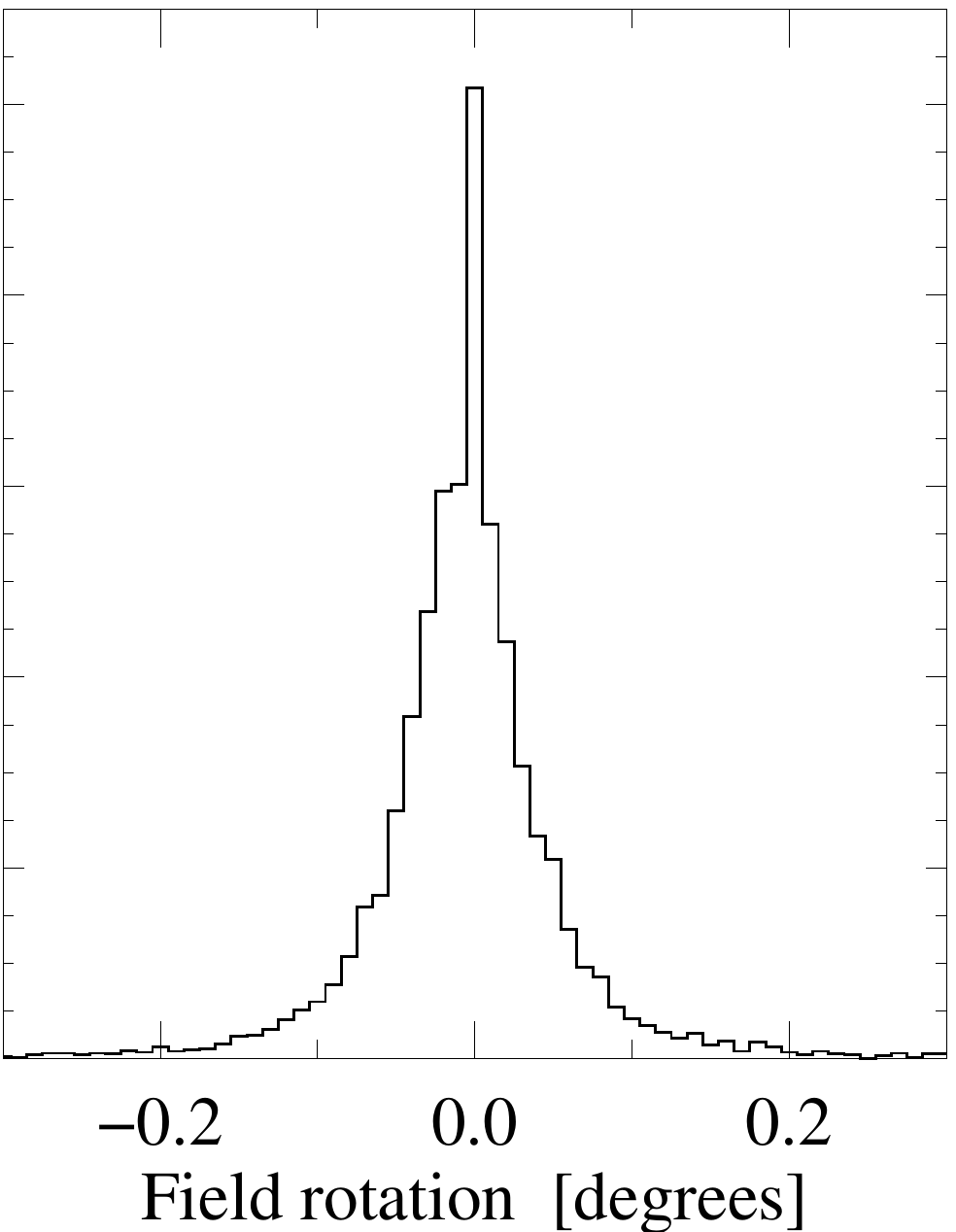}
    \caption{\texttt{catcorr} results for the input fields to 4XMM-DR9s which
      could be astrometrically corrected: histograms of the applied shifts
      along right ascension (\emph{left, red crosses,} $\Delta$R.A. $\times$
      cos(Dec.)) and declination (\emph{left, blue pluses,} $\Delta$Dec.), the
      additional field rotation \emph{(right)}, and the resulting absolute
      field shift \emph{(left, black)}.}
    \label{fig:catcorr}
  \end{figure}

  As described in \citetads{2016A&A...590A...1R}, celestial coordinates of
  sources emerging from the PSF fitting step of a given observation include a
  generally small systematic error arising from offsets in the spacecraft
  boresight position from the nominal pointing direction for the
  observation. Such systematics were removed from source lists of individual
  observations by applying the SAS task \texttt{catcorr} which reveals shifts
  in RA and DEC and a field rotation angle via probabilistic matching of X-ray
  sources with sources listed in SDSS \citepads{2012ApJS..203...21A},
  USNO-B1.0 \citepads{2003AJ....125..984M}, or 2MASS
  \citepads{2006AJ....131.1163S}. The procedure for 4XMM-DR9 is described in
  \citet{2020dr9}.

  When several overlapping observations are to be processed, such corrections
  can only be applied before simultaneous source detection and not at the end
  of the processing, because individual OBS\_IDs have their own
  \texttt{catcorr} corrections. We therefore introduced an event-based
  astrometric field rectification during pre-processing, shifting the recorded
  events to corrected positions. The \texttt{catcorr} results per observation,
  derived within the 4XMM-DR9 processing, were directly applied to the
  relevant header keywords of the attitude files and to the attitude
  coordinates, which were later used to project the event lists onto the
  common coordinate system of the stack. Stacked source detection was then
  performed on the individually modified event lists. The distribution of the
  shifts applied to the 4XMM-DR9s input observations is shown in
  Fig.~\ref{fig:catcorr}. The plots include the results from 2\,516 fields
  that were corrected against SDSS positions, 2\,267 fields against USNO-B1.0
  positions, and 842 against 2MASS positions. The astrometric accuracy of all
  three comparison catalogues is of the order of 0.1\arcsec\ to
  0.2\arcsec. For 4XMM-DR9 observations that could not be corrected, a mean
  systematic astrometric uncertainty of about 1.3\arcsec\ remains
  \citep{2020dr9}.

\begin{figure}
    \centering
    \includegraphics[width=.8\linewidth]{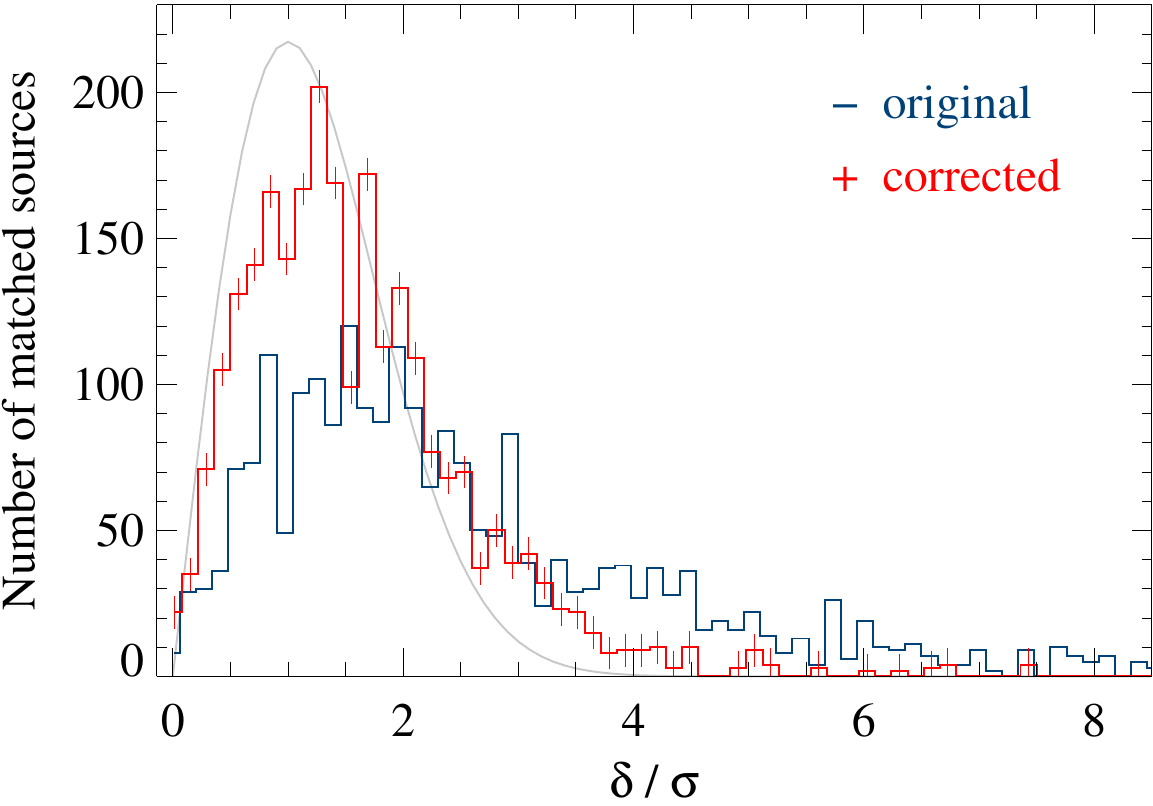}
    \caption{Distribution of the angular distances, given in units of the
      statistical error, between X-ray and optical SDSS positions before
      \emph{(blue)} and after \emph{(red)} event-based field rectification in
      a test sample of 25 stacks. The grey curve shows an ideal Rayleigh
      distribution.}
    \label{fig:astcorr}
  \end{figure}

  Before compiling the full 4XMM-DR9s, we verified the approach on a test
  sample of 25 stacks with large \texttt{catcorr} shifts. They comprised two
  to six observations with total offsets of up to 3.8\arcsec.
  Figure~\ref{fig:astcorr} shows the differences between X-ray and optical
  SDSS positions in the test sample with and without positional rectification
  (matching radius 3\arcsec). The success of the chosen approach becomes
  rather obvious (on the remaining deviations from an ideal Rayleigh
  distribution, cf.\ Sect.~\ref{sec:astrometry}). The number of matches and
  the positional accuracy of the X-ray sources from stacked source detection
  both increase. The improvement in positional accuracy was independently
  confirmed through a match with GAIA-DR2 \citepads{2018A&A...616A...1G} which
  was not used by \texttt{catcorr}, again within 3\arcsec.
 
\section{Field selection and catalogue processing}
\label{sec:obssel}

  \subsection{Determination of the background cut}
  \label{sec:bkgcut}

  \begin{figure}
    \includegraphics[width=\linewidth]{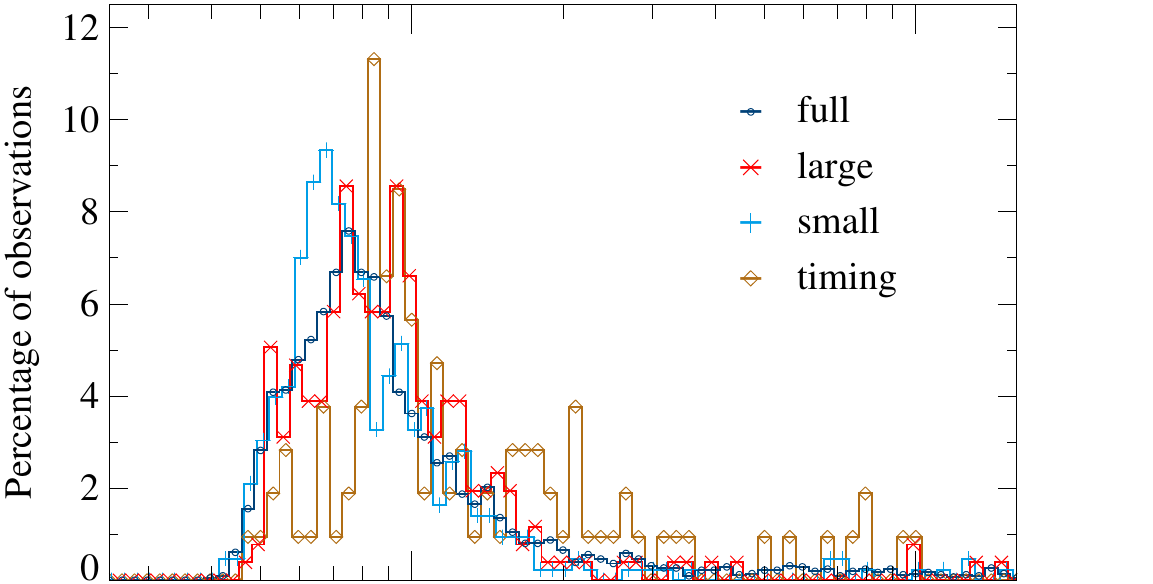}
    \includegraphics[width=\linewidth]{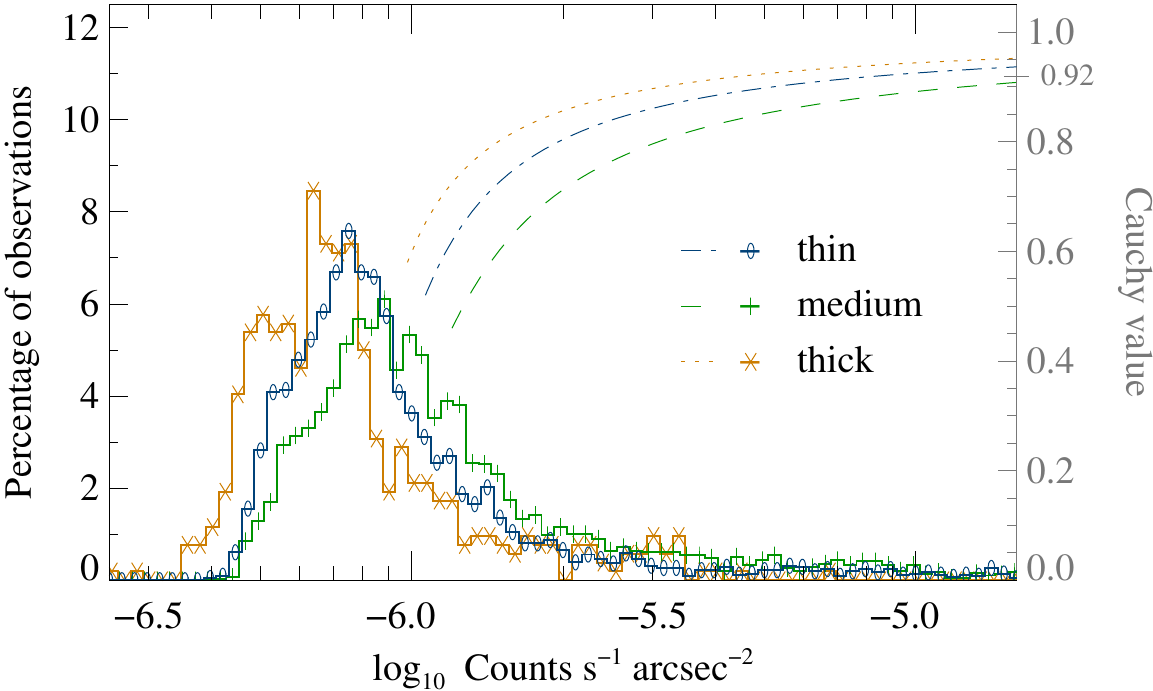}
    \caption{MOS2 background levels between 0.2 and 12.0\,keV for the
      observing modes full frame, large window, small window, and timing in
      thin filter \emph{(upper panel)} and for the thin, medium, and thick
      filter in full-frame mode \emph{(lower panel)}. The dashed and dotted
      lines show the upper part of the corresponding Cauchy probability values
      for the different filters.}
    \label{fig:bkglevels}
  \end{figure}

  Exposures affected by high background emission may reduce the detection
  reliability and the accuracy of the source parameters in a stack. In extreme
  cases, the number of tentative input sources to the maximum-likelihood
  fitting, which are selected by a sliding-box detection algorithm, may even
  become too large to process the stack at all. As for 3XMM-DR7s, we thus
  defined background cuts above which an exposure is discarded from stacked
  source detection. They were determined following the general method outlined
  in \citetads{2019A&A...624A..77T}, now derived individually for each camera,
  each filter, and each observation mode. This method was employed to select
  or discard exposures from the catalogue sample. The background maps used in
  source detection are described in Sect.~\ref{sec:bkgparams}.

  In the source-excised full-band images of all 4XMM-DR9 exposures, we
  determined the mean background count rate per unit area. For EPIC pn, the
  rate was calculated for each quadrant separately, where one quadrant
  comprises three CCDs. For EPIC MOS, it was calculated for each CCD
  separately except for the central one. Usually, the target of the
  observation is placed on this central CCD, and if a large area had to be cut
  out around a bright or extended target, the remaining area might be too
  small for a reliable background determination. Correspondingly, the other
  CCDs were considered in the analysis if at least 100 valid pixels remained
  after masking the sources. The maximum rate among all usable pn quadrants or
  MOS CCDs was chosen in each exposure and the distribution of background
  rates derived for each filter-mode combination. While the results for the
  three imaging modes, which were used in the catalogues, were similar within
  their uncertainties, clear discrepancies were found between imaging mode and
  timing mode and between exposures obtained with different filters
  (Fig.~\ref{fig:bkglevels}). An empirically chosen Lorentz function
  \citepads{2019A&A...624A..77T} was fitted to the rate distribution of all
  exposures taken in the same imaging mode and filter. In MOS timing mode, too
  few exposures were available to fit the filters individually, thus one fit
  was used for all of them. Since count rates are instrument-dependent and
  cannot be compared directly if taken in different technical setups, we opted
  for an independent quantity to characterise the background level. From the
  Lorentz functions, we thus derived the associated Cauchy cumulative
  probability distributions (examples shown in the lower panel of
  Fig.~\ref{fig:bkglevels}). The Cauchy probability attributed to an exposure
  then serves as a configuration-independent measure of the background level,
  which catalogue users can access to apply their own stricter restriction on
  the background level consistently without caring about technical details of
  the individual observations.

  To choose the final high-background cut, we selected sky areas which were
  observed repeatedly at very different background levels and performed source
  detection in the individual observations and their stacks. The usable
  sources at the different background levels, the sources in the stack, and
  those in the contributing observation with the lowest background were
  compared. From the results and from additional visual inspection of the
  images using a brightness scale depending on the exposure time, we chose a
  background cut of 0.92. Exposures with a higher Cauchy probability tend to
  degrade the source-detection quality and were rejected. The remaining
  observations are considered suitable for stacked source detection.

  \subsection{Exposure selection}
  \label{sec:finalfields}

  \begin{figure}
    \centering
    \includegraphics[width=\linewidth]{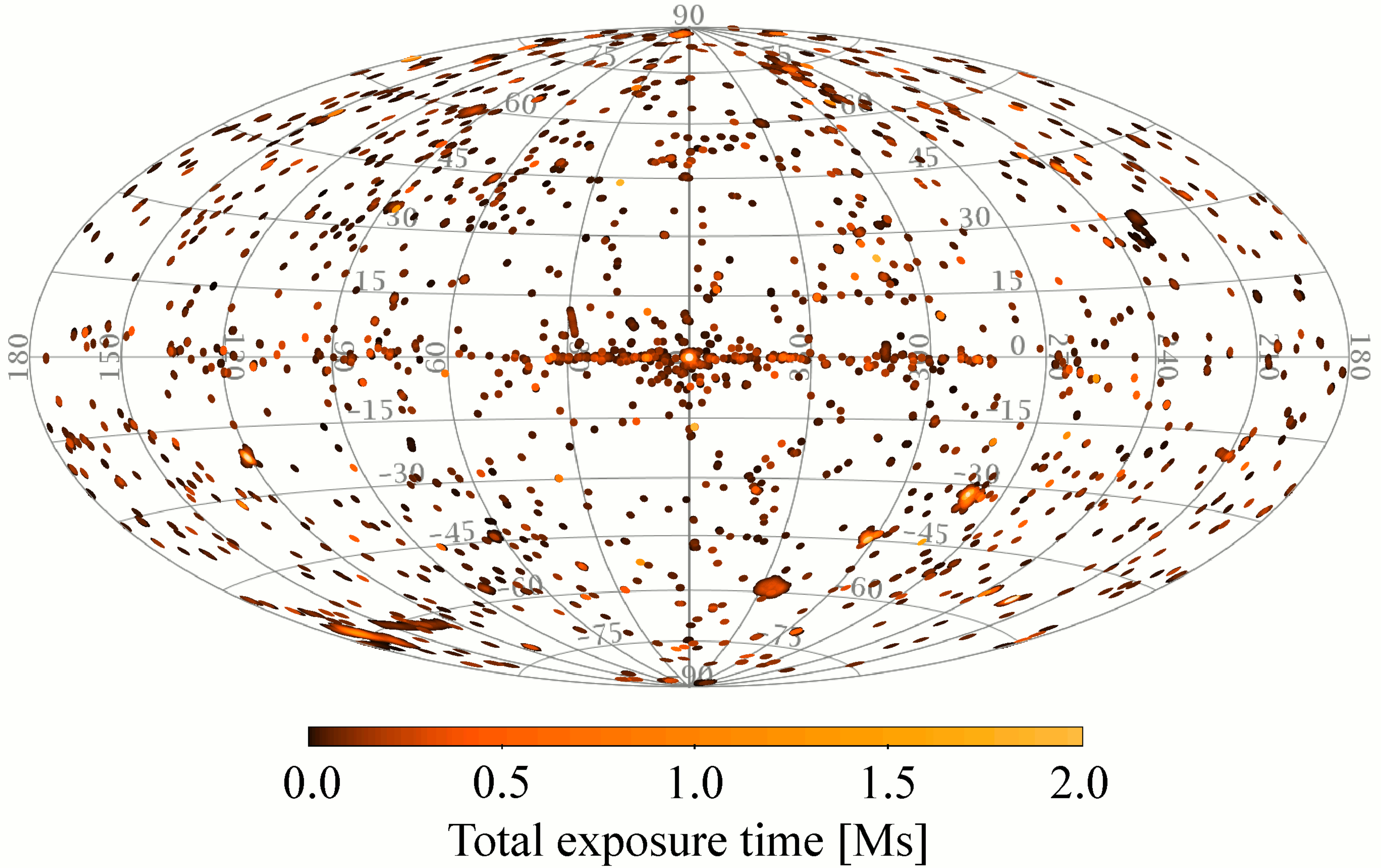}
    \caption{Sky map of the 4XMM-DR9s input observations in a Hammer-Aitoff
      projection in Galactic coordinates. The radius of the plot symbols
      corresponds to 2\degr, thus to four times the actual field-of-view of
      \textit{XMM-Newton}. Colour codes the total exposure time of the
      overlapping observations in each plotted bin.}
    \label{fig:skymap}
  \end{figure}

  4XMM-DR9s is based on a subset of all observations that entered 4XMM-DR9.
  Discarded from stacked source detection were: (1) observations with too
  small an overlap; (2) pn exposures in small window mode; (3) the central MOS
  CCDs of exposures in small window mode; and (4) exposures with too high a
  background (see Sect.~\ref{sec:bkgcut}).

  Observations were grouped into a stack if they overlap by at least
  1\arcmin\ in radius, which corresponds to the maximum radius of the fit
  region of a point source in the catalogue processing and ensures that the
  fit region is fully covered by the combined observations. Thus, observations
  were selected if their aim points have a maximum distance of 29\arcmin,
  referring to the mean field coordinates given in the headers of the 4XMM-DR9
  source lists. Because of the offset between the pn and MOS detectors, the
  area covered by EPIC instruments is slightly asymmetric. In terms of
  actually exposed area, 29 observations therefore overlap by less than
  1\arcmin\ with their neighbouring observations although formally meeting the
  29\arcmin\ criterion. These were de-selected manually. Another 88 candidate
  stacks which met the selection criteria (1) to (3) were discarded because
  the background level of all their observations or of all but one was too high.

\begin{figure}
    \centering
    \includegraphics[height=.505\linewidth]{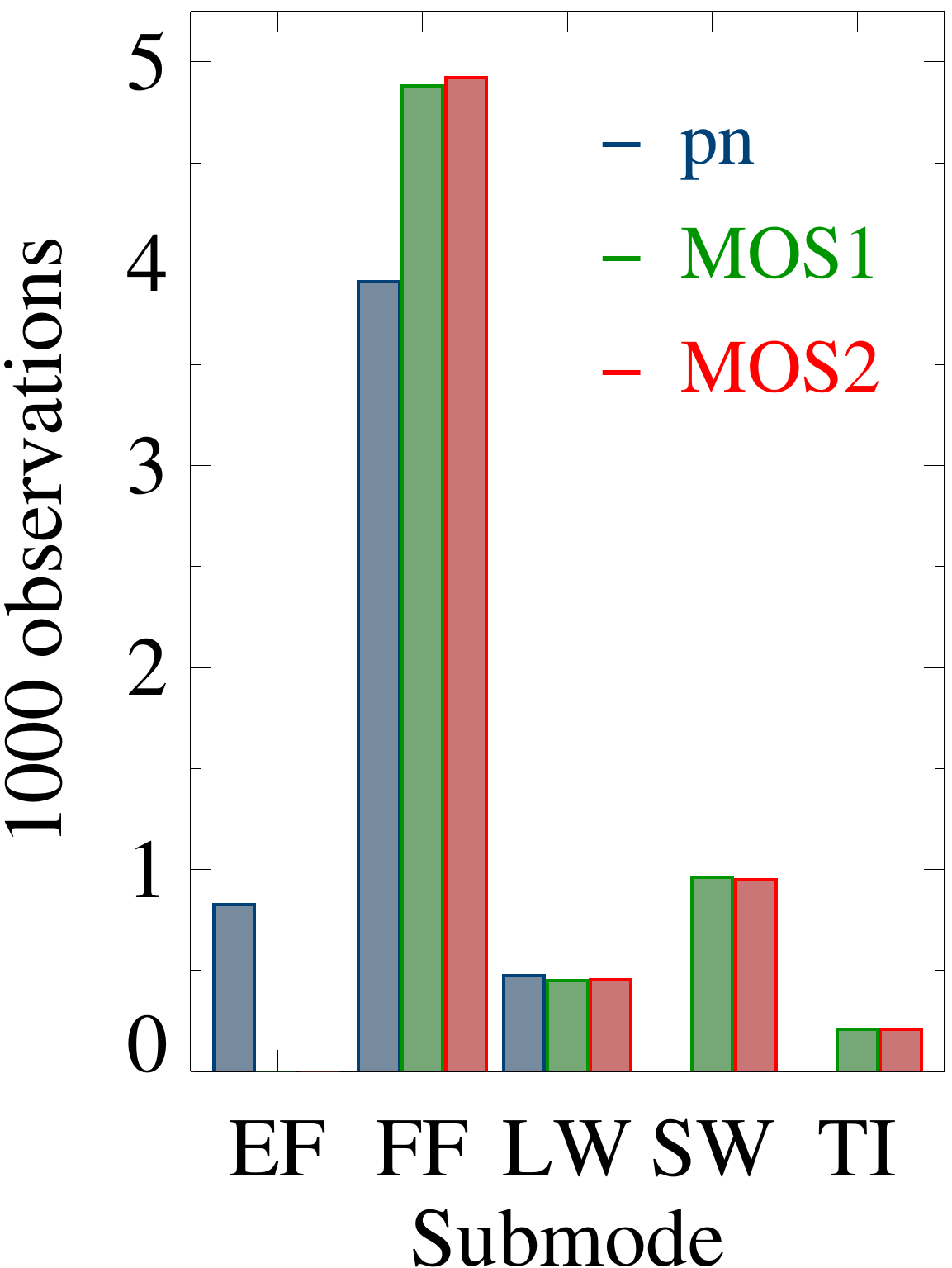}\hfill%
    \includegraphics[height=.505\linewidth]{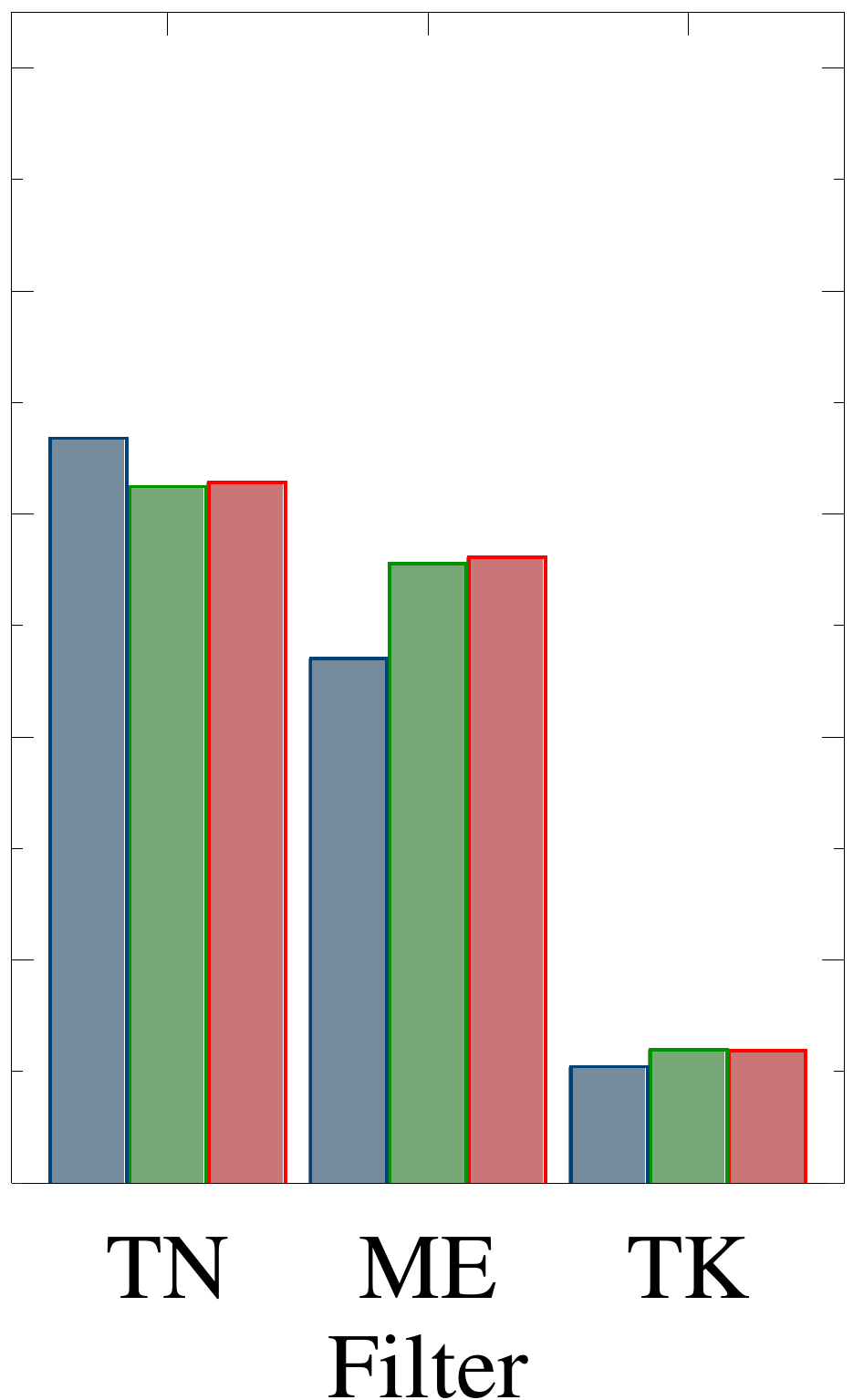}\hfill%
    \includegraphics[height=.505\linewidth]{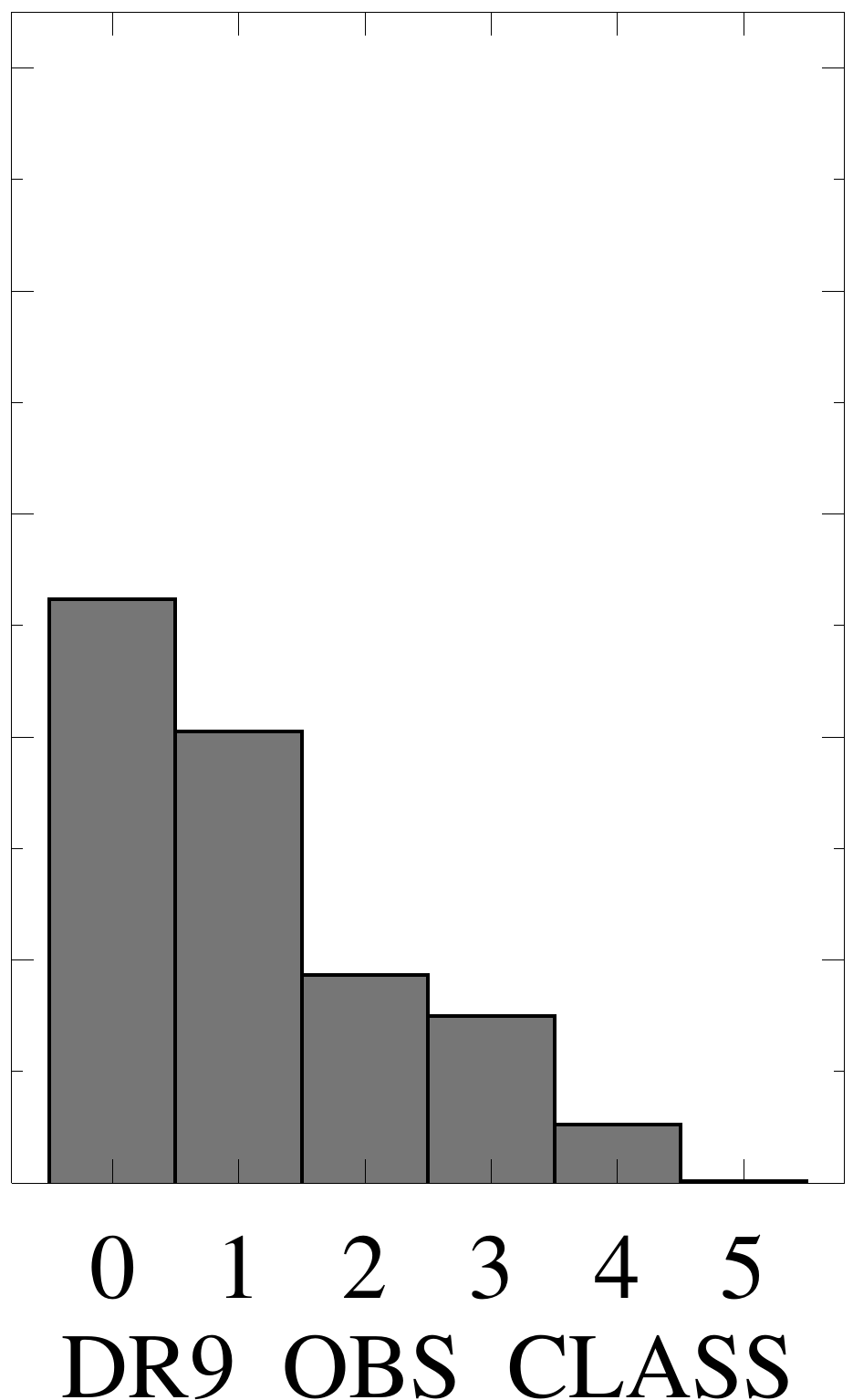}
    \caption{Distribution of observation modes, filters, and usable chip area
      OBS\_CLASS in 4XMM-DR9s. Possible modes are extended full frame (EF),
      full frame (FF), large window (LW), small window (SW), timing
      (TI). Filters include thin (TN), medium (ME), and thick (TK). Ten
      observations have the poorest OBS\_CLASS 5.}
    \label{fig:obsclasses}
  \end{figure}

\begin{figure}
    \centering
    \includegraphics[width=\linewidth]{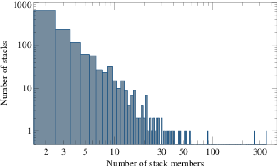}
    \caption{Number of observations in the 4XMM-DR9s stacks.}
    \label{fig:stacksizes}
  \end{figure}

\begin{figure}
    \centering
    \includegraphics[width=\linewidth]{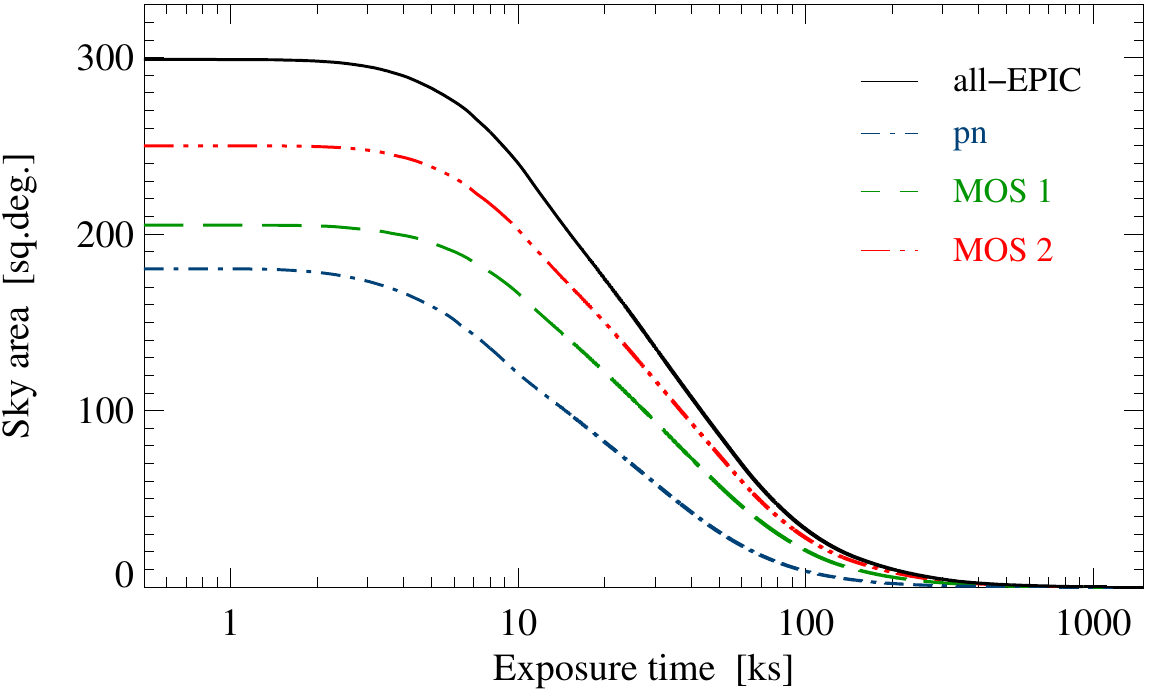}
    \caption{Cumulative sky area of multiply observed patches over exposure
      time in the input sample. `All-EPIC' shows the total exposure, for which
      the maximum exposure among the active EPIC instruments is determined in
      each observation and sky pixel and then summed over all observations.}
    \label{fig:skyareaexp}
  \end{figure}

  The final selection comprises 6\,604 observations with at least one usable
  exposure in 1\,329 stacks. Their depth and distribution over the sky are
  shown in Fig.~\ref{fig:skymap} in a Hammer-Aitoff representation and their
  observation modes, filters, and usable chip area according to their
  OBS\_CLASS classification in 4XMM-DR9 in Fig.~\ref{fig:obsclasses}. The
  number of observations in the stacks are given in
  Fig.~\ref{fig:stacksizes}. Two thirds of the stacks comprise two or three
  observations. The largest stacks are the XXL North and South regions with
  352 and 266 observations, of which up to eleven overlap directly at a given
  sky position. Figure \ref{fig:skyareaexp} shows the depth of all stacks as
  cumulative repeatedly covered sky area over total exposure time. Out of
  about 485 square degrees total sky area in 4XMM-DR9s, 300 square degrees
  were multiply observed with a maximum exposure time of about 1.9\,Ms.

  \subsection{Catalogue processing}
  \label{sec:processing}

  The data processing strategy, starting from the modified 4XMM-DR9 attitude
  files and event lists, and source detection were described in
  \citetads{2019A&A...624A..77T} and not changed. In addition to the standard
  products needed to run source detection, the pipeline produces coverage
  maps, mosaic images, and exposure maps for each stack, which are newly
  published together with the catalogue (Sect.~\ref{sec:access}).

\begin{figure}
    \centering
    \includegraphics[width=\linewidth]{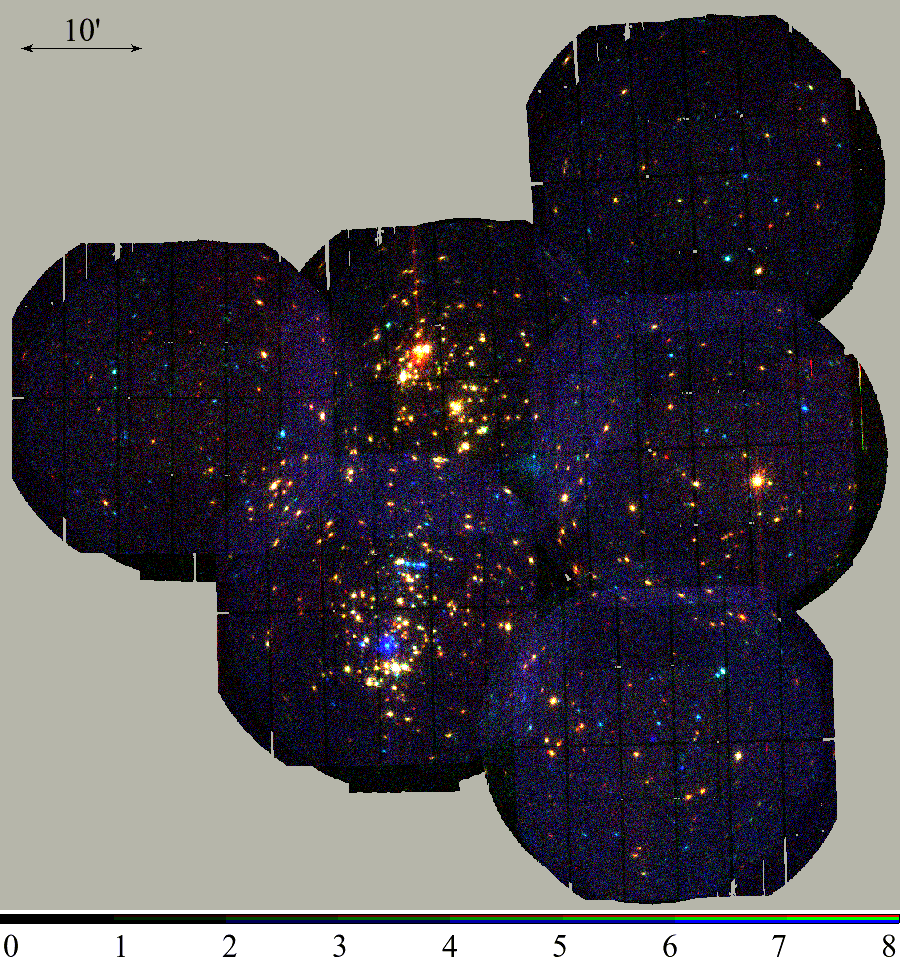}
    \includegraphics[width=\linewidth]{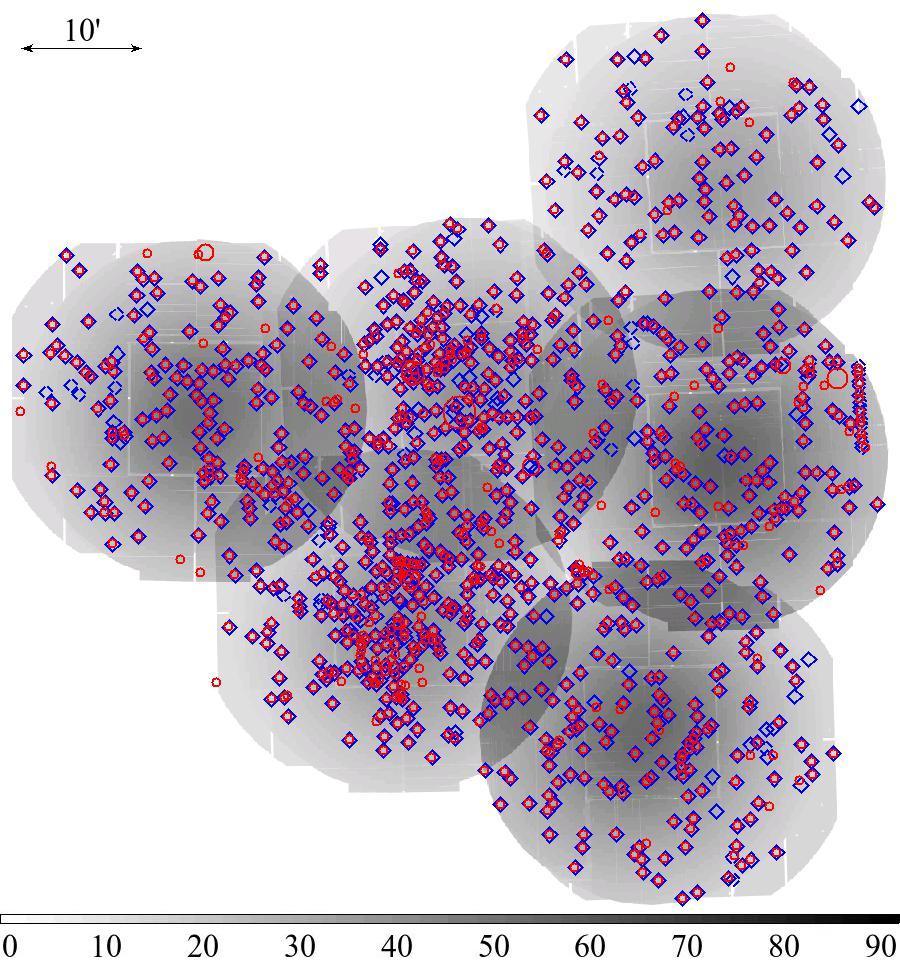}
    \caption{Example of a stack with six observations, targeting the star
      cluster NGC\,2264. \emph{Upper panel:} false-colour image in three
      energy bands $0.2-1.0$, $1.0-2.0$, and $2.0-12.0$\,keV to illustrate the
      different flux maxima of the detected objects. \emph{Lower panel:}
      exposure map, giving the total exposure in ks, taking vignetting into
      account, with 4XMM-DR9s sources as red circles and 4XMM-DR9 sources as
      blue diamonds. The symbol size scales with the core radius of the extent
      model. Several low-quality (flagged) DR9 detections were not recovered
      in DR9s.}
    \label{fig:ngc2264}
  \end{figure}

  While the largest stack in 3XMM-DR7s comprised 66 observations, the
  catalogue pipeline for 4XMM-DR9s was revised to be capable of processing
  several hundred observations and deep stacks with dozens of directly
  overlapping observations. The standard tasks published in SAS18 were
  modified to speed up preparation time by parallel pre-processing of the
  observations. Using these software versions, the catalogue could be produced
  on standard PCs and a server with higher memory
  capacity. Figure~\ref{fig:ngc2264} shows an example detection image and
  exposure map of a pipeline-processed stack comprising six observations.

\section{The second edition of the catalogue from overlapping observations: 4XMM-DR9s}
\label{sec:catalogue}

  \begin{table*}
    \caption{Sources in 4XMM-DR9s, compared to the first catalogue edition 3XMM-DR7s.}
    \label{tab:catalogue}
    \centering
    \begin{tabular}{lrr}
      \hline\hline\noalign{\smallskip}
      Description                                                                   & 4XMM-DR9s     & 3XMM-DR7s     \\
      \hline\noalign{\smallskip}
      Number of stacks                                                              &   1\,329      &      434      \\
      Number of observations                                                        &   6\,604      &   1\,789      \\
      \multicolumn{2}{l}{Time span first to last observation  \hfill Feb 03, 2000 $-$ Nov 13, 2018} & Feb 20, 2000 $-$ Apr 02, 2016 \\
      Approximate sky coverage                                                      & 480 sq.\ deg. & 150 sq.\ deg. \\
      Approximate multiply observed sky area                                        & 300 sq.\ deg. & 100 sq.\ deg. \\
      Total number of sources                                                       & 288\,191      &  71\,951      \\
      Sources with one contributing observation                                     &  69\,908      &  14\,286      \\
      Observed once with flag 0 or 1                                                &  65\,307      &  14\,076      \\
      Observed once and manually flagged                                            &   3\,668      & $/$\quad\strut \\
      Sources with several contributing observations                                & 218\,283      &  57\,665      \\
      Multiply observed sources with flag 0 or 1                                    & 191\,497      &  55\,450      \\
      Multiply observed and manually flagged                                        &  19\,224      & $/$\quad\strut \\
      Multiply observed with a total detection likelihood of at least six           & 181\,132      &  49\,935      \\
      Multiply observed with a total detection likelihood of at least ten           & 153\,487      &  42\,077      \\
      Multiply observed extended sources (extent radius $\geq 6\arcsec$) with flag 0 or 1 &   9\,234      &   2\,588      \\
      Multiply observed point sources with VAR\_PROB$\leq$10$^{-3}$ and flag 0 or 1 &  11\,327      &   3\,301      \\
      Multiply observed point sources with VAR\_PROB$\leq$10$^{-5}$ and flag 0 or 1 &   7\,182      &   1\,839      \\
      \hline
    \end{tabular}
  \end{table*}
  

  \subsection{Basic properties}
  \label{sec:properties}

  The serendipitous \textit{XMM-Newton} source catalogue from overlapping
  observations is composed from the source lists of all stacks. It consists of
  several rows for each individual source, which is identified through its IAU
  name 4XMMs\,J\emph{hhmmss.s}$\pm$\emph{ddmmss} and a unique identifier
  SRCID. A summary row lists the source parameters determined for the full
  stack of observations. The following rows give the source parameters for
  each contributing observation separately. They are also derived from the
  same stacked maximum-likelihood fit (cf.\ Sect.~\ref{sec:software}).
  Stack-specific columns can be used to restrict the catalogue to its summary
  rows, for example by choosing N\_CONTRIB$>$0 for all sources and
  N\_CONTRIB$>$1 for sources in overlap regions. Observation-specific rows can
  be selected for example via the observation identifier OBS\_ID. An overview
  of the catalogue layout including a column list and screen shots are given
  in the Appendix Section~\ref{sec:layout}. All detected sources from the
  input observations are transferred to the catalogue, whether located in
  overlap areas or not. A newly introduced boolean flag OVERLAP is set to true
  in all rows of the sources with at least two contributing observations and
  thus allows for direct selection of multiply observed sources.

\begin{figure}
    \centering
    \includegraphics[height=.545\linewidth]{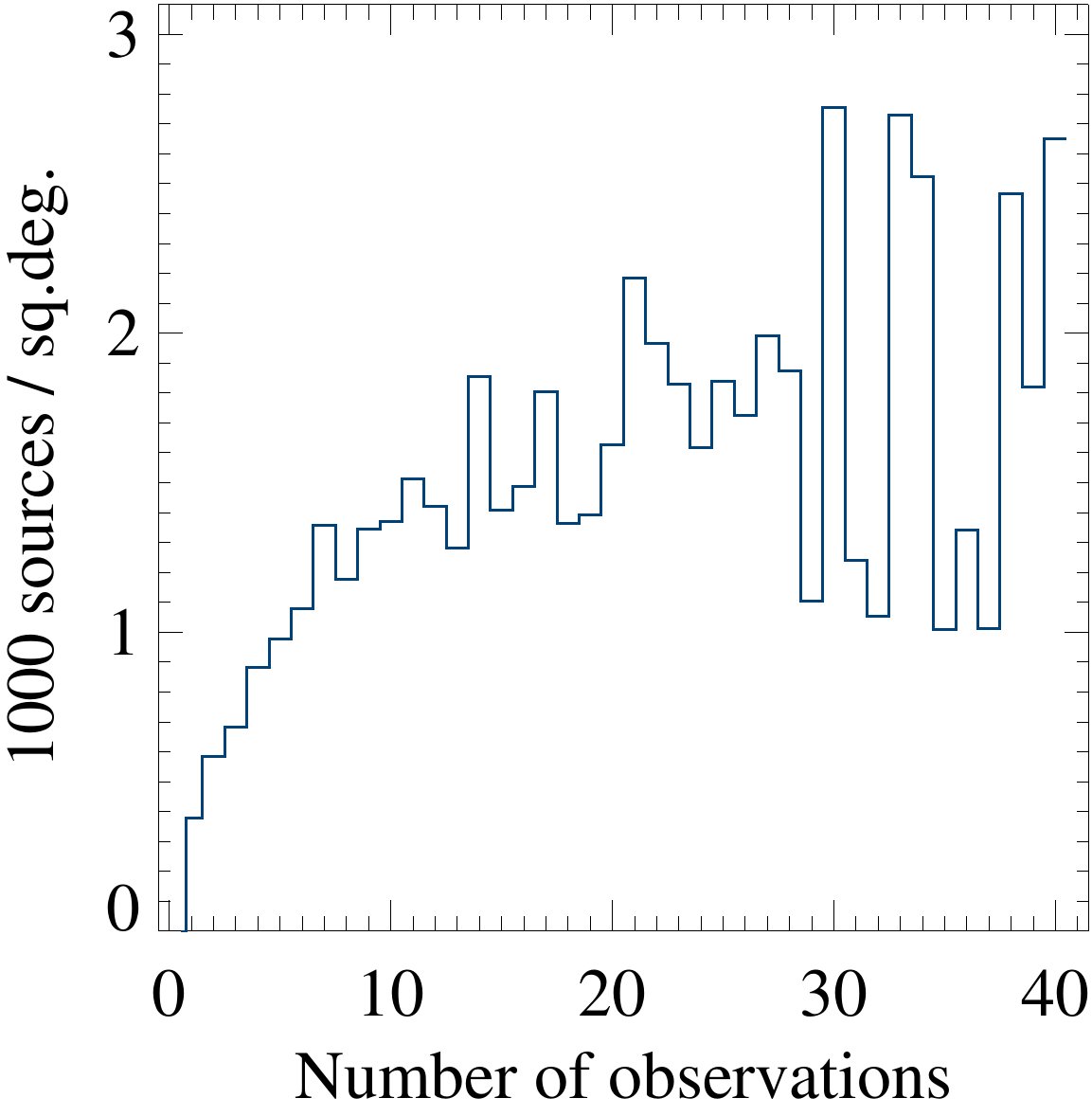}\hfill%
    \includegraphics[height=.545\linewidth]{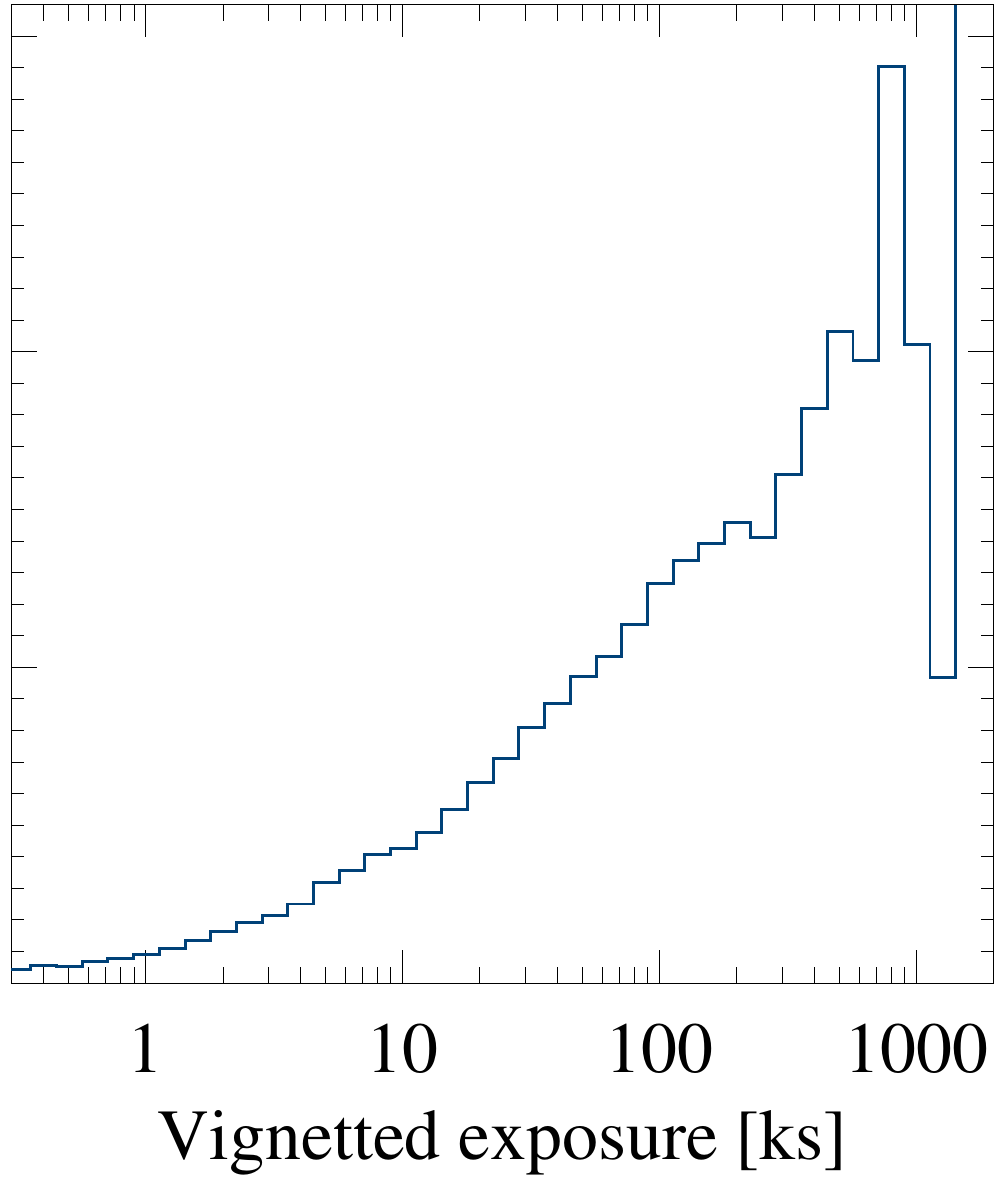}
    \caption{Source density in 4XMM-DR9s over number of contributing
      observations \emph{(left panel)} and total exposure time in all
      contributing observations, taking vignetting into account (\emph{right
        panel}). Only small sky areas are covered by more than 30 observations
      or for more than 0.5\,Ms (cf.\ Fig.~\ref{fig:skyareaexp}), causing a
      wider scatter in the distributions.}
    \label{fig:dense}
  \end{figure}

  4XMM-DR9s comprises 288\,191 unique sources from 6\,604 observations, and
  218\,283 of them were found in overlap areas. The other 69\,908 sources were
  observed only once. These are located in the outer parts of the stacks or in
  smaller regions missing from one of the observations like a chip gap, the
  gaps in MOS large-window mode observations, or on the de-activated MOS1
  CCDs. The higher source density through stacked source detection is plotted
  in Fig.~\ref{fig:dense} in terms of extrapolated source number per square
  degree over the number of contributing observations and over the cumulated
  exposure time, taking vignetting effects into account.


  Seven percent of the sources are detected as extended with a core radius of
  at least 6\arcsec, and 88\,\% have a good or very good quality flag
  (STACK\_FLAG$\leq$1, referred to as `un-flagged' throughout this paper,
  cf.\ Sect.~\ref{sec:screening}). The full-band median fluxes and flux errors
  of the catalogue sources are $1.7\times
  10^{-14}\,\textrm{erg\,cm}^{-2}\,\textrm{s}^{-1}$ and $6.1\times
  10^{-15}\,\textrm{erg\,cm}^{-2}\,\textrm{s}^{-1}$ for multiply observed
  sources and $2.1\times 10^{-14}\,\textrm{erg\,cm}^{-2}\,\textrm{s}^{-1}$ and
  $9.5\times 10^{-15}\,\textrm{erg\,cm}^{-2}\,\textrm{s}^{-1}$ in
  non-overlapping areas, respectively. About 4\,\% of the un-flagged multiply
  observed sources show signs of high long-term inter-observation variability
  according to $\textrm{VAR\_PROB}\leq 10^{-5}$. VAR\_PROB, one of the
  variability parameters which were introduced for 3XMM-DR7s
  \citepads{2019A&A...624A..77T}, is the probability that the fitted source
  fluxes are consistent with constant flux in all observations
  (cf.\ Sect.~\ref{sec:ltv}). An overview of the catalogue properties and a
  comparison to 3XMM-DR7s are given in Table~\ref{tab:catalogue}.

\begin{figure}
    \centering
    \includegraphics[width=.8\linewidth]{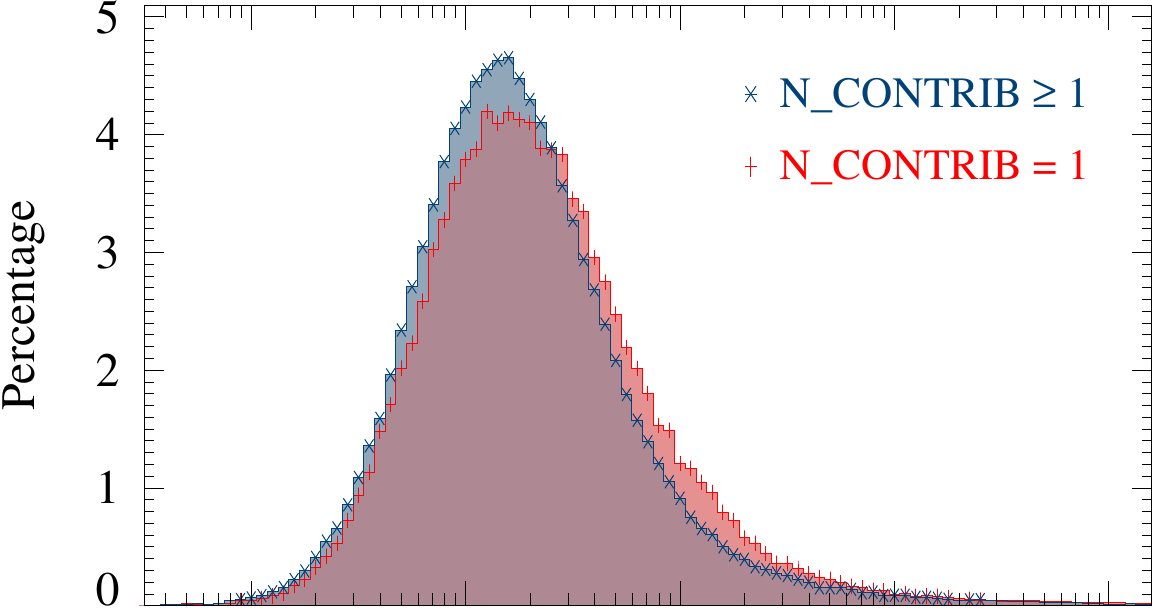}
    \includegraphics[width=.8\linewidth]{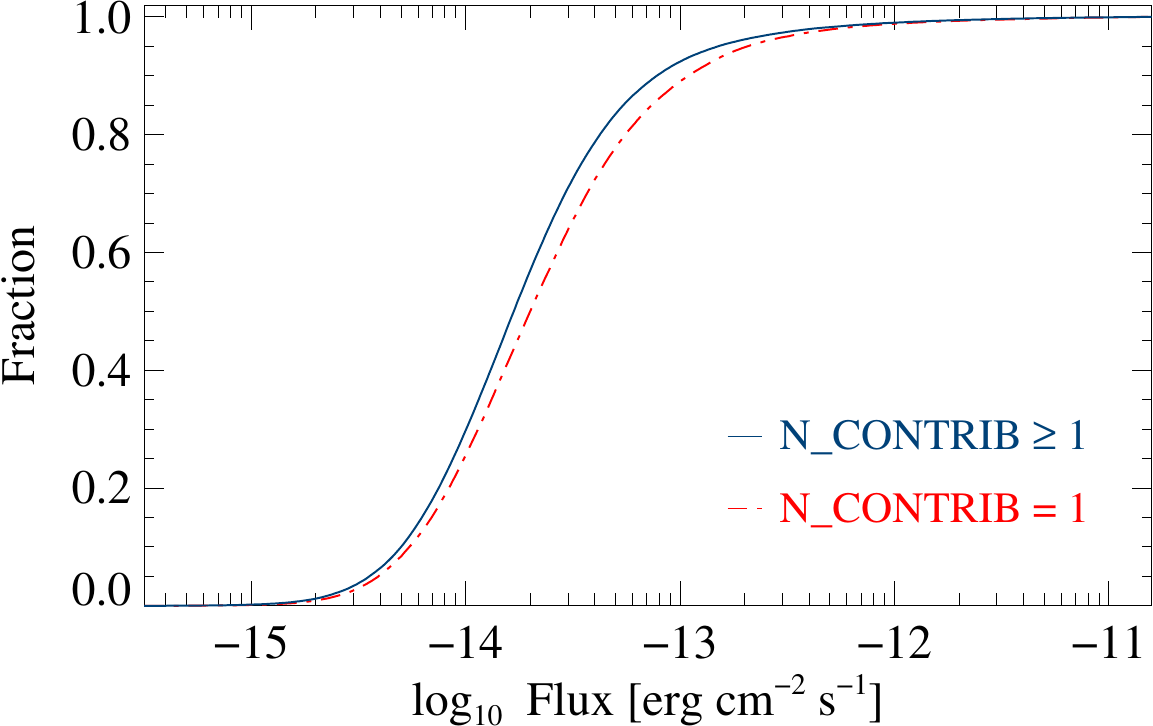}
    \caption{Flux distribution of multiply observed sources \emph{(blue)} and
      of sources covered once \emph{(red)} in 4XMM-DR9s. \emph{Upper panel:}
      histograms normalised to the sample size. \emph{Lower panel:} cumulative
      histograms.}
    \label{fig:fluxes}
  \end{figure}

\begin{figure}
    \centering
    \includegraphics[height=.545\linewidth]{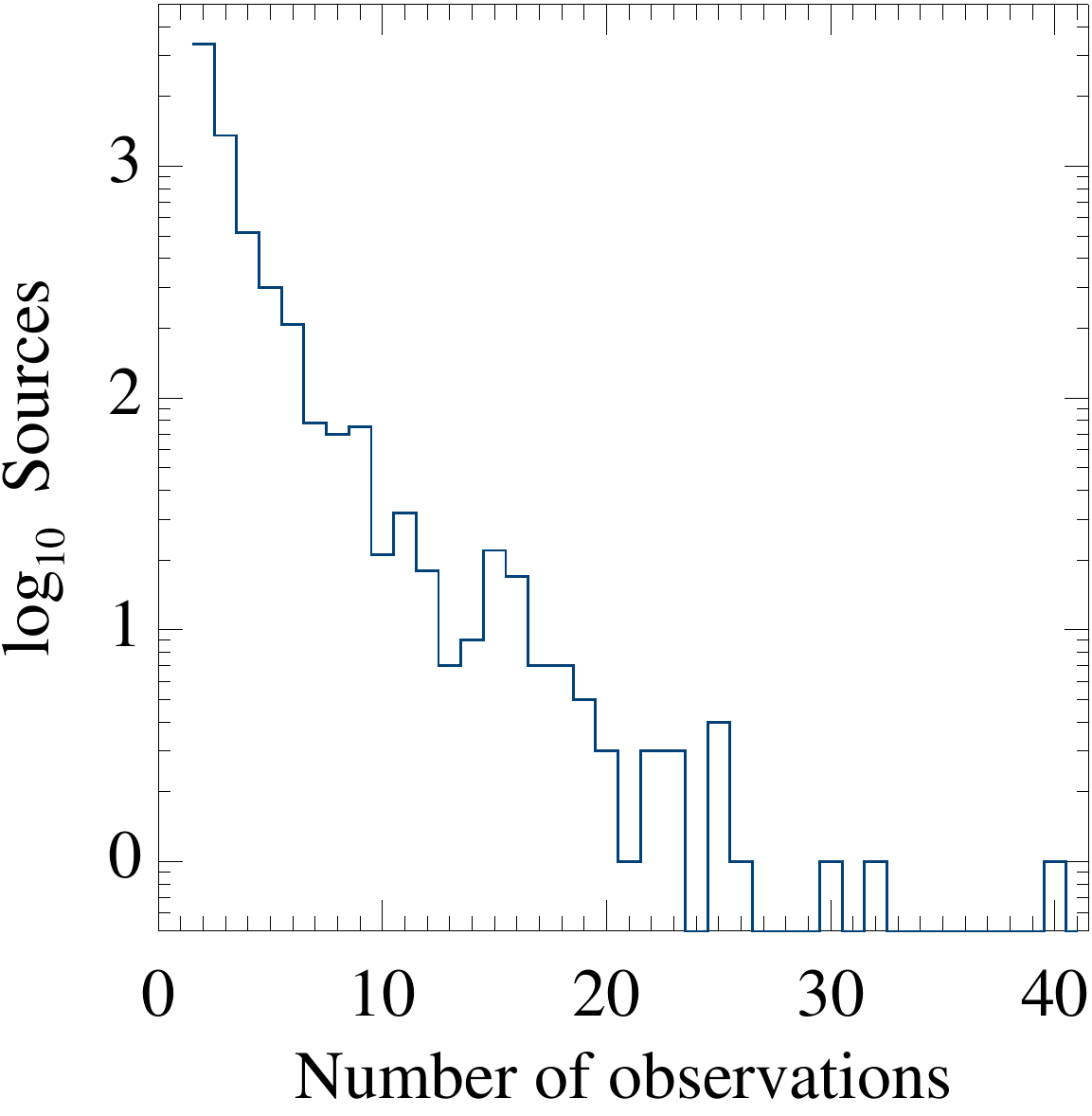}\hfill%
    \includegraphics[height=.545\linewidth]{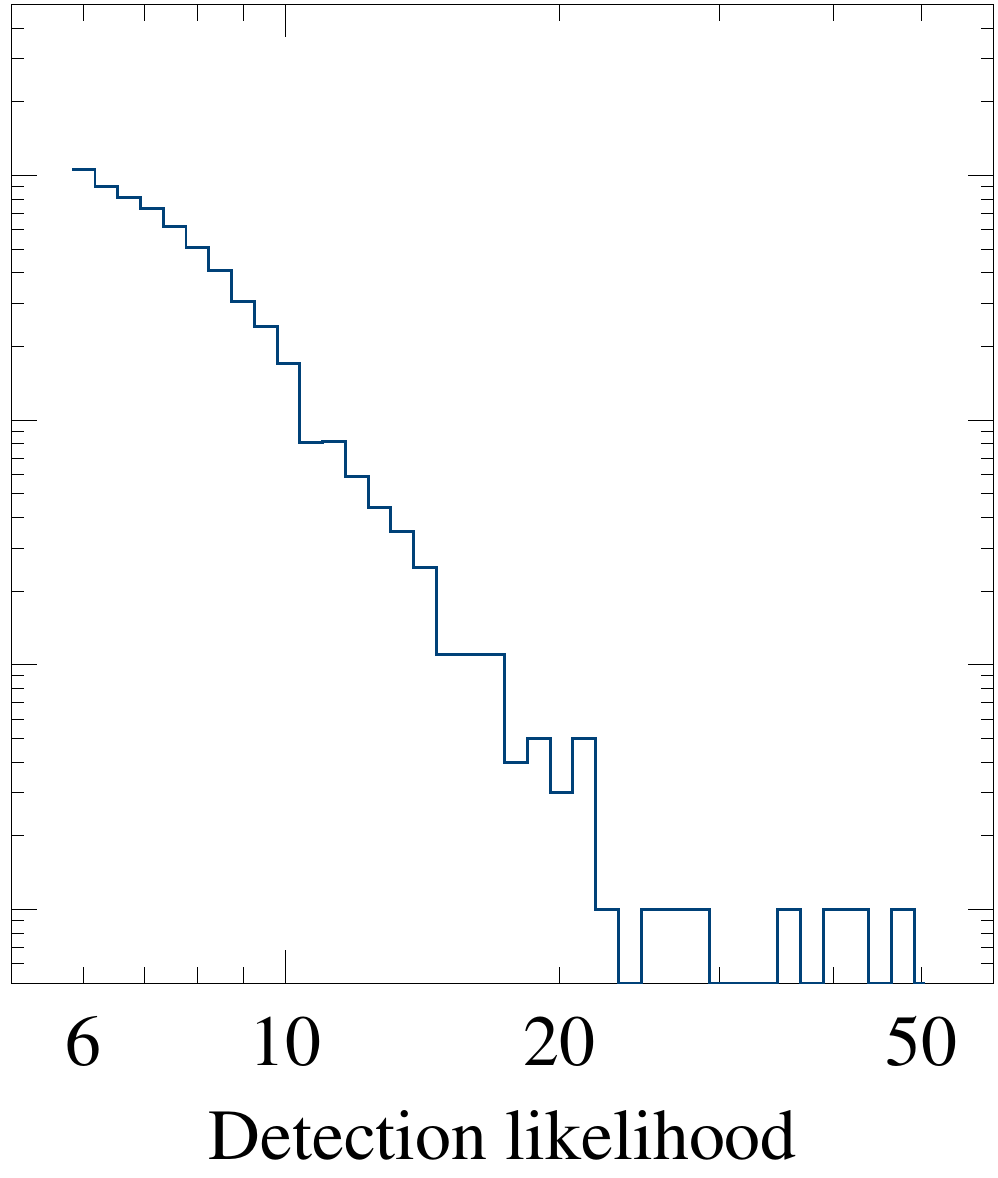}\vskip2mm
    \includegraphics[height=.545\linewidth]{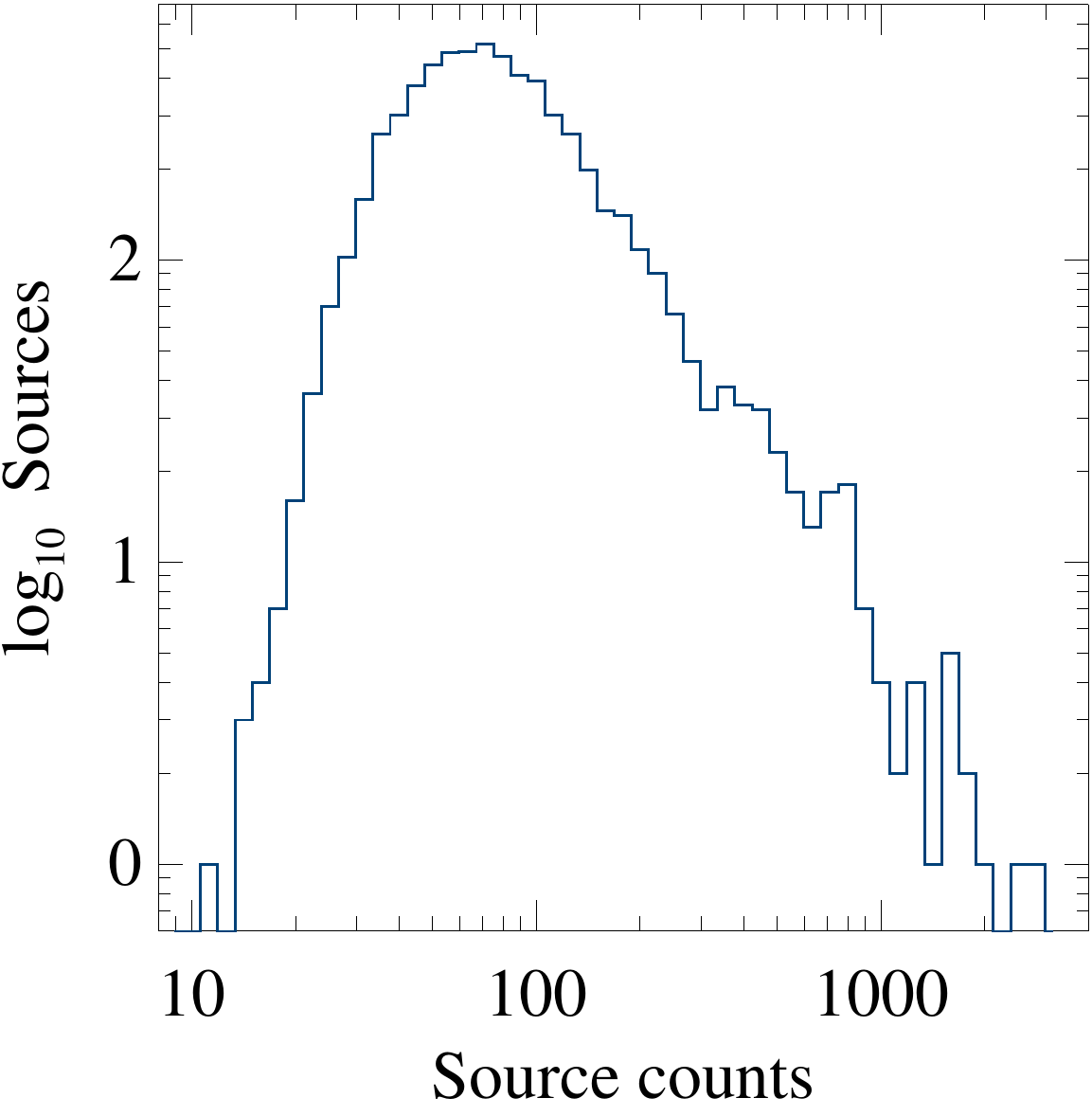}\hfill%
    \includegraphics[height=.545\linewidth]{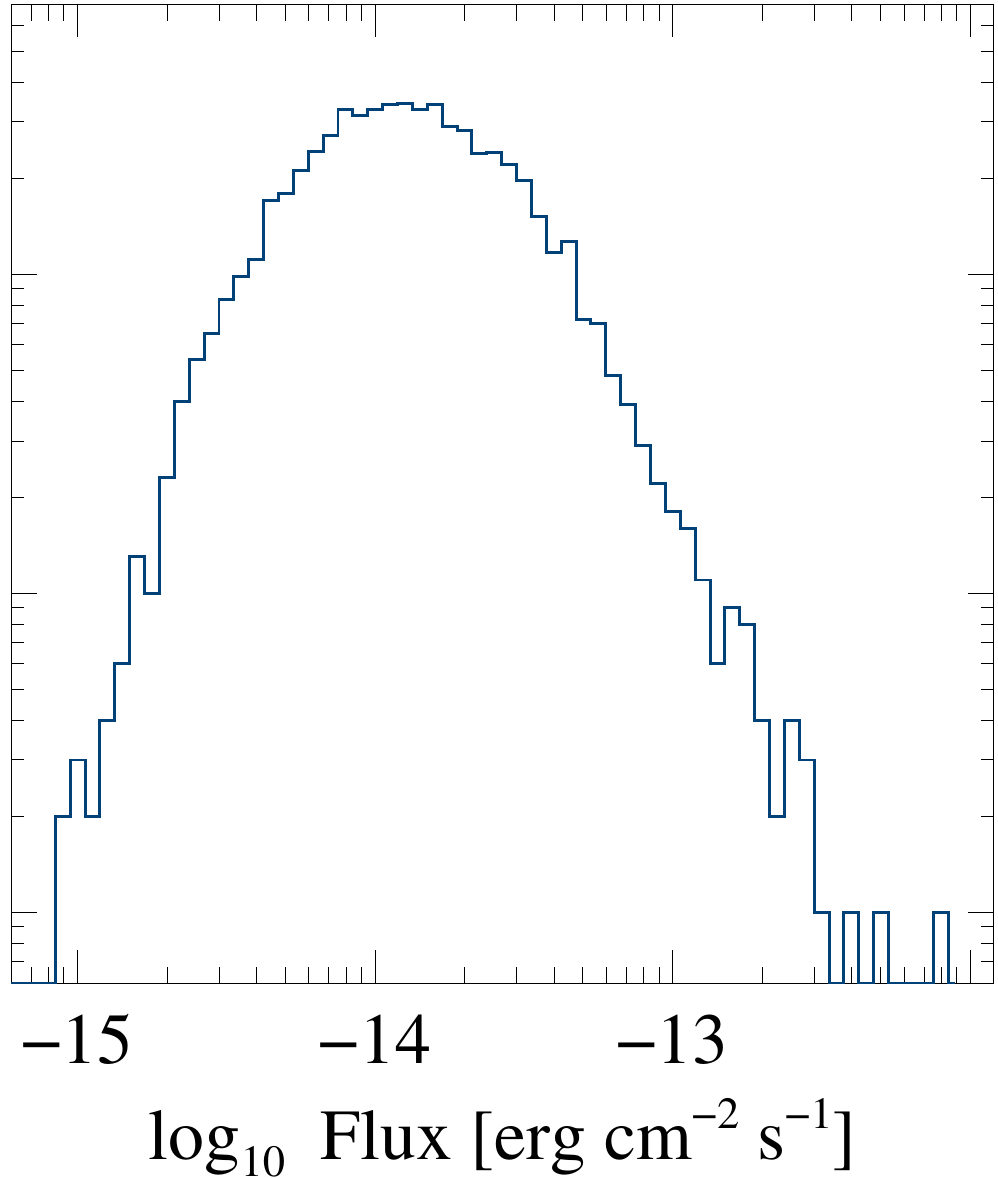}
    \caption{Parameters of sources that surpass the minimum detection
      likelihood only in the stack summary. Their observation-level detection
      likelihoods in the images of a specific observation, which are derived
      from the same simultaneous fit, are all below the limit of six.}
    \label{fig:stackonly}
  \end{figure}

  Figure~\ref{fig:fluxes} illustrates that through stacked source detection a
  larger fraction of low-flux sources is uncovered in the overlap areas
  (N\_CONTRIB$>$1) compared to sky areas observed once (N\_CONTRIB$=$1). The
  lower median flux in overlap than in non-overlap areas, given in the
  paragraph above, corresponds to the relative flux histogram peaking at lower
  fluxes (upper panel of Fig.~\ref{fig:fluxes}). The cumulative representation
  (lower panel of Fig.~\ref{fig:fluxes}) shows the higher number of detected
  low-flux sources in the overlap areas. 4XMM-DR9s contains 6\,142 sources
  that have low detection likelihood (EP\_DET\_ML$<$6) in all contributing
  observations and become significant (EP\_DET\_ML$\geq$6) only through
  stacked source detection. Their parameters are plotted in
  Fig.~\ref{fig:stackonly}: contributing observations, total detection
  likelihood, counts, and flux. New sources with up to 2\,000 counts and
  sources with a detection likelihood up to 50 are found that were missed when
  the observations would be processed individually. They are located in areas
  with eight to forty directly overlapping
  observations. Section~\ref{sec:dr9dr9s} discusses the source gain compared
  to the catalogue from individual observations 4XMM-DR9.

  \subsection{Quality assessment and visual screening}
  \label{sec:screening}

  All \textit{XMM-Newton} catalogues include information on the quality of a
  detection in terms of flags that indicate whether it is potentially spurious
  or located in a problematic region that could affect the reliability of its
  parameters. The initial quality assessment is performed automatically by the
  SSC-internal SAS tasks \texttt{dpssflag} within the processing pipeline,
  following the strategy of \citetads{2009A&A...493..339W}. It defines nine
  boolean flags for each exposure of a source, coding several poor observing
  conditions which potentially deteriorate the accuracy of the detection or
  its parameters. The flags enter the catalogue as nine-character strings
  $ii$\_FLAG, where $ii$ stands for one of the instrument abbreviations PN,
  M1, M2. An overview of their definition is given in
  Table~\ref{tab:dpssflag}. The all-EPIC string flag EP\_FLAG is true wherever
  an instrument flag is true. The integer STACK\_FLAG summarises the nine
  boolean flags and is `0' in case of no warnings, `1' for reduced detection
  quality in at least one image, `2' for potentially spurious sources
  (cf.\ Table~\ref{tab:dpssflag}), and `3' in the summary row for flags `2'
  during all contributing observations. The cleanest set can thus be chosen by
  applying the filtering expression `STACK\_FLAG$\leq$1', while detections
  with higher flags can still come from real sources, but may have uncertain
  fit parameters. Subsequent quality assessment is done through visual
  inspection of all source-detection results. The first catalogue from
  overlapping observations 3XMM-DR7s was based on a clean sample and used the
  automatically set flags only. 4XMM-DR9s includes significantly more
  observations taken under various and partly unfavourable
  conditions. Therefore, additional manual flags are introduced with this
  version, which resemble those in the series of catalogues from single
  observations.

\begin{figure*}
    \centering
    \includegraphics[width=\linewidth]{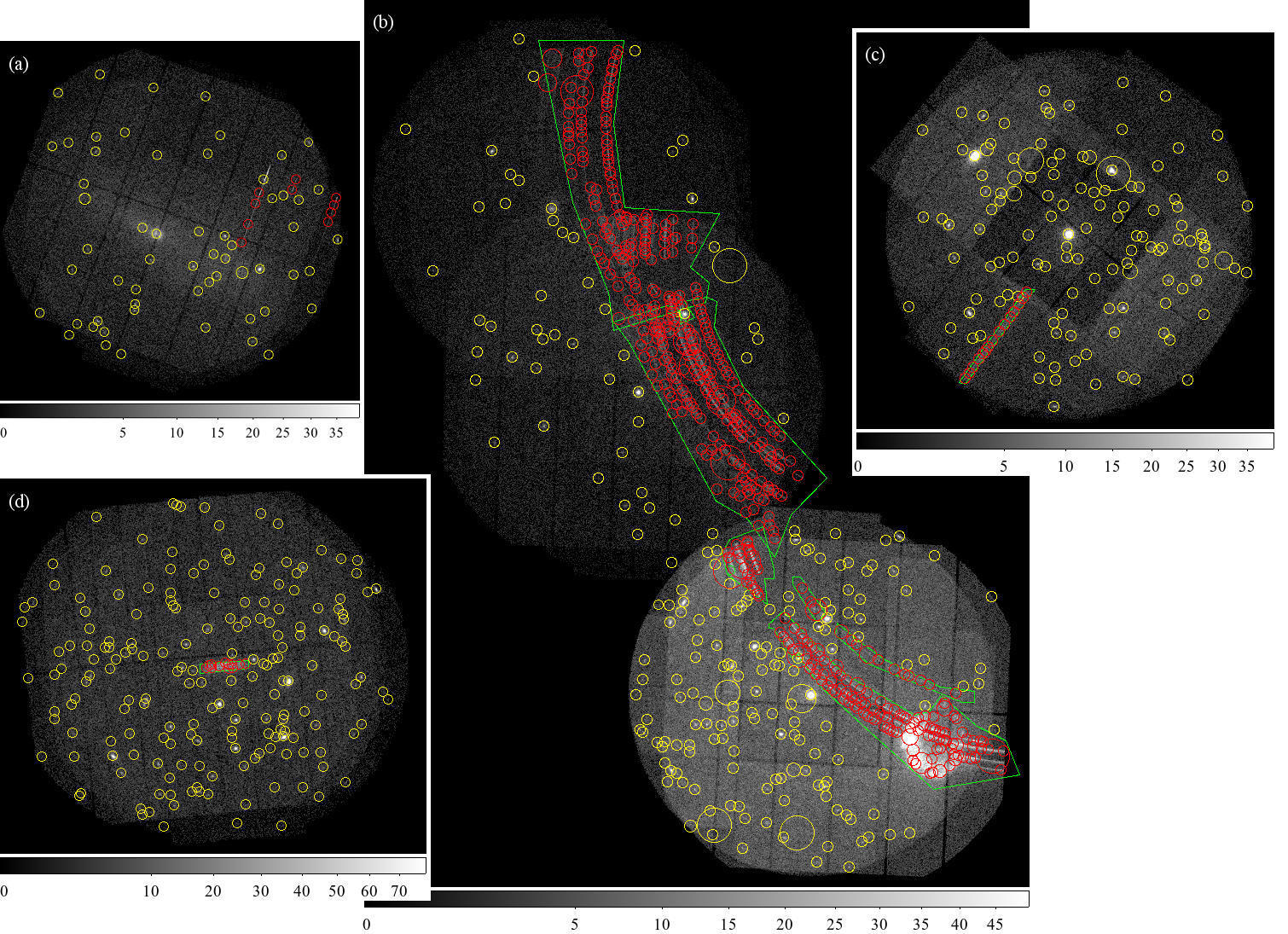}
    \caption{Examples from the manual screening process: bad detector features
      \emph{(panel a)}, single-reflection patterns \emph{(panel b)}, RGA
      diffraction pattern \emph{(panel c)}, path of a planet in a mosaic-mode
      observation \emph{(panel d)}. Yellow circles show all detections, where
      large radii correspond to the core radius of extended sources. Bad
      regions or detections are marked manually \emph{(green)}. The detections
      within them \emph{(red)} are then flagged in the catalogue.}
    \label{fig:screening}
  \end{figure*}

  The results of source detection in each stack were screened visually making
  use of SAOImage DS9 with XPA \citepads{2000ASPC..216...91J} and STILTS
  \citepads{2005ASPC..347...29T} commands. Obviously spurious sources are
  flagged manually, (a) if detections lie on single reflection patterns, (b)
  if several detections are found on detector features like clusters of bad
  pixels without any sign of a blended real source, (c) if they are heavily
  confused with bright extended emission and thus have unreliable source
  parameters, (d) if several detections lie directly on the PSF spikes of
  bright sources, (e) if a bright source is fitted with several off-set
  detections because of extreme pile-up, (f) if emission of a very bright
  target triggers spurious extended detections, or (g) if multiple overlapping
  sources are fitted to large extended emission or to the footprint of a Solar
  System body. The screening process is not meant to be complete, but shall
  help users to reject evidently not genuine sources. Ambiguous cases remain
  un-flagged in the visual screening to reduce the risk of erroneously
  excluding good detections, since they are prone to visual
  mis-classification, for example whenever a real source might be overlapping
  with a detector feature, if few non-overlapping extended detections cover
  large extended emission, or if two detections are used to describe the
  emission of probably only one real source. Sources can have automatically
  set flags and no manual flag, for example in the vicinity of bright or
  extended sources where their source parameters could be affected by higher
  uncertainties than in clean areas. Figure~\ref{fig:screening} shows examples
  from the screening process.

  Manual flags are indicated in the STACK\_FLAG and EP\_FLAG columns: EP\_FLAG
  is expanded from nine to ten characters by an additional flag (character) at
  the end of the string which is true (`T') for manually flagged sources and
  false (`F') for the rest. STACK\_FLAG is increased by ten if a manual flag
  was set. The results of the automated flagging are thus preserved in its
  units' digit, while the manual flag appears in its tens' digit. A filtering
  expression `STACK\_FLAG\,$\leq$\,1' removes both sources with the highest
  automatic flags 2 and 3 and sources with a manual flag. Eventually, 22\,892
  catalogue sources received a manual flag and additional 8\,495 have
  STACK\_FLAG 2 or 3.

  \subsection{Cross-matching with the 4XMM-DR9 catalogue}
  \label{sec:dr9matches}

  The catalogue 4XMM-DR9 from individual observations \citep{2020dr9} is
  distributed in two flavours: From the source lists of all individual
  observations, the catalogue of detections is created. Sources can thus be
  included several times in the 4XMM-DR9 catalogue of detections, if they were
  observed repeatedly. By a positional match, individual detections from
  different observations are merged into unique sources, and the parameters of
  the individual detections are merged into unique source parameters. In
  particular, an averaged unique source position and position error are
  calculated from the positions and errors of the individual detections. The
  results of the merging compose the 4XMM-DR9 catalogue of unique sources,
  also referred to as ``slim version'' in other publications.

  The catalogue 4XMM-DR9s from overlapping observations is matched with a
  subset of 4XMM-DR9 beginning with the catalogue of sources. The 473\,488
  unique DR9 sources which have at least one contributing detection with a
  good summary flag, SUM\_FLAG\,$\leq$\,1, are selected. The search radius is
  set to 2.2698 times their statistical and systematic position errors,
  corresponding to the 99.73\,\% confidence region of a Rayleigh
  distribution. In DR9, the combination of the errors is given in the column
  $\textrm{SC\_POSERR}\,/\sqrt{2}$. In DR9s, we use the statistical position
  error given in the column $\textrm{RADEC\_ERR}\,/\sqrt{2}$ and the
  additional error component derived from a fit to a Rayleigh distribution in
  Sect.~\ref{sec:astrometry}.

  For each 4XMM-DR9s source, the nearest associated DR9 source is included in
  the 4XMM-DR9s catalogue. In particular, if more than one DR9 source is found
  within the matching radius, the one with the closest position is chosen. The
  DR9 identifier, position, quality and variability flags are copied from the
  4XMM-DR9 catalogue of sources into the stack-level rows (for the structure
  of 4XMM-DR9s, cf.\ Fig.~\ref{fig:catview_topcat}). From the DR9 catalogue of
  detections, the parameters are copied into the DR9s observation-level rows
  of the associated source. The distances between the DR9s and the DR9
  positions are calculated and also listed. Columns with DR9 information are
  marked by the suffix \_4XMMDR9. If no association is found, the table cells
  remain empty. The matching results are further discussed in
  Sect.~\ref{sec:dr9dr9s}.
  
  \begin{table}
    \caption{Meaning of the automatic flags set by the task \texttt{dpssflag},
      position in the flag string, and integer summary
      STACK\_FLAG\tablefootmark{*}.}
    \label{tab:dpssflag}
    \centering
    \begin{tabular}{lp{53mm}l}
      \hline\hline\noalign{\smallskip}
        String position & \multicolumn{2}{l}{Description \hfill STACK\_FLAG} \\
      \hline\noalign{\smallskip}
      1 ~ TFFFFFFFF
        &  PSF coverage below 50\%
        &  1 \\
      2 ~ FTFFFFFFF
        &  detection close to a bright point-like detection
        &  1 \\
      3 ~ FFTFFFFFF
        &  detection close to an extended detection
        &  1 \\
      4 ~ FFFTFFFFF
        &  possibly spurious extended detection close to a bright point detection
        &  2 \\
      5 ~ FFFFTFFFF
        &  possibly spurious extended detection close to a bright extended detection
        &  2 \\
      6 ~ FFFFFTFFF
        &  possibly spurious extended detection which is significant in one band only
        &  2 \\
      7 ~ FFFFFFTFF
        &  summary: possibly spurious extended detection with at least one flag out of `4', `5', `6'
        &  2 \\
      8 ~ FFFFFFFTF
        &  detection on a bad pixel or CCD area
        &  2 \\
      9 ~ FFFFFFFFT
        &  detection close to a bad CCD area
        &  1 \\
      \hline
    \end{tabular}
    \tablefoot{\tablefoottext{*}\ STACK\_FLAG is zero if none of the listed
      flags is set in any of the contributing observations and takes the
      largest possible value triggered by the active boolean flags otherwise.}
  \end{table}
  
  \subsection{New and revised columns in 4XMM-DR9s}
  \label{sec:newcolumns}

  With 4XMM-DR9s, 26 columns are newly introduced with respect to
  3XMM-DR7s. The definitions of three more columns were revised, and all
  columns from the match with the catalogues from individual observations were
  renamed. These changes are listed in this Section. An overview of all
  4XMM-DR9s columns is given in Table~\ref{tab:columns} in the Appendix and
  more details on all the columns that were initially defined for 3XMM-DR7s in
  \citetads{2019A&A...624A..77T}.

  N\_EXP. Analogously to the number of contributing observations N\_CONTRIB,
  N\_EXP gives the number of exposures used in the fitting process, which is
  the sum of active instruments in all contributing observations. If all
  instruments were active during all observations, it is thus
  3\,$\times$\,N\_CONTRIB in the summary rows. Exposures may be missing
  because an instrument was operated in timing mode for example, because of
  technical problems, or because they were excluded from the catalogue
  processing due to a high background level (Sect.~\ref{sec:bkgcut}).

  EXTENT\_ML. The likelihood of a source being extended is included for
  extended and newly also for point-like sources in all 4XMM catalogues. It is
  derived from a fit with a $\beta$-profile broadened PSF and calculated with
  respect to the null hypothesis that the source is point-like as the
  log-likelihood difference between the extended and the point-like fits using
  Cash statistics. The minimum allowed core radius of extended sources is
  6\arcsec. For negative values, thus, the source is detected with higher
  significance in the fit with a point-like PSF.

  PN\_PILEUP, M1\_PILEUP, M2\_PILEUP indicate whether the source might be
  affected by pile-up in the pn, MOS1, MOS2 exposure. The method to derive the
  values is described in \citet{2020dr9}.

  N\_BLEND gives the number of simultaneously fitted sources in
  \texttt{emldetect}. In the catalogue pipeline, it is limited to two.

  DIST\_REF is the distance of a source to the reference coordinates in the
  centre of a stack. It is given in arcminutes.

  The flag OVERLAP indicates whether the centre of the source is located in
  the overlap area of two or more observations. It corresponds to
  N\_CONTRIB\,$>$\,1 in the summary row and is set in all rows of a source. In
  rare cases, for instance on chip gaps, the centre of the source may be
  covered by one observation only, triggering the flag to be false, while the
  outer regions are covered by several observations.

  EP\_FLAG. The first nine characters of the all-EPIC quality flag include the
  automatically set flags as in the previous catalogue. A tenth character adds
  the manual flag from visual screening (Sect.~\ref{sec:screening}), being
  `T' if the source has been regarded obviously spurious and `F'
  otherwise.

  STACK\_FLAG. The integer summary of the quality assessment takes the values
  0 to 3 from the automatically set flags as introduced by
  \citetads{2019A&A...624A..77T} and summarised in
  Sect.~\ref{sec:screening}. If a manual flag has been set, STACK\_FLAG is
  increased by ten.

  ASTCORR. This flag indicates whether an exposure was astrometrically
  corrected before performing stacked source detection. It is thus set in the
  observation-level rows and undefined in the summary rows. If it is true, the
  following astrometry-related columns are set, whose prefix CC\_ stands for
  the task \texttt{catcorr}:

  CC\_RAOFFSET, CC\_DEOFFSET give the field shift in arcseconds as derived by
  the SAS task \texttt{catcorr} for 4XMM-DR9 (\citealt{2020dr9} and
  Sect.~\ref{sec:astcorr}).  CC\_ROT\_CORR gives the \texttt{catcorr} field
  rotation in degrees.  CC\_POFFSET is the absolute offset between the
  original and the corrected source position in arcseconds.  CC\_RAOFFERR,
  CC\_DECOFFERR, CC\_ROT\_ERR, and CC\_POFFERR are the corresponding 1$\sigma$
  errors.

  CC\_REFCAT is the name of the reference catalogue used by \texttt{catcorr}.
  CC\_NMATCHES gives the number of matches between 4XMM-DR9 detections in the
  field and sources in the reference catalogue which were used by
  \texttt{catcorr} to determine the field rotation and shift.
  
  The following six columns inform on the background level of an observation
  (Sect.~\ref{sec:bkgcut}). They are thus set in the observation-level rows
  and undefined in the summary rows:
  
  PN\_BKG\_CRAREA, M1\_BKG\_CRAREA, M2\_BKG\_ CRAREA give the
  instrument-specific background rate per square arcsecond in the source-free
  regions. They are derived per exposure and thus identical for all sources in
  a given exposure.

  PN\_BKG\_CPROB, M1\_BKG\_CPROB, M2\_BKG\_CPROB are the respective Cauchy
  probabilities. They are derived from the rate distributions per observing
  mode and filter (Sect.~\ref{sec:bkgcut}).

  *\_4XMMDR9. Columns copied from 4XMM-DR9 have the suffix \_4XMMDR9 and are
  set if a 4XMM-DR9 source has been associated to a DR9s source
  (Sect.~\ref{sec:dr9matches}). They correspond to the 3XMM-DR7 columns
  described by \citetads{2019A&A...624A..77T}. A new column
  SUM\_FLAG\_MIN\_4XMMDR9 gives the lowest -- best -- summary flag of all
  individual detections which are merged into the 4XMM-DR9 source.

  The integer OBS\_CLASS indicates the usable area of an observation in
  4XMM-DR9. It is thus set in the observation-level rows and undefined in the
  summary rows. In 3XMM-DR7s, it was included in the list of observations
  only. The definition of the OBS\_CLASSes is detailed by \citet{2020dr9}.

  \subsection{Auxiliary products and catalogue access}
  \label{sec:access}

  The catalogue is published in the Flexible Image Transport System (FITS)
  format at the web pages of the \textit{XMM-Newton} SSC at IRAP,
  Toulouse\footnote{\url{http://xmmssc.irap.omp.eu/Catalogue/4XMM-DR9s/4XMM_DR9stack.html}},
  at AIP,
  Potsdam\footnote{\url{https://xmmssc.aip.de/cms/catalogues/4xmmdr9s/}}, and
  at the VizieR ftp
  archive\footnote{\url{ftp://cdsarc.u-strasbg.fr/pub/cats/J/A+A/TBD/TBD/}
    \emph{(will be added)}} hosted by the Centre de Donn\'ees astronomiques de
  Strasbourg (CDS). Searchable interfaces are provided by
  VizieR\footnote{\url{https://vizier.u-strasbg.fr/viz-bin/VizieR?-source=IX/TBD}
    \emph{(will be added)}}, the \textit{XMM-Newton} Science Archive
  (XSA)\footnote{\url{https://www.cosmos.esa.int/web/xmm-newton/xsa}}, and the
  HEASARC
  service\footnote{\url{https://heasarc.gsfc.nasa.gov/W3Browse/xmm-newton/xmmstack.html}}.
  The online
  documentation\footnote{\url{https://xmmssc.aip.de/cms/users-guide/}} and the
  list of observations in FITS format and as an HTML table are available from
  the IRAP and AIP web pages.

  Several auxiliary products are produced alongside the catalogue and
  published online. For each stack, all-EPIC mosaic images in the five
  \textit{XMM-Newton} standard energy bands and in the full 0.2$-$12.0\,keV
  energy range, two full-band exposure maps, and three types of coverage maps
  are published at the AIP web
  page\footnote{\url{https://xmmssc.aip.de/cms/stacks/}}. For each catalogue
  source, three auxiliary images are provided: An 0.2$-$12.0\,keV full-band
  X-ray image covering a 10\arcmin$\times$10\arcmin\ region centred at the
  source position, a false-colour X-ray image of the same region, and an
  optical finding chart covering a 2\arcmin$\times$2\arcmin\ region. For each
  source with at least two observations, an X-ray light curve is provided. All
  auxiliary images and light-curve plots are available from the XSA
  interface. Information on their production, the changes for 4XMM-DR9s with
  respect to the previous versions, and example images are included in
  Sects.~\ref{sec:auximas} and \ref{sec:auxdata} in the Appendix.

  Upper flux limits for various high energy missions are provided by the
  \textit{XMM-Newton} SOC in an upper-limit service on the \textit{XMM-Newton}
  web pages\footnote{\url{http://xmmuls.esac.esa.int/upperlimitserver/}}. The
  upper limit to the source flux at a given position is calculated from
  pipeline processed images, exposure maps, and background maps
  \citep{2020hiligt}. Stacked data for 4XMM-DR9s have been made available to
  include upper limit fluxes in overlapping images.

\section{Catalogue characterisation and long-term variability of sources}
\label{sec:results}

  \subsection{Astrometric accuracy}
  \label{sec:astrometry}

\begin{figure}
    \centering
    \includegraphics[width=.8\linewidth]{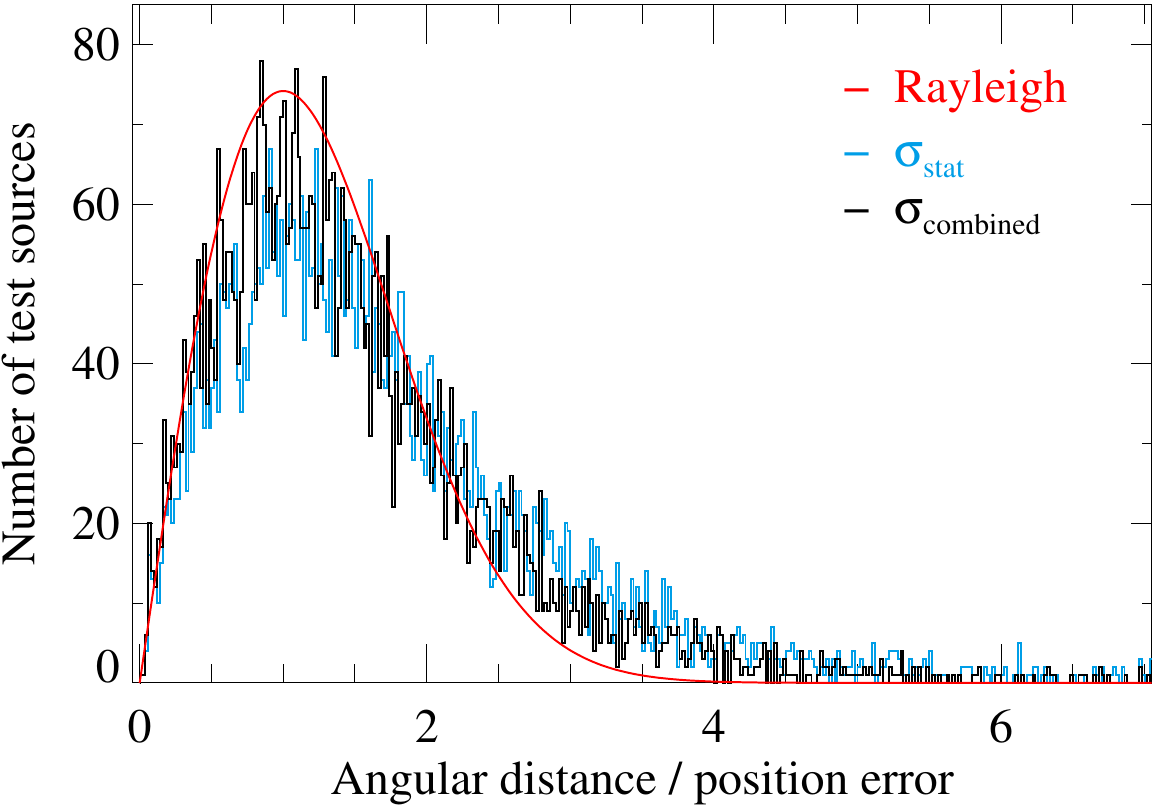}
    \caption{Error-normalised position offsets between 4XMM-DR9s and 6\,118
      SDSS-selected quasars compared to an ideal Rayleigh distribution
      \emph{(red)}. The histograms show the distributions for the pure
      statistical position error $\sigma_\textrm{stat}$ \emph{(blue)} and for
      $\sigma_\textrm{combined}$ \emph{(black)} which includes the additional
      error component $\sigma_\textrm{add}$=0.227\arcsec.}
    \label{fig:rayleighfit}
  \end{figure}

  The median statistical position error $\textrm{RADEC\_ERR}\,/\sqrt{2}$ of
  4XMM-DR9s is 0.8\arcsec\ for point-like sources, 1.6\arcsec\ for extended
  sources in overlap areas, and 2.3\arcsec\ for extended sources in uniquely
  observed areas. To validate their positional accuracy, we compare them to
  the better constrained positions of SDSS-DR12 quasars
  (\citealtads{2015ApJS..219...12A}, for the method
  cf.\ \citealtads{2019A&A...624A..77T}). From the best matches within a
  radius of 15\arcsec\ between all sources in both catalogues, 6\,118
  spectroscopically confirmed quasars are selected. Their offsets in units of
  the 1$\sigma$-error are shown in Fig.~\ref{fig:rayleighfit}. Ideal positions
  and position errors are expected to be Rayleigh distributed. Deviations from
  a Rayleigh distribution indicate that the position errors are not purely
  statistical, but include systematics for example from uncertainties in the
  boresight calibration. An additional error component is thus added
  quadratically to the statistical position errors as
  $\sigma_\textrm{combined}=(\sigma_\textrm{stat}^2+\sigma_\textrm{add}^2)^{0.5}$
  and varied to find the best agreement with a Rayleigh distribution. The best
  fit is achieved with $\sigma_\textrm{add}$=0.227\arcsec. To compare with
  3XMM-DR7s, we also determine a linearly added error component
  $\sigma_\textrm{stat}+\sigma_\textrm{lin}$. In a slightly poorer fit than
  the quadratic term, it evaluates to 0.064\arcsec. The results are
  essentially the same when reducing the matching radius to 5\arcsec. The
  error-normalised offsets peak neatly at one, but show similar deviations
  from an ideal Rayleigh distribution as the other serendipitous catalogues
  \citepads[e.g.][]{2016A&A...590A...1R,2020dr9}: a lower maximum and a
  broader tail, suggesting that part of the EPIC error might be
  underestimated.
  
  When working with large images in tangential projections, potential
  distortions have to be taken into account. The catalogue provides the
  distance of a source to the reference coordinates of the stack in the column
  DIST\_REF. The largest image in the catalogue is almost 10\degr\ wide, and
  the largest offset of a source from the stack centre is 4.8\degr. Since the
  PSFs are chosen for azimuth and off-axis angle with respect to the aim point
  of the individual observation and fitted in sub-images, projection effects
  are negligible anyway. Source positions in 4XMM-DR9 and DR9s show no
  systematic deviation from each other. In particular there is no offset as a
  function of DIST\_REF.

  \subsection{Comparison to 4XMM-DR9}
  \label{sec:dr9dr9s}

  In the observations that were used to compile 4XMM-DR9s, 4XMM-DR9 records
  282\,804 unique sources\footnote{This number includes 3\,192 detections of
    1\,320 unique sources discovered solely on the central CCD of exposures in
    small-window mode. These CCDs were discarded from the 4XMM-DR9s
    processing.}. Almost 85\,\% of the DR9 sources have at least one
  contributing detection with a good quality flag 0 or 1. Applying the
  matching strategy of Sect.~\ref{sec:dr9matches}, we find DR9 counterparts
  for 214\,170 sources in 4XMM-DR9s. For 193 DR9s sources, more than one DR9
  source was located within the matching radius and the closest match
  chosen. A total of 74\,021 4XMM-DR9s sources have no association. Part of
  this increase is regarded a benefit of the depth of the repeated
  observations. Another part, typically close to the detection limit, appears
  as a consequence of the astrometric correction applied prior to stacked
  source detection and the different image binning. We address these aspects
  by firstly analysing the source content of the two catalogues in non-overlap
  areas to then estimate the increase of sources in the overlapping areas
  thanks to stacking them.

\begin{figure}
    \centering
    \includegraphics[height=.52\linewidth]{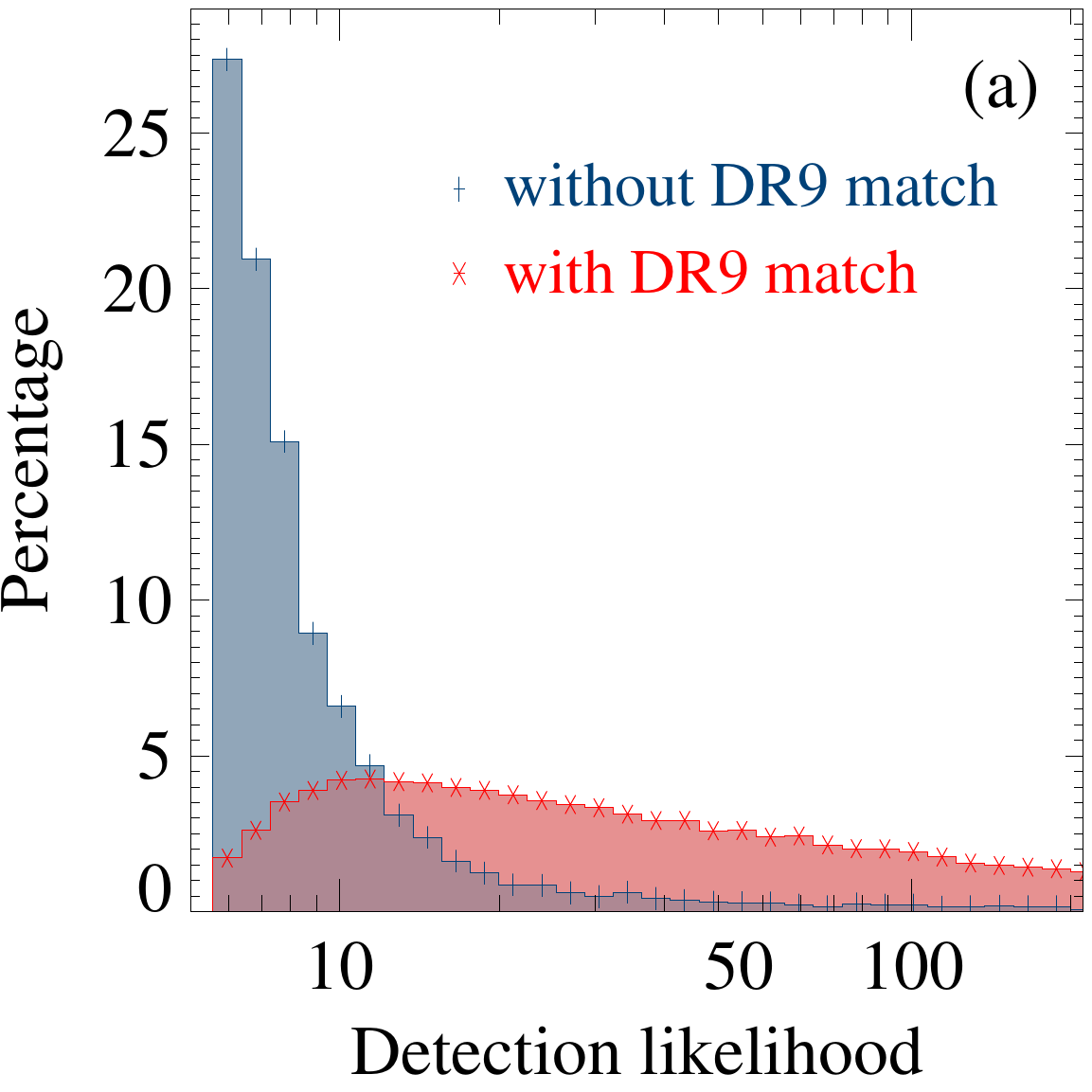}\hfill%
    \includegraphics[height=.52\linewidth]{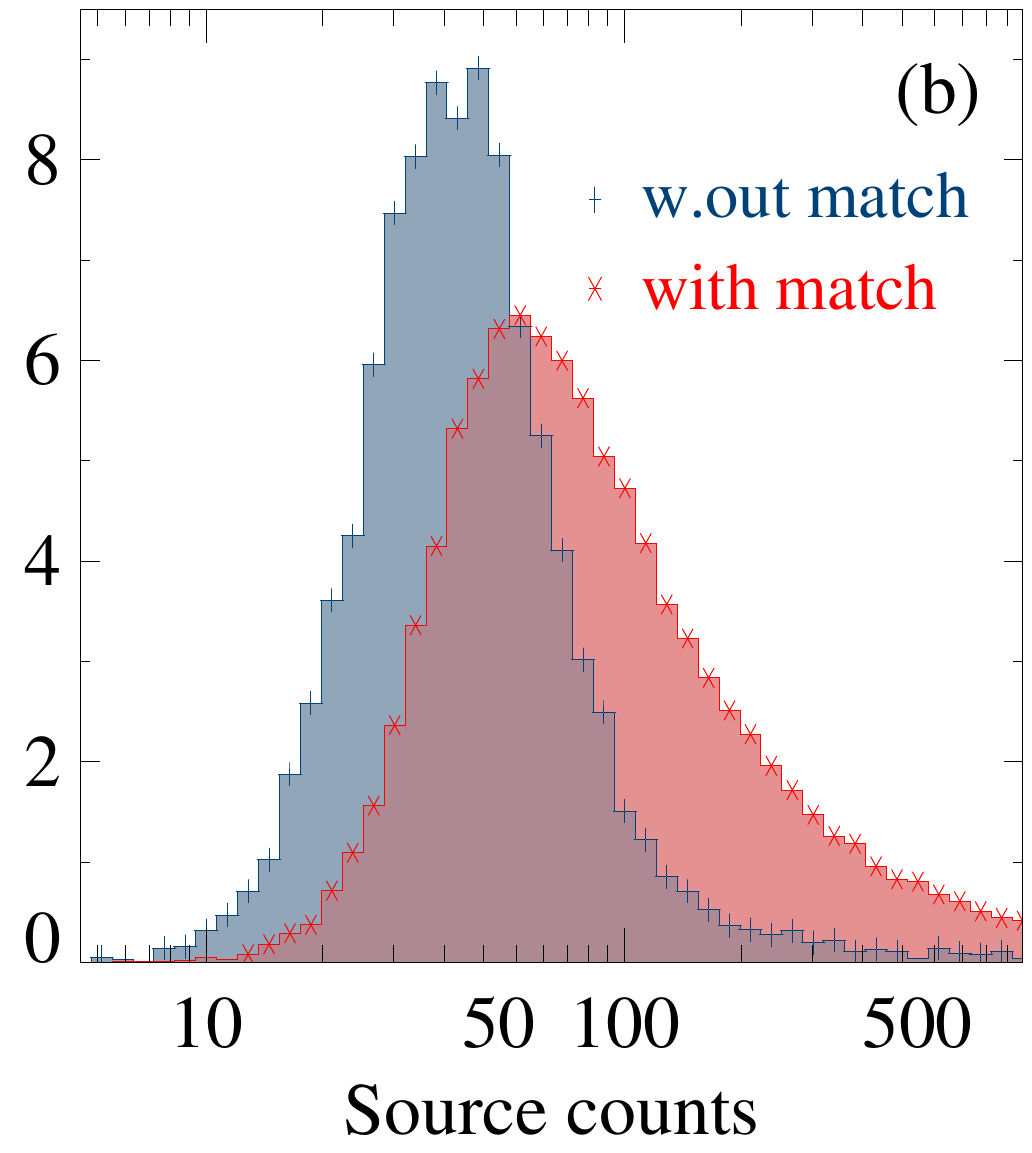}\vskip1mm
    \includegraphics[height=.52\linewidth]{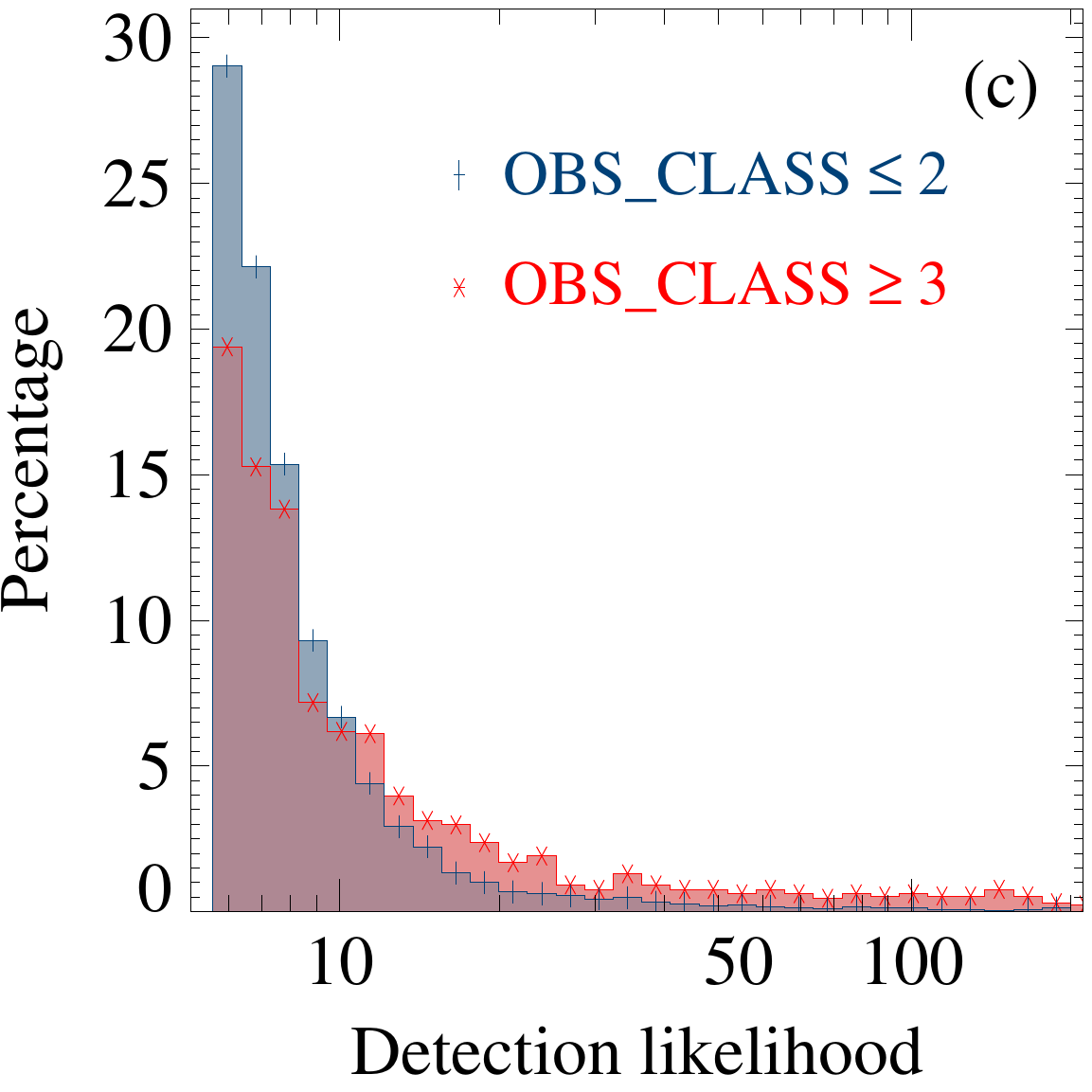}\hfill%
    \includegraphics[height=.52\linewidth]{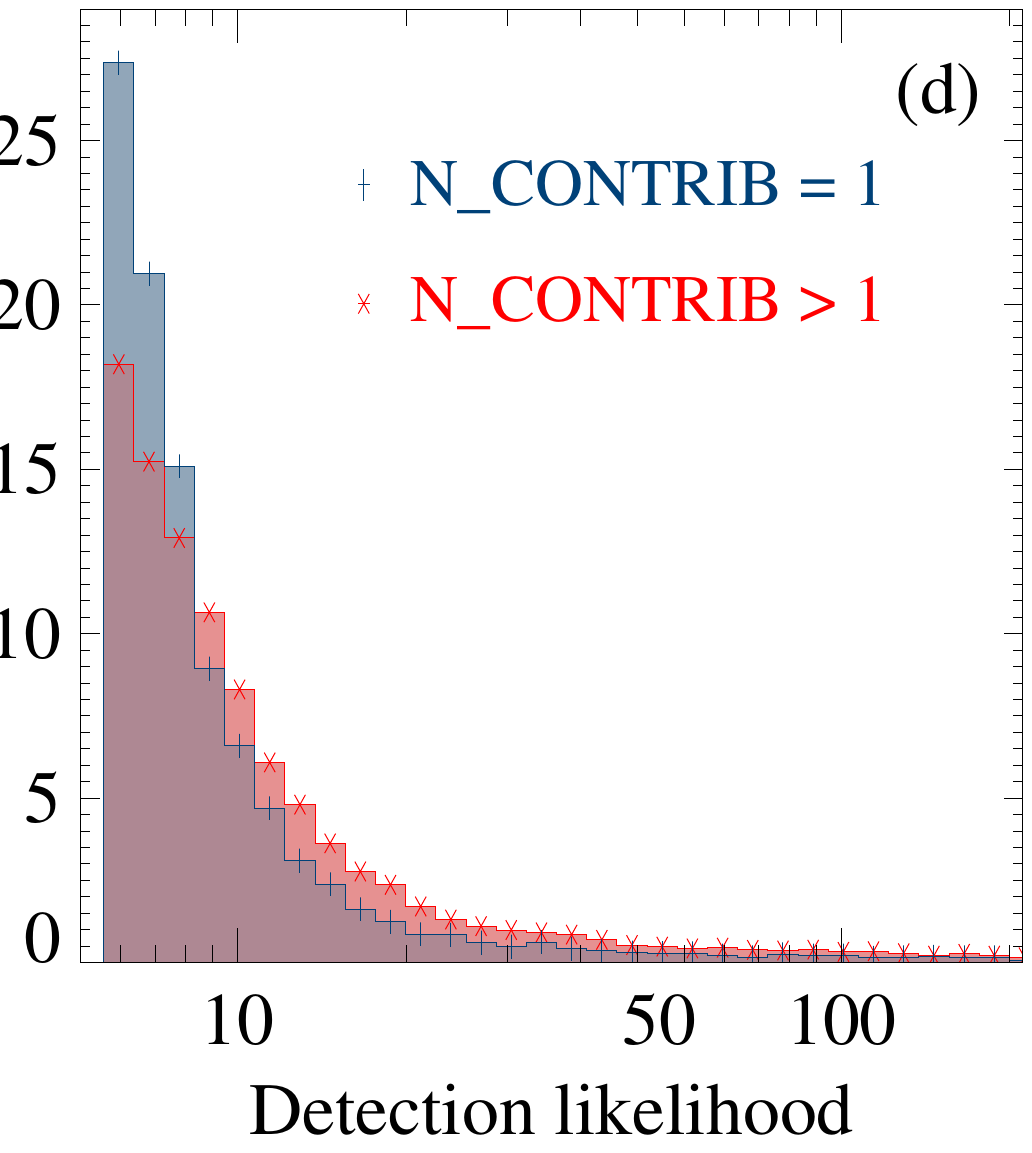}
    \caption{Parameters of 4XMM-DR9s-only sources with one contributing
      observation. \emph{Upper panels:} Detection likelihoods \emph{(panel a)}
      and source counts \emph{(panel b)} of N\_CONTRIB=1-sources without
      \emph{(blue)} and with \emph{(red)} a DR9 association. \emph{Lower
        panels: Panel c:} Detection likelihoods of DR9s-only sources with
      N\_CONTRIB=1 in good \emph{(blue)} and complex \emph{(red)} sky areas
      according to the DR9 quality index OBS\_CLASS. \emph{Panel d:} Detection
      likelihoods of DR9s-only sources with N\_CONTRIB=1 \emph{(blue)}
      compared to DR9s-only sources in overlap areas \emph{(red)}. All
      distributions are normalised to the sample size.}
    \label{fig:nodr9match}
  \end{figure}

  In regions observed only once, DR9s and DR9 represent two source detection
  runs on the shifted and un-shifted event lists. Out of the 65\,307
  un-flagged DR9s sources with just one contributing observation, 9\,018 were
  not associated to a 4XMM-DR9 source. 7\,620 of them are point-like and
  1\,758 are extended sources with a larger positional uncertainty than
  point-like sources. The detection likelihoods and counts of the point-like
  un-flagged sources are shown in Fig.~\ref{fig:nodr9match}. Most of the
  non-matches are close to the detection limit with low detection likelihood
  and few counts: 66\,\% have detection likelihoods below 10 and 85\,\% below
  20. Close to the detection limit, differences between DR9s and DR9 source
  lists mainly arise from the astrometric correction, which is applied to the
  observations before running stacked source detection for DR9s and only after
  source detection on the individual observations for DR9. The different
  reference coordinates for the image binning, which are common coordinates
  for all observations in a stack for DR9s versus the individual aim point of
  each observation for DR9, may also slightly influence the fit results in
  particular near the detection limit. Both astrometric correction and
  reference coordinates alter the centring of the image bins and cause a
  slightly different distribution of the photons over the pixels. This affects
  the PSF fit and thus the detection likelihood. Changes in the likelihood can
  result in gain or loss of sources close to the detection limit.

  A further reason for non-matches between DR9 and DR9s lies in the
  pre-selection of sources before matching them. DR9 sources with a poor
  quality flag SUM\_FLAG$\geq$2 were not included in the matching exercise
  described in Sect.~\ref{sec:dr9matches} and thus not listed in 4XMM-DR9s. We
  investigate the effect of this restriction in a second match between all
  un-matched DR9s and DR9 sources, now using all DR9 sources irrespective of
  their quality flag. The distance between matches is now limited to a maximum
  of 1\arcmin\ to reduce false associations of sources with large positional
  errors. More than 80\,\% of the DR9s-only sources with STACK\_FLAG$\leq$1
  are close to a DR9 source with SUM\_FLAG$\geq$2.

  In addition, a significantly higher percentage of un-matched than of matched
  DR9s sources is subject to source confusion, and many un-matched sources are
  located in problematic observations for example with high X-ray background,
  bright single-reflection patterns, or very complex extended structures,
  where the source position can be determined less precisely. A non-quantified
  fraction of sources will thus be lost through the matching
  radius. Figure~\ref{fig:nodr9match}c shows the likelihood distributions for
  different observation qualities. At detection likelihoods above 20, about
  40\,\% of the un-matched sources are in OBS\_CLASS$\leq$3 observations
  compared to 9\,\% of the matched sources. Visual inspection confirms that
  bright sources without a DR9 association, which can be seen in
  Fig.~\ref{fig:nodr9match}b, are typically located in such problematic
  regions, or their potential counterpart in DR9 was flagged with
  SUM\_FLAG$\geq$2 and therefore not used in the match. Thus, sources in
  non-overlapping areas come without a DR9 match through various reasons and
  partly combinations of them: different image binning and the corresponding
  fitting effects, poor observing conditions, complex X-ray structures, source
  confusion, asymmetric use of flagged sources in the match not to contaminate
  DR9s with questionable DR9 associations.


  Finally, we investigate the DR9s-only sources in overlap regions
  (N\_CONTRIB$>$1), hence in genuine stacking areas. Compared to
  non-overlapping areas, the source density is increased thanks to the higher
  cumulated exposure time (cf.\ Fig.~\ref{fig:dense}). Through stacked source
  detection, the source positions are better constrained, which gives more
  reliable matching results. In those sky areas, about 20\,\% of the DR9s
  sources with the best quality flag STACK\_FLAG=0 have no counterpart in
  DR9. The fraction of DR9s-only sources is thus about twice as high in
  overlapping areas than in non-overlap areas. The un-flagged DR9s-only
  sources tend to have higher detection likelihoods than in non-overlap areas
  (Fig.~\ref{fig:nodr9match}d), a larger fraction is found in regions without
  source confusion, and the fraction of DR9s-only sources in observations with
  poor OBS\_CLASS$\geq$3 is lower than in non-overlap areas. Eighty-two
  percent of the DR9s-only sources involve at least one good observation with
  OBS\_CLASS$\leq$2, compared to 94\,\% of the sources with a DR9
  association. If including 4XMM-DR9 sources in the match which only have
  detections with problematic quality flags SUM\_FLAG$\geq2$ as described in
  the previous paragraph, another 3\,426 potential DR9 counterparts to DR9s
  sources with STACK\_FLAG$\leq1$ are found, still leaving 18\,\% of them
  un-matched (22\,\% of all DR9s sources and quality flags). The high fraction
  of DR9s-only sources in overlap areas is thus regarded being mostly an
  effect of the higher sensitivity through stacking.

  \subsection{Long-term variability}
  \label{sec:ltv}

  \begin{figure}
    \centering
    \includegraphics[width=\linewidth]{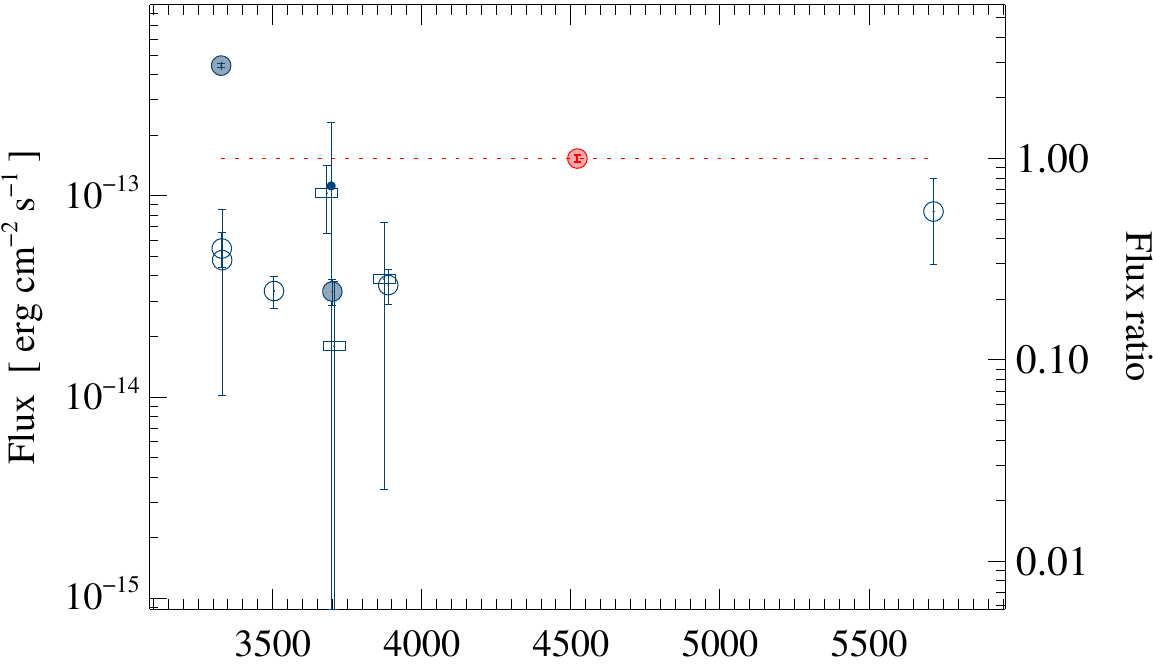}\vskip1pt
    \includegraphics[width=\linewidth]{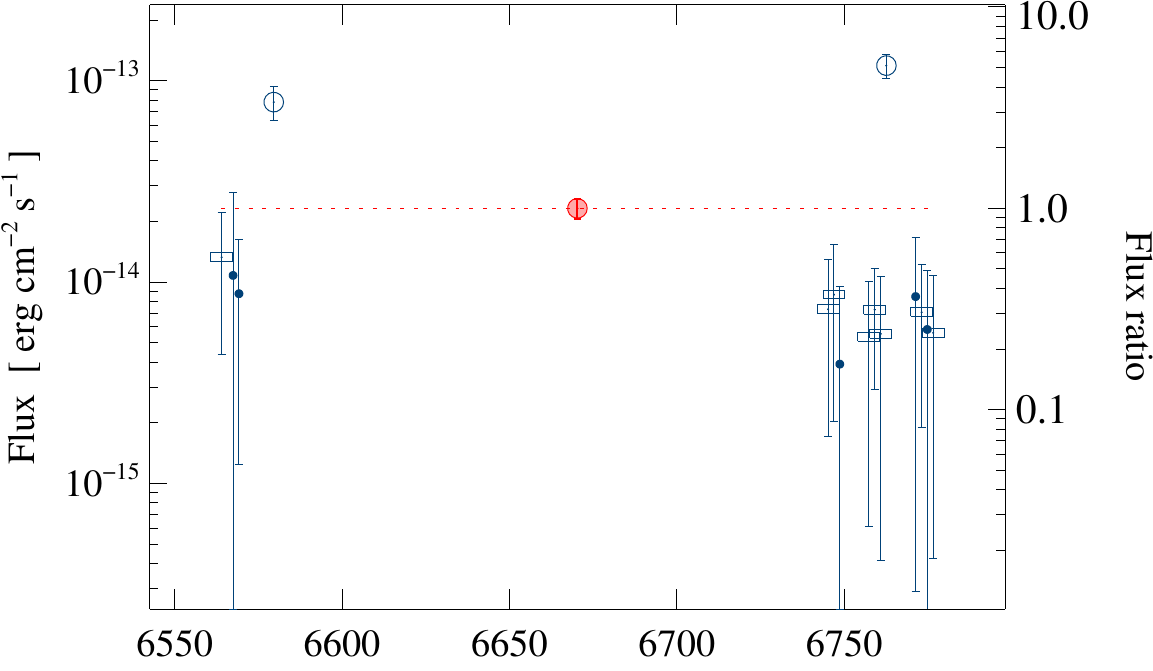}\vskip1pt
    \includegraphics[width=\linewidth]{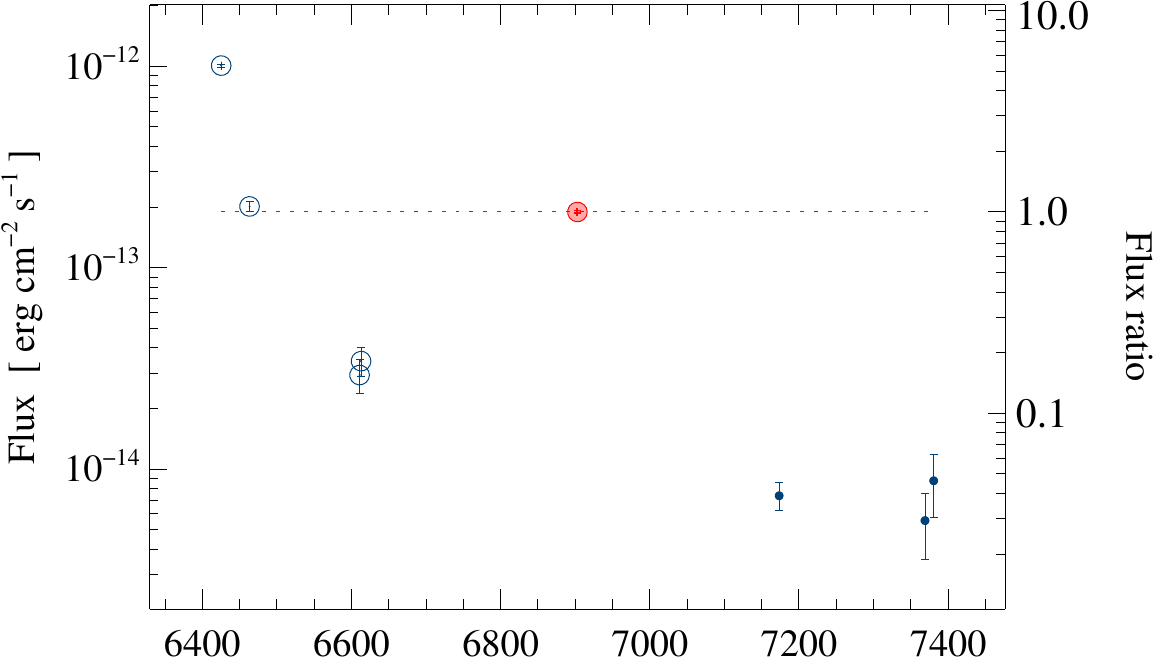}\vskip1pt
    \includegraphics[width=\linewidth]{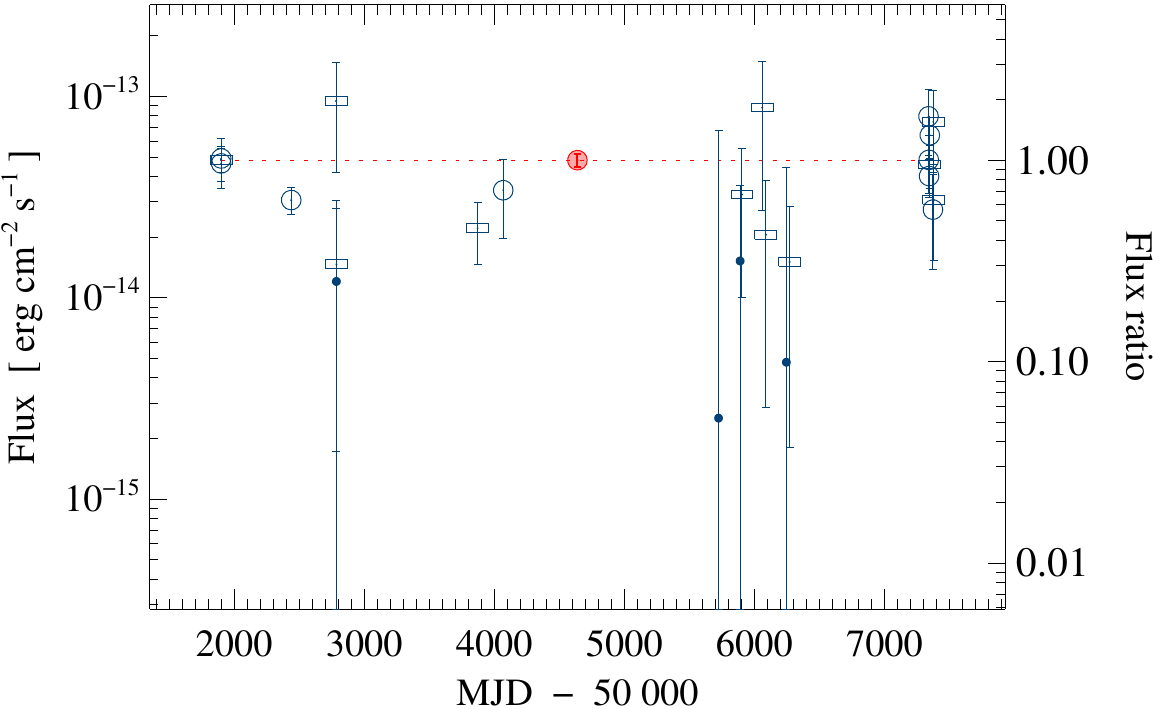}
    \caption{Example long-term light curves of variable sources in
      4XMM-DR9s. \emph{Red:} mean all-EPIC flux and flux error. \emph{Blue:}
      observation-level fluxes. The plot symbols code the
      variability. Information on short-term variability within an observation
      is inherited from 4XMM-DR9. \emph{Filled circle:}
      variability. \emph{Open circle:} no variability. \emph{Box:} short-term
      variability unknown. \emph{Dot:} non-detection in 4XMM-DR9. From top to
      bottom, the panels show a flaring M dwarf, a star of unknown optical
      variability, a gamma-ray burst afterglow, which became undetectable in
      single \textit{XMM-Newton} observations after about two years, and a
      quasar.}
    \label{fig:examplelcs}
  \end{figure}

  Variability studies directly from stacked source detection have major
  advantages over individual detections: Fluxes are determined at the same
  position in all exposures. The total flux is derived in the simultaneous fit
  without need to match single detections, which might involve false
  associations. And lastly, the fluxes are determined for any detection
  likelihood in the individual observation. About 12\,\% of the
  observation-level fluxes in 4XMM-DR9s are derived for a source likelihood
  below the detection limit, thus for a non-detection in the respective
  individual observation. Fake variability however can still rarely arise, for
  example from hot pixels during an observation or spurious detections close
  to a bright source. Quality filtering based on the flags provided in the
  catalogue is thus essential and visual inspection of the auxiliary source
  images delivered with the catalogue is recommended in searches for so far
  unknown long-term variability between observations.

  For each catalogue source with at least two valid measurements, the
  variability parameters defined by \citetads{2019A&A...624A..77T} are
  provided. They are calculated from the stacked and observation-level fluxes
  in each energy band $F_k$, where $k$ runs from 1 to $n$ observations
  (columns EP\_$e$\_FLUX in the catalogue, $e=1..5$), from the total flux
  $F_\textrm{EPIC}$ (column EP\_FLUX in the catalogue), and from their
  respective 1$\sigma$ errors: the reduced $\chi^2$ of long-term flux changes
    \begin{equation}
      \textrm{VAR\_CHI2} = \frac{1}{n-1}\sum_{k=1}^n
      \left(\frac{F_k-F_\textrm{EPIC}}{\sigma_k}\right)^2 ,  
    \end{equation}
    the associated cumulative chi-square probability of the flux measurements
    being consistent with constant flux
    \begin{equation}
      \textrm{VAR\_PROB} = \int_{\chi^2}^\infty
      \frac{x^{\nu/2-1}e^{-x/2}}{2^{\nu/2}\Gamma(\nu/2)} dx ,
    \end{equation}
    where smaller values indicate a higher chance that the source is variable
    and $\Gamma$ denotes the gamma function, the ratio between the highest and
    lowest observation-level flux
    \begin{equation}
      \textrm{FRATIO} = F_\textrm{max}/F_\textrm{min} ,
    \end{equation}
  the associated 1$\sigma$ error
    \begin{equation}
    \textrm{FRATIO\_ERR} =
    \left(\frac{\sigma_{F\textrm{min}}^2}{F_\textrm{min}^2}+
    \frac{\sigma_{F\textrm{max}}^2}%
         {F_\textrm{max}^2}\right)^{0.5}\,\frac{F_\textrm{max}}{F_\textrm{min}} ,
    \end{equation}
    and the largest flux difference between any combination of the
    observation-level fluxes in terms of $\sigma$
    \begin{equation}
      \textrm{FLUXVAR} =
      \max_{k,l\in[1,n]}\frac{|F_k-F_l|}{\sqrt{\sigma_k^2+\sigma_l^2}}
    \end{equation}
  where both $k$ and $l$ cover the indices of the $n$ observations of the
  source.

  For a thorough variability analysis, several variability parameters should
  be studied jointly. Here, we concentrate on the all-EPIC parameter VAR\_PROB
  only, which is the probability that the mean fluxes per observation are
  consistent with constant long-term behaviour. 7\,182 un-flagged sources are
  likely long-term variable with $\textrm{VAR\_PROB}\leq 10^{-5}$ (3.75\,\% of
  all multiply observed un-flagged sources), and 11\,327 (5.91\,\%) have
  $\textrm{VAR\_PROB}\leq 10^{-3}$. Almost 90\,\% of the variability
  candidates in 4XMM-DR9s have an association in 4XMM-DR9, but about 20\,\% of
  them with only one valid DR9 detection although covered by several
  observations. For the latter ones, stacked source detection is the only way
  to investigate the long-term evolution of their fluxes, which are derived
  for all stacked observations and do not depend on the detectability of a
  source in the individual observation as in DR9. In DR9, short-term
  intra-observation variability is determined by means of a $\chi^2$
  variability test on the light curves of a detection in each instrument
  against the null-hypothesis that the brightness fluctuations are consistent
  with constant source flux \citepads{2016A&A...590A...1R,2020dr9}. A source
  is considered short-term variable if the probability derived from the
  $\chi^2$ test is $10^{-5}$ or lower in at least one instrument in at least
  one observation. The light curves of 26.5\,\% of the unique 4XMM-DR9 sources
  have enough good time bins to perform the $\chi^2$ test. Among the DR9
  associations of long-term variable 4XMM-DR9s sources with
  $\textrm{VAR\_PROB}\leq 10^{-5}$, 14\,\% are marked as short-term variable
  as well, while 54\,\% show no sign of short-term variability. For the rest,
  the short-term probability could not be determined.

  The long-term light curves of the most variable 4XMM-DR9s objects show
  various kinds of variability: for example flaring or on-off behaviour, flux
  changes of the order of days, continuous brightening or darkening of
  sources. Figure~\ref{fig:examplelcs} includes example light curves from
  4XMM-DR9s. In the following subsections, we illustrate the potential of DR9s
  variability in case studies of object classes and X-ray surveys.

  \subsubsection{Variability of SIMBAD- and SDSS-classified sources}

\begin{figure}
    \centering
    \includegraphics[width=\linewidth]{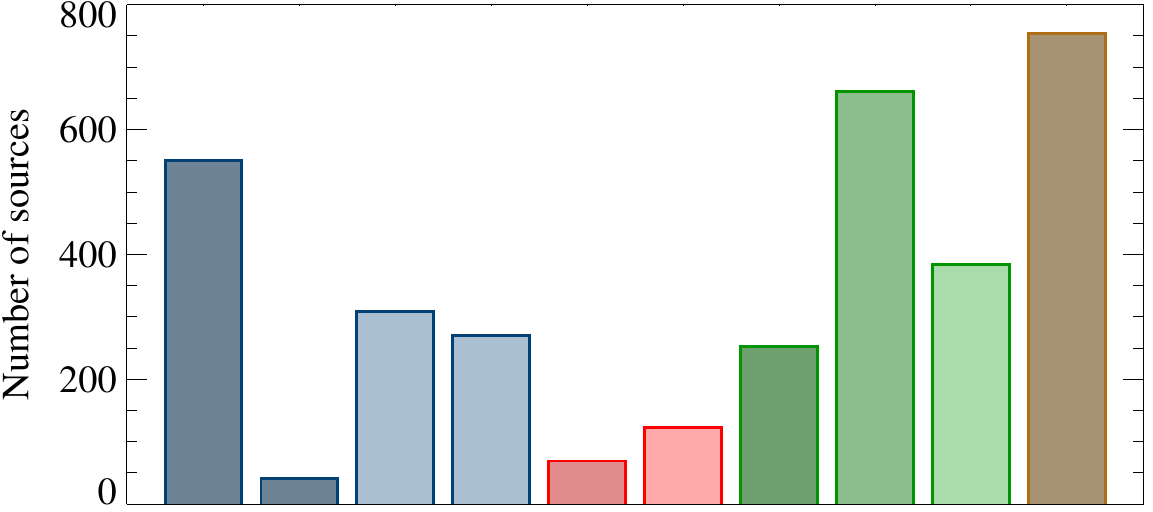}
    \includegraphics[width=\linewidth]{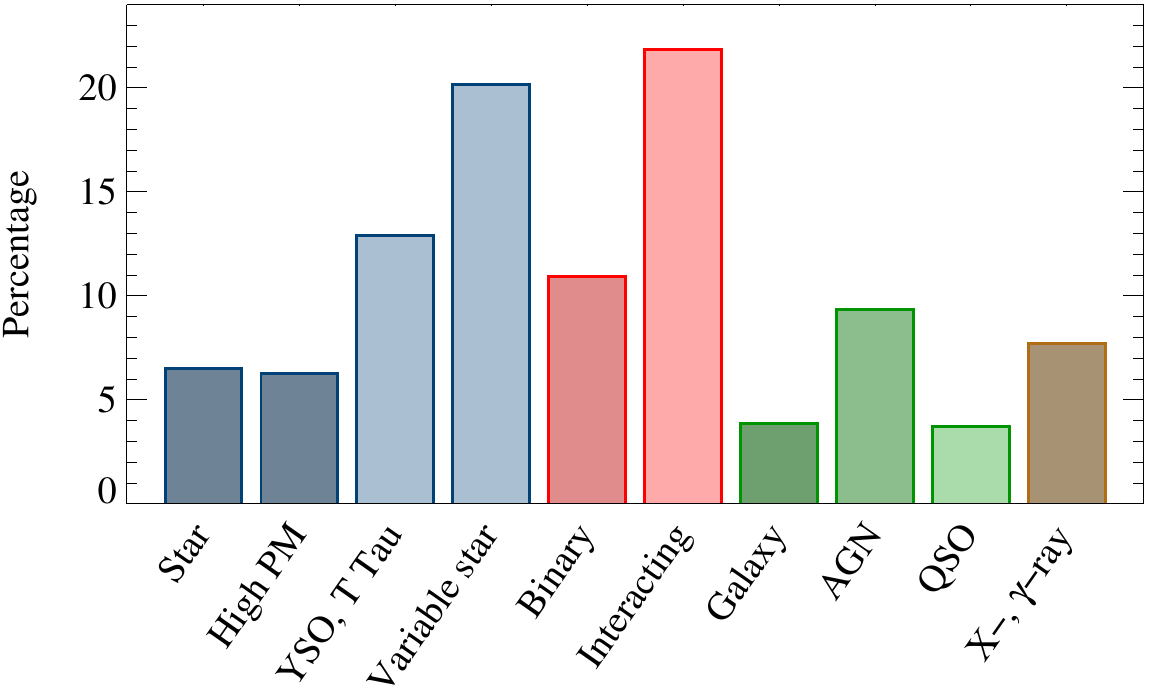}
    \caption{SIMBAD and SDSS categories of 3\,343 tentative counterparts to
      4XMM-DR9s variables with $\textrm{VAR\_PROB}\leq 10^{-5}$. \emph{Upper
        panel:} total number of matches. \emph{Lower panel:} percentage of
      variables among all matched variable and non-variable class members.}
    \label{fig:varclasses}
  \end{figure}

  More than 80\,000 DR9s un-flagged point sources in 4XMM-DR9s have a
  tentative counterpart in SIMBAD \citepads{2000A&AS..143....9W} or SDSS-DR12
  \citepads{2015ApJS..219...12A} within 3\arcsec. Among them, 3\,779 have
  $\textrm{VAR\_PROB}\leq 10^{-5}$ in 4XMM-DR9s, compared to a total of 7\,182
  un-flagged variable DR9s sources (Table~\ref{tab:catalogue}). The pure
  position match results in about 5\,\% false associations among the variable
  sources \citepads[for the method cf.][]{2019A&A...624A..77T}. We use the
  matched sources to investigate the variable content of different source
  classes. For ten source classifications in SIMBAD and -- if available -- in
  SDSS, the number of variables and their share among all matched class
  members are shown in Figure~\ref{fig:varclasses}: stars without the
  following sub-categories, high-proper motion stars, young stellar objects
  and T\,Tauri stars, non-interacting binaries, interacting binaries, galaxies
  without AGN and QSOs, AGN without QSOs, QSOs, and un-classified X-ray and
  $\gamma$-ray sources. In few cases, high-proper motion stars can mimic
  long-term variability in 4XMM-DR9s, if they are fitted as several detections
  at different positions. In total, 652 4XMM-DR9s sources are matched with a
  high-proper motion star in SIMBAD, and 41 have $\textrm{VAR\_PROB}\leq
  10^{-5}$, which is the same proportion as for all stars.

\begin{figure}
    \centering
    \includegraphics[width=\linewidth]{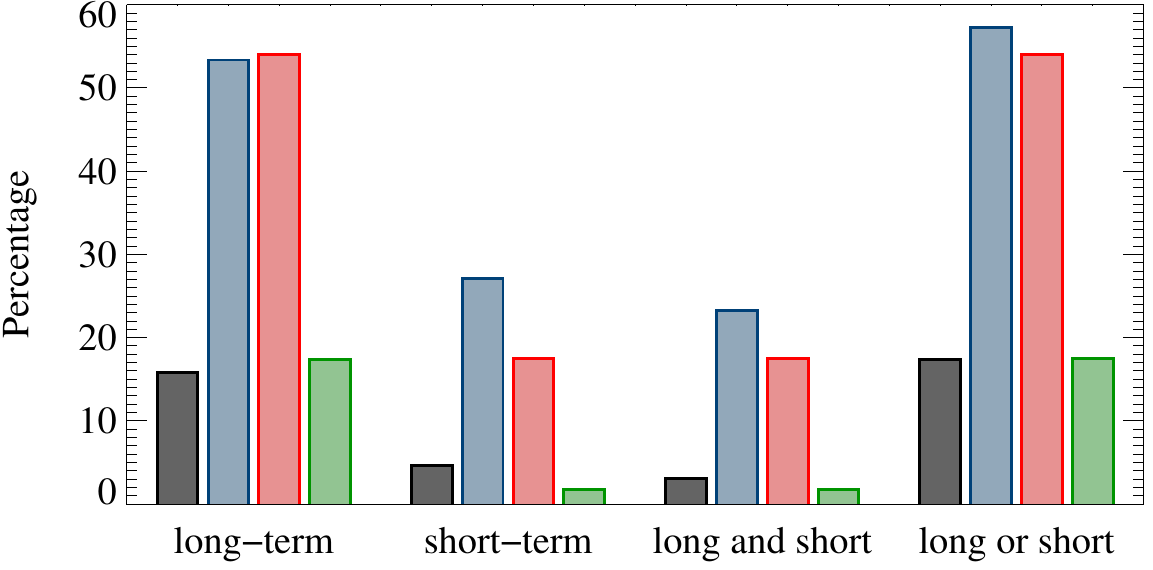}
    \caption{Long-term inter-observation and short-term intra-observation
      variability for SIMBAD- and SDSS- selected source classes: from left to
      right all classified sources \emph{(black)}, all variable stars without
      binaries \emph{(blue)}, interacting binaries \emph{(red)}, and quasars
      \emph{(green)}.}
    \label{fig:varsl}
  \end{figure}

  From the matching sample, we select sources with information on long-term
  and on short-term variability. They have at least three 4XMM-DR9s
  observations that cover a minimum time span of 30 days and a 4XMM-DR9
  association with a sufficient number of counts to determine
  intra-observation variability. Analogously to 4XMM-DR9, sources are
  considered short-term variable for $\textrm{CHI2PROB\_4XMMDR9}\leq 10^{-5}$
  and long-term variable for $\textrm{VAR\_PROB}\leq
  10^{-5}$. Figure~\ref{fig:varsl} gives the share of long- and short-term
  variables among their classes in the subsample. Three source classes are
  included, which are expected to have different fractions of long- and
  short-term variable members: stars, which can be variable on all time
  scales; interacting binaries, for part of which orbital modulations can be
  recovered within an observation and for part of which long-term flux changes
  for example due to accretion-rate changes or nova eruptions are detected;
  and quasars, which are predominantly variable on longer time scales, if at
  all.

  \subsubsection{Variable 4XMM-DR9s sources in the \textit{XMM-Newton} slew
    and eROSITA surveys}

  The \textit{XMM-Newton} slew survey \citepads{2006ESASP.604..913F},
  performed while slewing the telescope between targets, has reached a sky
  coverage of about 84\,\% as of March 2017 and an EPIC pn sensitivity of
  about $6\times 10^{-13}\,\textrm{erg\,cm}^{-2}\,\textrm{s}^{-1}$ in the soft
  energy band\footnote{In the slew catalogue, the soft energy band is defined
    as 0.2$-$2.0\,keV, and the hard band as 2$-$12\,keV. The full band is
    0.2$-$12.0\,keV as in the other catalogues.}, $3\times
  10^{-12}\,\textrm{erg\,cm}^{-2}\,\textrm{s}^{-1}$ in the hard energy band,
  and $1\times 10^{-12}\,\textrm{erg\,cm}^{-2}\,\textrm{s}^{-1}$ in total
  \citepads{2008A&A...480..611S,2012A&A...548A..99W}. To investigate its
  potential for combined variability studies, we derive the fluxes of the
  4XMM-DR9s sources in the slew energy bands and select the bright objects
  whose soft, hard, or total all-EPIC or instrument-level flux is above the
  slew sensitivity. For them, slew observations can contribute to the
  long-term variability information. Among 1\,596 bright enough point-like and
  un-flagged sources, more than 700 are in the footprint of the slew survey
  and covered up to 13 times by slew exposures. 497 have a tentative
  counterpart within a 15\arcsec\ matching radius in the full Slew Survey
  Source Catalogue XMMSL2 and 444 in its so-called clean edition, from which
  detections with low detection likelihood and detections with poor quality
  flags were removed. 358 matches with the full XMMSL2 are located in overlap
  regions of DR9s and 139 in regions observed once, increasing the pool of
  sources that can be used in \textit{XMM-Newton} variability studies. From
  all DR9s and XMMSL2 fluxes of a source, we derive a combined VAR\_PROB\_comb
  that all measurements are consistent with constant flux. Additional 28
  sources show signs of long-term variability according to
  $\textrm{VAR\_PROB\_comb}\leq 10^{-5}$, which are not detected as long-term
  variable in DR9s alone: 22 with only one DR9s observation and 6 with two
  DR9s observations, which were taken between one day and 12 years apart from
  each other.

  The Russian-German Spectrum Roentgen Gamma mission (SRG) was launched in
  July 2019 and surveys the X-ray sky with its two telescopes eROSITA and
  ART-XC for four years. eROSITA \citepads{2012arXiv1209.3114M} covers a
  similar energy range as \textit{XMM-Newton}/EPIC and commenced its survey
  end of 2019. Synergies with \textit{XMM-Newton} include in particular
  analyses of the long-term behaviour of X-ray sources targeted by both
  missions. To estimate the common source content of 4XMM-DR9s and future
  eROSITA catalogues, we derive an all-sky sensitivity map from the exposure
  forecast for eROSITA's first six-months survey (eRASS:1) and for the total
  four-year survey (eRASS:8), which are available to the eROSITA
  consortium. We then compare the EPIC fluxes of each 4XMM-DR9s source to the
  map values at the source position. Thirty-four percent the sources have a
  flux above the eRASS:1 limiting sensitivity in total or at least in one
  contributing observation. The fraction increases to 75\,\% in eRASS:8 after
  completing eROSITA's four-year survey, forming a valuable resource for
  cross-mission studies.

\section{Summary and conclusions}
\label{sec:summary}

  4XMM-DR9s is the second serendipitous source catalogue from overlapping
  observations, based on the so far largest sample of exposures. They were
  selected if they overlap by at least 1\arcmin\ in radius and have a
  background level below a 92\,\% Cauchy probability as derived in
  Sect.~\ref{sec:bkgcut}. Thanks to an event-based astrometric correction,
  which was applied at the beginning of the processing, the positional accuracy
  of the sources has been clearly improved compared to source detection in the
  uncorrected observations.

  In addition to the automated source flagging by the task \texttt{dpssflag},
  the catalogue source lists were screened visually and obviously spurious
  detections marked manually. Both processes cannot be complete, and an
  un-flagged catalogue sample cannot be expected to be free from bad
  detections. But source selection based on the quality flags reduces the
  spurious content of the catalogue significantly.

  From 6\,604 observations in 1\,329 stacks, 288\,191 unique sources were
  extracted in total, 218\,283 of them multiply observed. Additional sources
  were detected compared to source detection on individual observations, and
  the source parameters can be better constrained in overlap areas. Long-term
  inter-observation variability is investigated directly based on the
  source-detection fit without need to match detections from different
  observations. Thanks to the simultaneous fit to all observations in a stack,
  320\,590 new flux determinations and flux errors are available for 106\,127
  sources with a 4XMM-DR9 association, which could not be detected in part of
  the individual DR9 observations. 9\,912 4XMM-DR9s sources still come without
  a measurable flux in one or more contributing observations because of zero
  counts in the fit region. Upper flux limits can be retrieved from the
  upper-limit server of the SOC.

  The flux determinations from the simultaneous source detection fits let us
  directly derive variability parameters for all catalogue sources with at
  least two valid measurements. About 6\,\% of them show signs of at least
  moderate (VAR\_PROB$\leq$10$^{-3}$) and 4\,\% of high
  (VAR\_PROB$\leq$10$^{-5}$) long-term variability. Only part of them are
  known to be short-term variable as well, and about a third have no detection
  or only one detection in 4XMM-DR9. Their fluxes and long-term variability
  were measured for the very first time in 4XMM-DR9s. The catalogue from
  overlapping observations thus serves as a large data base for cross-matching
  and for long-term studies of X-ray emitting objects, also in the context of
  new and future missions like eROSITA and \textsl{Athena}, ESA's proposed
  future X-ray observatory carrying a high-resolution spectrograph and a
  wide-field imager, which was selected by ESA within its Cosmic Vision
  programme \citepads{2013arXiv1306.2307N}.
  
  The next catalogue releases in the 4XMM series will concentrate on new
  public observations. While the standard 4XMM catalogue of detections can be
  incremented by adding the new detections, a refined strategy is needed for
  the stacked catalogue. The additional observations may form new stacks which
  can be included directly, they may become part of existing stacks, or may
  even combine previously independent stacks. Therefore, source detection will
  be (re-)run on all old and new observations contributing to the modified sky
  area of the newly designed stacks. Sources from those sky areas contained in
  the current catalogue will be replaced with those from the new run. It is
  foreseen to publish the catalogue updates on an approximately yearly basis.


\begin{acknowledgements}

  We thank our anonymous referee for the useful comments. The support of SSC
  work at AIP by Deutsches Zentrum f\"ur Luft- und Raumfahrt (DLR) through
  grants 50\,OX\,1701 and 50\,OX\,1901 is gratefully acknowledged. We
  particularly appreciate the close collaboration with the colleagues at ESA's
  \textit{XMM-Newton} Science Operations Centre (SOC) and the support by the
  CDS team. The French teams are grateful to Centre National d'\'Etudes
  Spatiales (CNES) for their outstanding support for the SSC activities. FJC
  acknowledges financial support through grant AYA2015-64346-C2-1P
  (MINECO/FEDER). FJC and MTC acknowledges financial support from the Spanish
  Ministry MCIU under project RTI2018-096686-B-C21 (MCIU/AEI/FEDER/UE),
  cofunded by FEDER funds and from the Agencia Estatal de Investigaci\'on,
  Unidad de Excelencia Mar\'ia de Maeztu, ref.\ MDM-2017-0765.

  This project has made use of CDS services and the SIMBAD database, operated
  at CDS, Strasbourg, France, of FTOOLS by NASA's HEASARC
  \citepads{1995ASPC...77..367B}, of TOPCAT/STILTS
  \citepads{2005ASPC..347...29T}, and of SDSS-III. Funding for SDSS-III has
  been provided by the Alfred P. Sloan Foundation, the Participating
  Institutions, the National Science Foundation, and the U.S.\ Department of
  Energy Office of Science. The SDSS web site is www.sdss.org. SDSS is managed
  by the Astrophysical Research Consortium for the Participating Institutions
  of the SDSS Collaboration including the Brazilian Participation Group, the
  Carnegie Institution for Science, Carnegie Mellon University, the Chilean
  Participation Group, the French Participation Group, Harvard-Smithsonian
  Center for Astrophysics, Instituto de Astrof\'isica de Canarias, The Johns
  Hopkins University, Kavli Institute for the Physics and Mathematics of the
  Universe (IPMU) / University of Tokyo, the Korean Participation Group,
  Lawrence Berkeley National Laboratory, Leibniz Institut f\"ur Astrophysik
  Potsdam (AIP), Max-Planck-Institut f\"ur Astronomie (MPIA Heidelberg),
  Max-Planck-Institut f\"ur Astrophysik (MPA Garching), Max-Planck-Institut
  f\"ur Extraterrestrische Physik (MPE), National Astronomical Observatories
  of China, New Mexico State University, New York University, University of
  Notre Dame, Observat\'ario Nacional / MCTI, The Ohio State University,
  Pennsylvania State University, Shanghai Astronomical Observatory, United
  Kingdom Participation Group, Universidad Nacional Aut\'onoma de M\'exico,
  University of Arizona, University of Colorado Boulder, University of Oxford,
  University of Portsmouth, University of Utah, University of Virginia,
  University of Washington, University of Wisconsin, Vanderbilt University,
  and Yale University.

\end{acknowledgements}


\bibliographystyle{aa}
\bibliography{37706}

\begin{appendix}
\label{sec:appendix}

\section{Auxiliary information on the catalogue}

  \begin{figure}
    \centering
    \includegraphics[width=\linewidth]{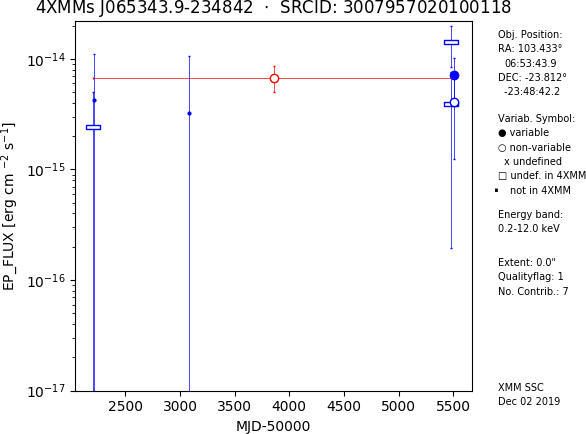}
    \caption{Example of the auxiliary light curves produced for 4XMM-DR9s
      sources and published in the XSA.}
    \label{fig:auxlc}
  \end{figure}

  \begin{figure}
    \centering
    \includegraphics[width=\linewidth]{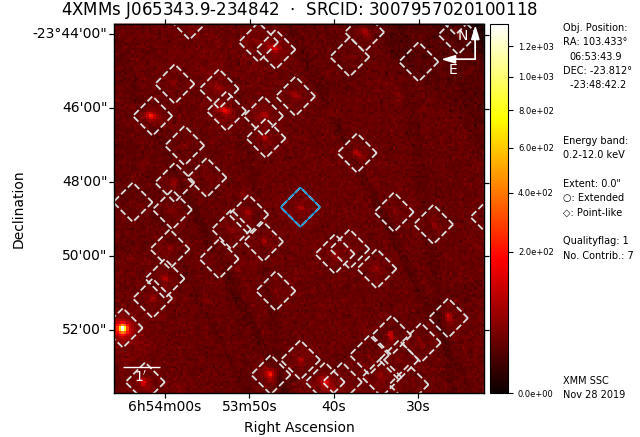}\vskip4mm
    \includegraphics[width=\linewidth]{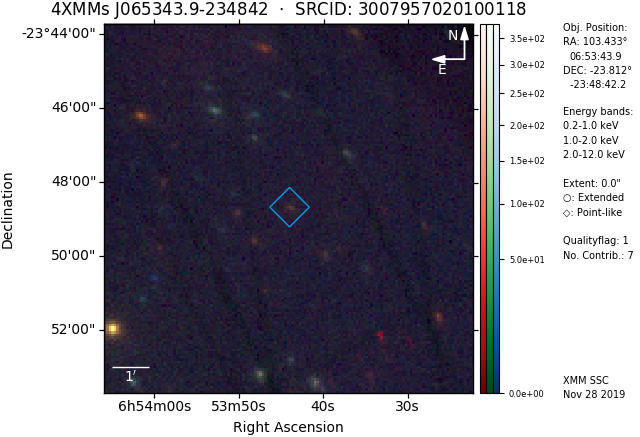}\vskip4mm
    \includegraphics[width=\linewidth]{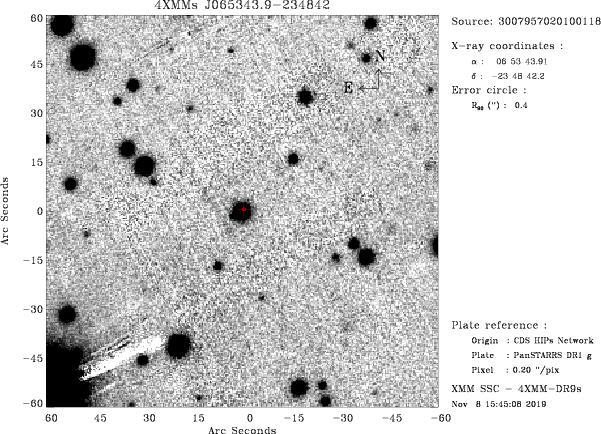}
    \caption{Example of the auxiliary images produced for 4XMM-DR9s sources
      for the object of Fig.~\ref{fig:auxlc}: full-band X-ray image
      \emph{(upper panel)}, false-colour X-ray image \emph{(middle panel)},
      and optical finding chart \emph{(lower panel)}.}
    \label{fig:auximas}
  \end{figure}


  \begin{figure*}
    \centering
    \includegraphics[width=.33\linewidth]{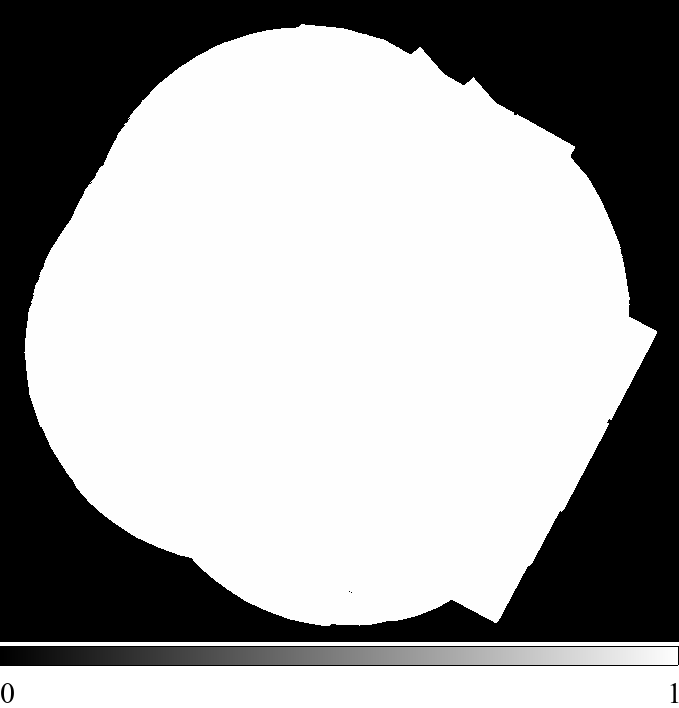}\hfill%
    \includegraphics[width=.33\linewidth]{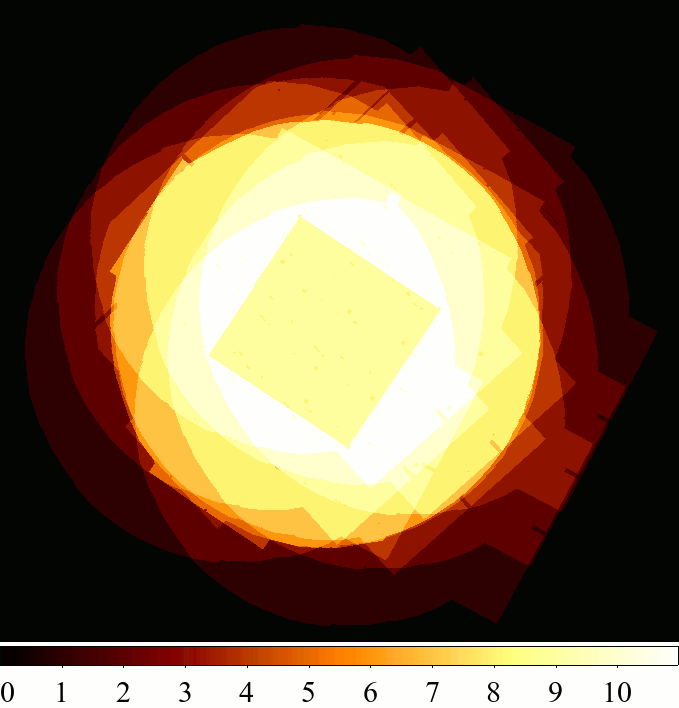}\hfill%
    \includegraphics[width=.33\linewidth]{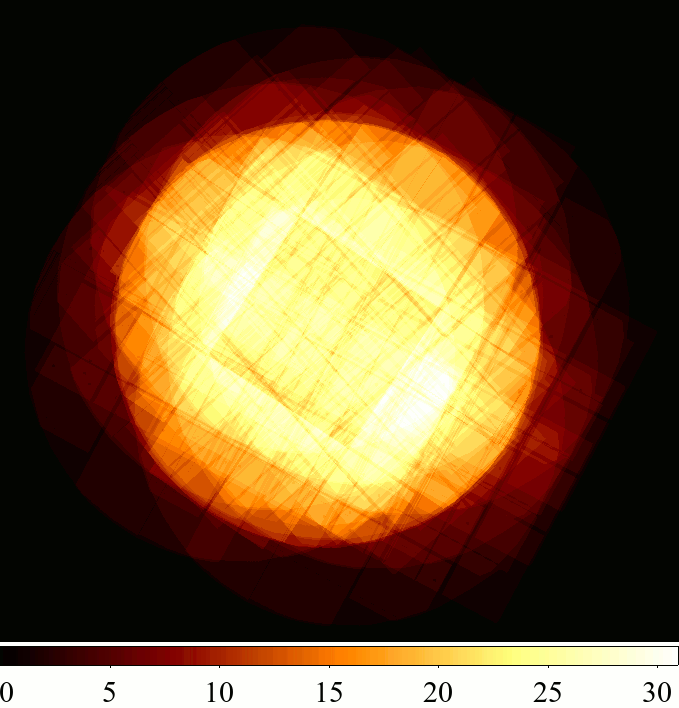}\vskip2mm
    \includegraphics[width=.33\linewidth]{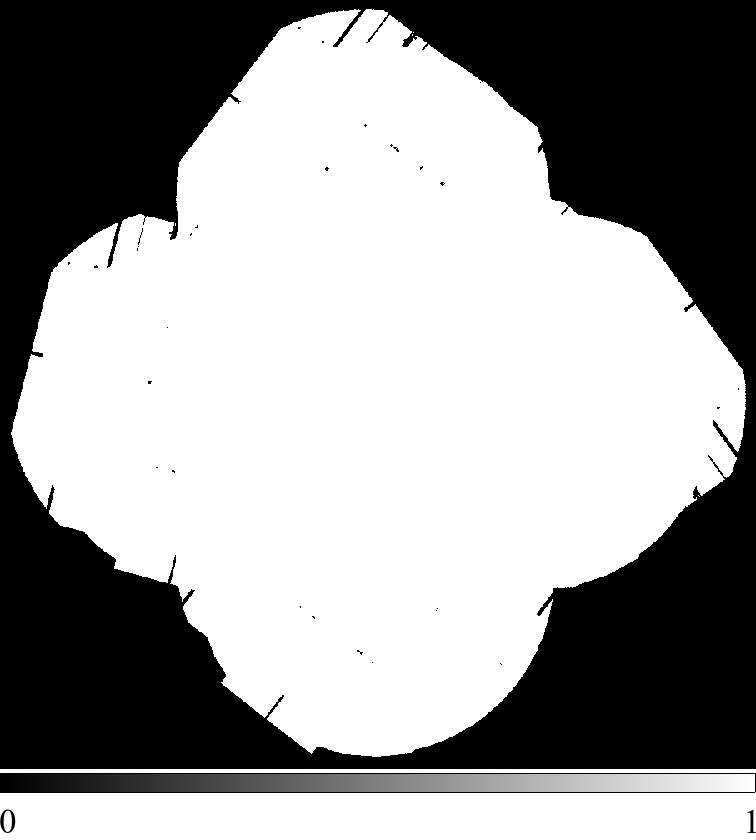}\hfill%
    \includegraphics[width=.33\linewidth]{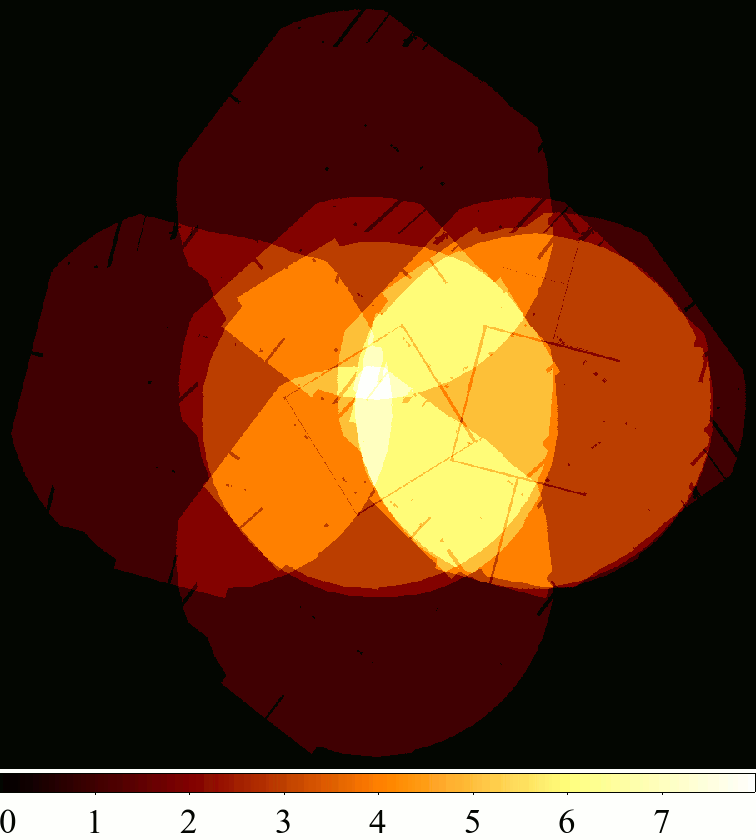}\hfill%
    \includegraphics[width=.33\linewidth]{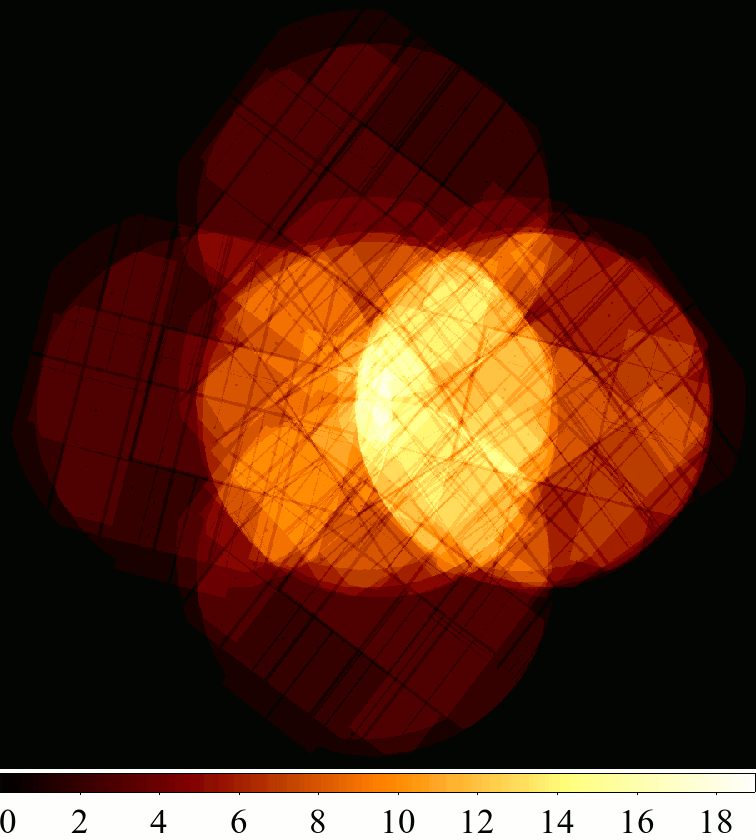}\vskip2mm
    \includegraphics[width=.33\linewidth]{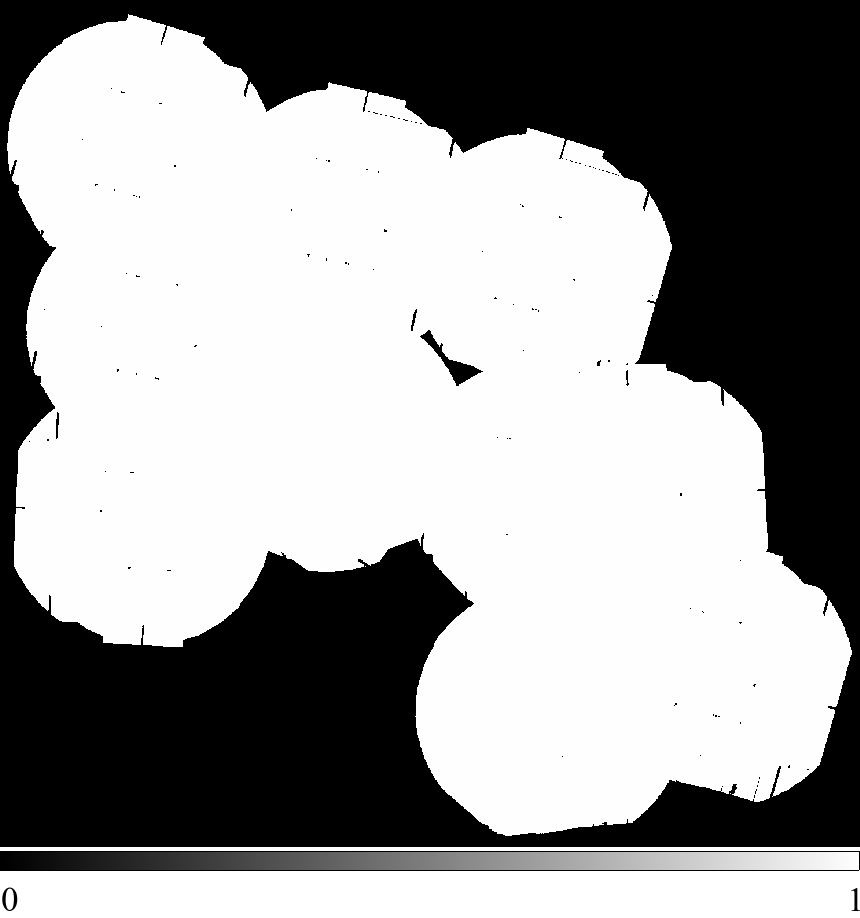}\hfill%
    \includegraphics[width=.33\linewidth]{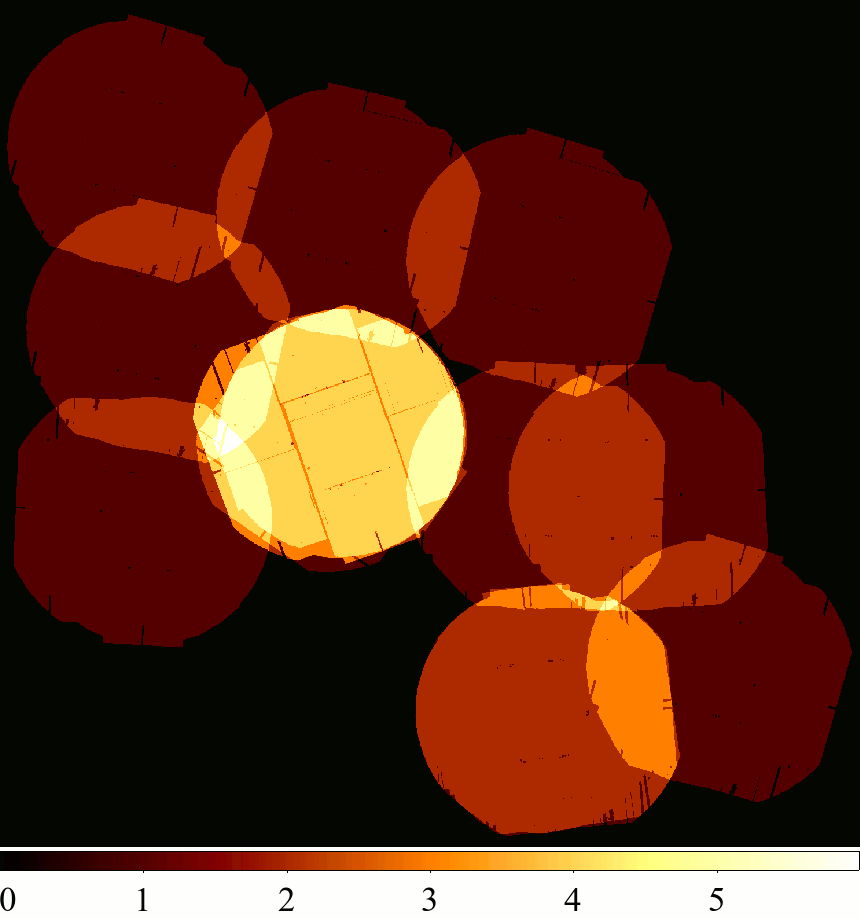}\hfill%
    \includegraphics[width=.33\linewidth]{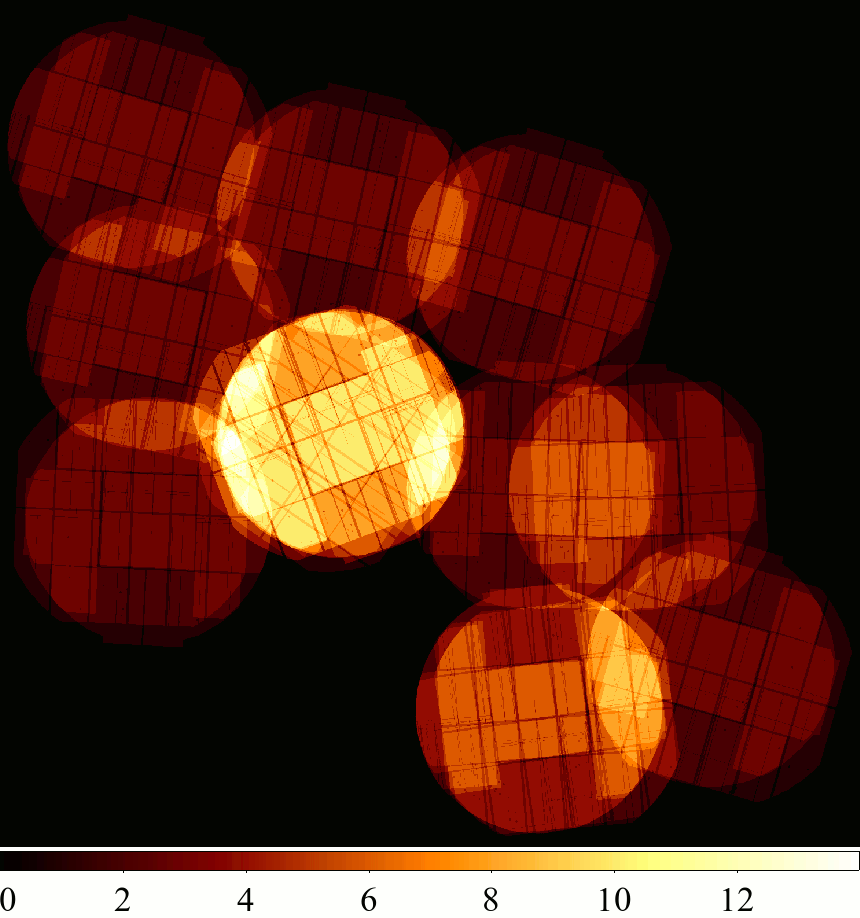}\vskip2mm
    \caption{Examples of the coverage maps of 4XMM-DR9s stacks: exposed and
      un-exposed pixels \emph{(left panels)}, number of overlapping
      observations \emph{(middle panels)}, and number of overlapping
      exposures \emph{(right panels)}.}
    \label{fig:covmaps}
  \end{figure*}

  \begin{figure*}
    \includegraphics[width=\linewidth]{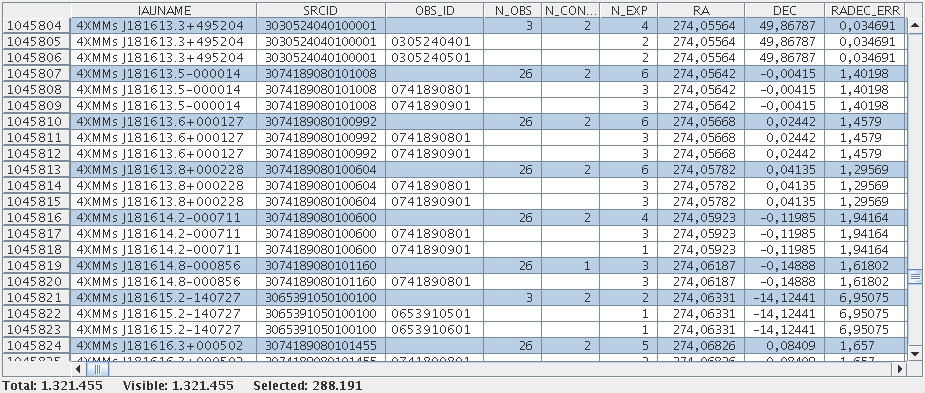}
    \caption{Screen shot of the 4XMM-DR9s FITS table as displayed by
      TOPCAT. Stack-summary rows have been selected using the expression
      `N\_CONTRIB$>$0' and are marked in blue. Observation-level rows can be
      identified through their OBS\_ID for example.}
    \label{fig:catview_topcat}
  \end{figure*}

  \begin{figure*}
    \includegraphics[width=\linewidth]{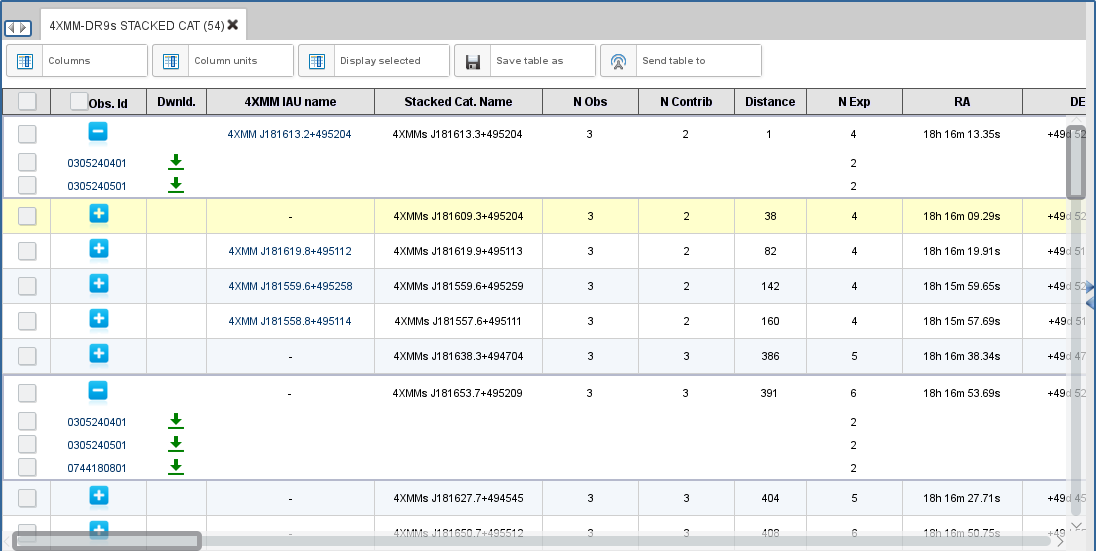}
    \caption{Screen shot of a 4XMM-DR9s search result in the XSA. The sources
      are listed with their 4XMMs IAU name (``Stacked Cat.\ Name'') and, if
      available, with the 4XMM IAU name of their counterpart in 4XMM-DR9. A
      mouse click on the plus sign opens the list of contributing observations
      with an option to download their XSA data, as shown for the first and
      seventh source in this example.}
    \label{fig:catview_xsa}
  \end{figure*}

  \subsection{Auxiliary images and light curves}
  \label{sec:auximas}

  Auxiliary images in PNG format are published for each 4XMM-DR9s source:
  X-ray images from the 4XMM-DR9s pipeline and an optical finding chart chosen
  from ESO Online Digitized Sky Survey, Pan-STARRS G, and skyMapper G. An
  X-ray light-curve plot is published for each multiply observed source. The
  production of these data is detailed in \citetads{2019A&A...624A..77T}. The
  following updates were applied for 4XMM-DR9s: The full-band X-ray image now
  shows all detected sources in the plotted region with dotted white marks in
  addition to the central source which is marked in blue. As before, diamonds
  with a fixed size are used for point-like sources and circles with the same
  radius as the core of the extent model for extended sources. In the light
  curves, an additional plot symbol has been introduced besides the open
  circle for sources with a VAR\_PROB above 0.01 and a filled circle for
  VAR\_PROB\,$\leq$\,0.01: If a source was fitted with zero counts for one out
  of two observations, VAR\_PROB remains undefined. The single valid flux
  value is marked by an x in these cases. If more than two observations are
  involved, the variability parameters are derived from those with defined
  fluxes. The flux axis is adjusted to minimum and maximum flux of the source
  including error bars, covers at least one magnitude, and has an absolute
  minimum of $10^{-17}\,\textrm{erg\,cm}^{-2}\,\textrm{s}^{-1}$.
  Figures~\ref{fig:auxlc} and ~\ref{fig:auximas} include an example set of the
  light curves and images that are produced for each 4XMM-DR9s catalogue
  source and available from the XSA interface (Sect.~\ref{sec:access}).

  \subsection{Auxiliary data}
  \label{sec:auxdata}

  Our catalogue pipeline generates auxiliary data in FITS format for each
  4XMM-DR9s stack, including composite images, exposure maps, and coverage
  maps of the stack. They are published via xmmssc.aip.de. All-EPIC mosaic
  images are generated for the five \textit{XMM-Newton} standard energy bands
  and for the full 0.2$-$12.0\,keV band. They are the sum of all observation-
  and instrument-level images.

  Two exposure maps include the total exposure at a given pixel, which is
  summed over all contributing observations. For each observation, the maximum
  exposure time of the three EPIC instruments per pixel is chosen. One
  exposure map gives the total un-vignetted exposure. The other map is created
  from exposure maps that are multiplied with the respective vignetting
  factor, which depends on the instrument, the detector coordinates, and the
  energy band.

  Three coverage maps are created from the detection masks of the individual
  exposures. The maps named "cov" show the valid pixels in a stack, "1"
  indicating the exposed pixels and "0" the un-exposed pixels. The maps
  ``nobs'' give the number of overlapping observations and ``nexp'' the number
  of overlapping exposures in each pixel,
  respectively. Figure~\ref{fig:covmaps} shows three example sets of coverage
  masks for different patterns of overlaps.

  \subsection{Catalogue layout}
  \label{sec:layout}

  Several sets of source parameters are derived from the simultaneous
  source-detection fits (Sect.~\ref{sec:software}). They are listed in the
  catalogue in 310 parameter columns and in several rows for each detected
  source. The first row includes the stack summary, i.e.\ the parameters
  derived from all images and all observations involved. In the following
  rows, source parameters are given for each contributing observation that was
  considered in the fit. They are derived from the subset of images taken
  during this specific observation. Since some parameter columns are only
  defined in the stack-summary rows and some only in the observation-level
  rows, these columns can be used to identify them. Stack-summary rows can be
  selected for example through their valid N\_OBS and N\_CONTRIB
  values. Observation-level rows are identified for example through their
  OBS\_ID and REVOLUT entries. Sources with a match in 4XMM-DR9 are
  characterised by a valid SRCID\_4XMMDR9 or their URL\_4XMMDR9. At
  observation level, they have a valid
  DETID\_4XMMDR9. Figures~\ref{fig:catview_topcat} and \ref{fig:catview_xsa}
  show usage examples of the FITS table in a FITS viewer (here: TOPCAT) and of
  the information in the online interface XSA.
  
  Table~\ref{tab:columns} summarises the 310 catalogue columns. Longer
  descriptions of the columns introduced with 3XMM-DR7s can be found in
  \citetads{2019A&A...624A..77T} and of the new and updated 4XMM-DR9s columns
  in Sect.~\ref{sec:newcolumns} of this paper.

  
  \longtab[1]{
    \centering
    \begin{longtable}{c@{~~}l@{}c@{~}c@{~~~}p{100mm}}
    \caption{Overview of the columns in 4XMM-DR9s.}
    \label{tab:columns}\\
    \hline\hline\noalign{\smallskip}
       No &  Column name              &  Units            &  Format   &  Description \\
       \hline\noalign{\smallskip}\endfirsthead
    \caption{Continued.} \\
    \hline\hline\noalign{\smallskip}
       No~~   &  Column name              &  Units            &  Format   &  Description \\
       \hline\noalign{\smallskip}
    \endhead
    \endfoot
    \hline
    \endlastfoot
        1~\,  &  IAUNAME                  &                   &  string   &  IAU name of the source                                                                             \\
        2~\,  &  SRCID                    &                   &  long     &  Identifier of the source                                                                           \\
        3$^o$ &  OBS\_ID                  &                   &  string   &  \textit{XMM-Newton} observation identification (observation-specific)                              \\
        4$^s$ &  N\_OBS                   &                   &  integer  &  Number of observations involved in the stack (stack-specific)                                      \\
        5$^s$ &  N\_CONTRIB               &                   &  integer  &  Number of observations in which the source was fitted (stack-specific)                             \\
        6~\,  &  N\_EXP                   &                   &  integer  &  Number of exposures for which the source was fitted                                                \\          
        7~\,  &  RA                       &  deg              &  double   &  Right ascension (J2000)                                                                            \\
        8~\,  &  DEC                      &  deg              &  double   &  Declination (J2000)                                                                                \\
        9~\,  &  RADEC\_ERR               &  arcsec           &  float    &  Square root of squared sum of 1$\sigma$ errors in RA and DEC                                       \\
       10~\,  &  LII                      &  deg              &  double   &  Galactic longitude                                                                                 \\
       11~\,  &  BII                      &  deg              &  double   &  Galactic latitude                                                                                  \\
       12~\,  &  X\_IMA                   &  pixel            &  float    &  X image coordinate                                                                                 \\
       13~\,  &  X\_IMA\_ERR              &  pixel            &  float    &  1$\sigma$ error on X\_IMA                                                                          \\
       14~\,  &  Y\_IMA                   &  pixel            &  float    &  Y image coordinate                                                                                 \\
       15~\,  &  Y\_IMA\_ERR              &  pixel            &  float    &  1$\sigma$ error on Y\_IMA                                                                          \\
       16~\,  &  DIST\_NN                 &  arcsec           &  float    &  Distance to the nearest neighbouring detection                                                     \\
       17~\,  &  N\_BLEND                 &                   &  integer  &  Number of simultaneously fitted sources                                                               \\
       18~\,  &  \multicolumn{2}{@{}l}{IAUNAME\_4XMMDR9}         &  string   &  IAU name assigned to the nearest unique 4XMM-DR9 source                                            \\
       19~\,  &  SRCID\_4XMMDR9           &                   &  long     &  Source identifier of the nearest unique source in 4XMM-DR9                                         \\
       20$^o$ &  DETID\_4XMMDR9           &                   &  long     &  Identifier of the associated 4XMM-DR9 detection in this OBS\_ID (observation-specific)             \\
       21~\,  &  RA\_4XMMDR9              &  deg              &  double   &  Mean right ascension (SC\_)RA of the associated 4XMM-DR9 source / detection                        \\
       22~\,  &  DEC\_4XMMDR9             &  deg              &  double   &  Mean declination (SC\_)DEC of the associated 4XMM-DR9 source / detection                           \\
       23~\,  &  POSERR\_4XMMDR9          &  arcsec           &  float    &  Statistical and systematic 4XMM-DR9 position error                                                 \\
       24~\,  &  DIST\_4XMMDR9            &  arcsec           &  double   &  Distance to the associated 4XMM-DR9 source                                                         \\
       25$^s$ &  \multicolumn{2}{@{}l}{NDETECT\_4XMMDR9}         &  short    &  Number of DR9 detections of the associated 4XMM-DR9 source                                     \\
       26~\,  &  EP\_FLUX                 &  erg/cm$^2$/s     &  float    &  All-EPIC flux                                                                                      \\
       27~\,  &  EP\_FLUX\_ERR            &  erg/cm$^2$/s     &  float    &  1$\sigma$ error on EP\_FLUX                                                                        \\
   28..37~\,  &  EP\_$n$\_FLUX            &  erg/cm$^2$/s     &  float    &  Total flux in energy band $n$                                                                      \\
              &  EP\_$n$\_FLUX\_ERR       &  erg/cm$^2$/s     &  float    &  1$\sigma$ error on EP\_$n$\_FLUX                                                                     \\
   38..73~\,  &  $II$\_FLUX               &  erg/cm$^2$/s     &  float    &  Total EPIC pn, MOS1, MOS2 flux                                                                     \\
              &  $II$\_FLUX\_ERR          &  erg/cm$^2$/s     &  float    &  1$\sigma$ error on $II$\_FLUX                                                                      \\
              &  $II$\_$n$\_FLUX          &  erg/cm$^2$/s     &  float    &  EPIC pn, MOS1, MOS flux in energy band $n$                                                         \\
              &  $II$\_$n$\_FLUX\_ERR     &  erg/cm$^2$/s     &  float    &  1$\sigma$ error on $II$\_$n$\_FLUX                                                                 \\
       74~\,  &  EP\_RATE                 &  counts/s         &  float    &  All-EPIC count rate                                                                                \\
       75~\,  &  EP\_RATE\_ERR            &  counts/s         &  float    &  1$\sigma$ error on EP\_RATE                                                                        \\
  76..111~\,  &  $II$\_RATE               &  counts/s         &  float    &  Total EPIC pn, MOS1, MOS2 count rate                                                               \\
              &  $II$\_RATE\_ERR          &  counts/s         &  float    &  1$\sigma$ error on $II$\_RATE                                                                      \\
              &  $II$\_$n$\_RATE          &  counts/s         &  float    &  EPIC pn, MOS1, MOS2 count rate in energy band $n$                                                  \\
              &  $II$\_$n$\_RATE\_ERR     &  counts/s         &  float    &  1$\sigma$ error on $II$\_$n$\_RATE                                                                 \\
      112~\,  &  EP\_CTS                  &  counts           &  float    &  All-EPIC number of counts                                                                          \\
      113~\,  &  EP\_CTS\_ERR             &  counts           &  float    &  1$\sigma$ error on EP\_CTS                                                                         \\
 114..119~\,  &  $II$\_CTS                &  counts           &  float    &  EPIC pn, MOS1, MOS2 number of counts                                                               \\
              &  $II$\_CTS\_ERR           &  counts           &  float    &  1$\sigma$ error on $II$\_CTS                                                                       \\
      120~\,  &  EP\_DET\_ML              &                   &  float    &  All-EPIC equivalent maximum detection likelihood                                                   \\
 121..138~\,  &  $II$\_DET\_ML            &                   &  float    &  EPIC pn, MOS1, MOS2 equivalent maximum detection likelihood                                        \\
              &  $II$\_$n$\_DET\_ML       &                   &  float    &  EPIC pn, MOS1, MOS2 detection likelihood in energy band $n$                                        \\
      139~\,  &  EXTENT                   &  arcsec           &  float    &  Extent radius                                                                                      \\
      140~\,  &  EXTENT\_ERR              &  arcsec           &  float    &  1$\sigma$ error on EXTENT                                                                          \\
      141~\,  &  EXTENT\_ML               &                   &  float    &  Likelihood of the detection being extended                                                         \\
 142..149~\,  &  EP\_HR$i$                &                   &  float    &  All-EPIC hardness ratio of energy bands $i$ and $i+1$                                              \\
              &  EP\_HR$i$\_ERR           &                   &  float    &  1$\sigma$ error on EP\_HR$i$                                                                       \\
 150..173~\,  &  $II$\_HR$i$              &                   &  float    &  EPIC pn, MOS1, MOS2 hardness ratio of energy bands $i$ and $i+1$                                   \\
              &  $II$\_HR$i$\_ERR         &                   &  float    &  1$\sigma$ error on $II$\_HR$i$                                                                     \\
\hline\\ \multicolumn{5}{@{}l@{}}{\tablefoot{$II$ denotes one of the EPIC instruments, abbreviated by PN, M1, M2. Energy band numbers $n$ run from one to five, $i$ from one to four. If a source parameter is defined for part of the rows only, it is indicated in the description and by a superscript `s' to the column number for stack-specific and `o' for observation-specific information.}\vspace*{-3mm}}\\
 174..191~\,  &  $II$\_EXP                &  s                &  float    &  PSF-weighted exposure in EPIC pn, MOS1, MOS2                                                       \\
              &  $II$\_$n$\_EXP           &  s                &  float    &  PSF-weighted exposure in energy band $n$ of EPIC pn, MOS1, MOS2                                    \\
 192..209~\,  &  $II$\_BG                 &  counts/pixel     &  float    &  EPIC pn, MOS1, MOS2 background map at the source position                                          \\
              &  $II$\_$n$\_BG            &  counts/pixel     &  float    &  EPIC pn, MOS1, MOS2 background map in energy band $n$                                              \\
      210~\,  &  EP\_ONTIME               &  s                &  float    &  Total good exposure time all-EPIC                                                                  \\
 211..213~\,  &  $II$\_ONTIME             &  s                &  float    &  Total good exposure time in EPIC pn, MOS1, MOS2                                                    \\
 214..216~\,  &  $II$\_PILEUP             &                   &  float    &  Estimate of the pile-up level in EPIC pn, MOS1, MOS2                                               \\
 217..219~\,  &  $II$\_MASKFRAC           &                   &  float    &  PSF-weighted detector coverage in EPIC pn, MOS1, MOS2                                              \\
      220~\,  &  DIST\_REF                &  arcmin           &  float    &  Distance to the reference coordinates of the field                                                 \\
      221$^o$ &  EP\_OFFAX                &  arcmin           &  float    &  Off-axis angle between source position and aim point all-EPIC (observation-specific)               \\
 222..224$^o$ &  $II$\_OFFAX              &  arcmin           &  float    &  Off-axis angle between source position and aim point in EPIC pn, MOS1, MOS2 (observation-specific) \\
 225..239$^o$ &  $II$\_$n$\_VIG           &                   &  float    &  EPIC pn, MOS1, MOS2 vignetting factor in energy band $n$ for the observation (observation-specific)\\
      240~\,  &  OVERLAP                  &                   &  boolean  &  Flag marking repeatedly observed sources                                                           \\
      241~\,  &  STACK\_FLAG              &                   &  short    &  Integer representation of the stack detection flags                                                \\
      242~\,  &  EP\_FLAG                 &                   &  string   &  All-EPIC detection flags                                                                           \\
 243..245~\,  &  $II$\_FLAG               &                   &  string   &  EPIC pn, MOS1, MOS2 detection flags                                                                \\
      246~\,  &  VAR\_CHI2                &                   &  float    &  Reduced $\chi^2$ of EPIC inter-observation variability                                             \\
 247..251~\,  &  VAR\_CHI2\_$n$           &                   &  float    &  Reduced $\chi^2$ of EPIC band $n$ inter-observation variability                                    \\
      252~\,  &  VAR\_PROB                &                   &  double   &  Probability that VAR\_CHI2 is consistent with constant EPIC flux                                   \\
 253..257~\,  &  VAR\_PROB\_$n$           &                   &  double   &  Probability that VAR\_CHI2\_$n$ is consistent with constant band $n$ flux                          \\
      258~\,  &  FRATIO                   &                   &  float    &  EPIC flux ratio                                                                                    \\
      259~\,  &  FRATIO\_ERR              &                   &  float    &  1$\sigma$ error on FRATIO                                                                          \\
 260..269~\,  &  FRATIO\_$n$              &                   &  float    &  EPIC band $n$ flux ratio                                                                           \\
              &  FRATIO\_$n$\_ERR         &                   &  float    &  1$\sigma$ error on FRATIO\_$n$                                                                     \\
      270~\,  &  FLUXVAR                  &                   &  float    &  Largest EPIC flux difference in terms of $\sigma$                                                  \\
 271..275~\,  &  FLUXVAR\_$n$             &                   &  float    &  Largest EPIC band $n$ flux difference in terms of $\sigma$                                         \\
      276~\,  &  \multicolumn{2}{@{}l}{CHI2PROB\_4XMMDR9}        &  double   &  Probability of the nearest unique 4XMM-DR9 source to be consistent with constant flux              \\
      277~\,  &  \multicolumn{2}{@{}l}{FVAR\_4XMMDR9}            &  double   &  Fractional intra-observation excess variance of the nearest unique 4XMM-DR9 source                 \\
      278~\,  &  \multicolumn{2}{@{}l}{FVARERR\_4XMMDR9}         &  float    &  1$\sigma$ error on FVAR\_4XMMDR9                                                                   \\
      279~\,  &  \multicolumn{2}{@{}l}{VAR\_FLAG\_4XMMDR9}       &  boolean  &  Intra-observation variability flag of the associated 4XMM-DR9 source                               \\
      280~\,  &  \multicolumn{2}{@{}l}{SUM\_FLAG\_4XMMDR9}       &  short    &  Summary quality flag of the associated 4XMM-DR9 source / detection                                 \\
      281$^s$ &  \multicolumn{2}{@{}l}{SUM\_FLAG\_MIN\_4XMMDR9}  &  short    &  Best quality flag of the detections of the nearest unique 4XMM-DR9 source (stack-specific)         \\
      282~\,  &  MJD\_FIRST               &  days             &  double   &  Modified Julian Date JD-2400000.5 of the observation start                                         \\
      283~\,  &  MJD\_LAST                &  days             &  double   &  Modified Julian Date JD-2400000.5 of the observation end                                           \\
      284$^o$ &  REVOLUT                  &  orbit            &  short    &  \textit{XMM-Newton} revolution number (observation-specific)                    \\
      285$^o$ &  PA\_PNT                  &  deg              &  float    &  Mean position angle of the spacecraft (observation-specific)                                       \\
      286$^o$ &  OBS\_CLASS               &                   &  short    &  Observation quality from 4XMM-DR9 screening (observation-specific)                                 \\
 287..289$^o$ &  $II$\_SUBMODE            &                   &  string   &  EPIC pn, MOS1, MOS2 submode (observation-specific)                                                 \\
 290..292$^o$ &  $II$\_FILTER             &                   &  string   &  EPIC pn, MOS1, MOS2 filter (observation-specific)                                                  \\
      293$^o$ &  ASTCORR                  &                   &  boolean  &  Flag: observation was astrometrically corrected (observation-specific)                             \\
      294$^o$ &  CC\_RAOFFSET             &  arcsec           &  double   &  \texttt{catcorr} shift of the right ascension (observation-specific)                               \\
      295$^o$ &  CC\_RAOFFERR             &  arcsec           &  double   &  1$\sigma$ error on CC\_RAOFFSET (observation-specific)                                             \\
      296$^o$ &  CC\_DEOFFSET             &  arcsec           &  double   &  \texttt{catcorr} shift of the declination (observation-specific)                                   \\
      297$^o$ &  CC\_DEOFFERR             &  arcsec           &  double   &  1$\sigma$ error on CC\_DEOFFSET (observation-specific)                                             \\
      298$^o$ &  CC\_ROT\_CORR            &  deg              &  double   &  \texttt{catcorr} shift of the position angle in the field (observation-specific)                   \\
      299$^o$ &  CC\_ROT\_ERR             &  deg              &  double   &  1$\sigma$ error on CC\_ROT\_CORR (observation-specific)                                            \\
      300$^o$ &  CC\_POFFSET              &  arcsec           &  double   &  \texttt{catcorr} total position shift of the field (observation-specific)                          \\
      301$^o$ &  CC\_POFFERR              &  arcsec           &  double   &  1$\sigma$ error on CC\_POFFSET (observation-specific)                                              \\
      302$^o$ &  CC\_REFCAT               &                   &  string   &  \texttt{catcorr} reference catalogue (observation-specific)                                        \\
      303$^o$ &  CC\_NMATCHES             &                   &  short    &  \texttt{catcorr} number of usable matches with the reference catalogue                             \\
 304..306$^o$ &  $II$\_BKG\_CRAREA        &  cts/s/arcsec$^2$ &  double   &  EPIC pn, MOS1, MOS2 background rate per area (observation-specific)                                \\
 307..309$^o$ &  $II$\_BKG\_CPROB         &                   &  double   &  EPIC pn, MOS1, MOS2 Cauchy probability derived from $II$\_BKG\_CRAREA (observation-specific)       \\
      310~\,  &  URL\_4XMMDR9             &                   &  string   &  Web-page URL of the nearest unique 4XMM-DR9 source                                                 \\
      \hline
    \end{longtable}
  }

\end{appendix}

\end{document}